\documentclass[a4paper]{article}

\usepackage[left=2.25cm, right=2.25cm, top=2cm, bottom=2cm, marginparwidth=1.7cm, marginparsep=2mm]{geometry}
\usepackage[utf8]{inputenc}
\usepackage[T1]{fontenc}
\usepackage{cmbright}
\usepackage{graphicx}
\usepackage[table, dvipsnames]{xcolor}
\usepackage{amsfonts}
\usepackage{amsmath}
\usepackage{amsthm}
\usepackage{mathtools, nccmath}
\usepackage{url}
\usepackage{booktabs}
\usepackage{tabularx}
\usepackage{nicefrac}
\usepackage{microtype}
\usepackage{booktabs}
\usepackage[hidelinks]{hyperref}
\hypersetup{
    colorlinks=true,
    linkcolor=blue,
    urlcolor=blue,
    citecolor=blue,
    anchorcolor=blue}
\usepackage[inline]{enumitem}
\usepackage{float}
\usepackage{wrapfig}
\usepackage[font=footnotesize, labelfont=bf]{caption}
\usepackage{subcaption}
\usepackage{adjustbox}
\usepackage{tikz}
\usetikzlibrary{shapes.geometric}
\newcounter{theo}[section] 
\renewcommand{\thetheo}{\arabic{theo}}

\usepackage{changepage}
\usepackage{etoc}
\usepackage{ragged2e}
\usepackage[framemethod=TikZ]{mdframed}
\usepackage[super, comma, numbers]{natbib}
\usepackage[symbol]{footmisc}
\usepackage{multicol}
\usepackage{lipsum}
\usepackage{pifont}
\usepackage{longtable}
\usepackage{multicol}
\usepackage{booktabs}
\usepackage{orcidlink}
\usepackage{rotating} 
\usepackage{makecell} 
\usepackage{multirow} 
\definecolor{orcidlogocol}{HTML}{000000}
\definecolor{changecolor}{HTML}{000000} 
\newcommand{\new}[1]{\textcolor{changecolor}{#1}}

\newcommand{\aref}[1]{\hyperref[#1]{Supplementary \autoref*{#1}}}
\newcommand{\norm}[1]{\left\lVert#1\right\rVert}

\definecolor{darkblue}{HTML}{1A254B}
\definecolor{lightblue}{HTML}{A7BED3}
\definecolor{blue}{HTML}{114083}
\definecolor{green}{HTML}{81B5AE}
\definecolor{pink}{HTML}{F2545B}
\definecolor{red}{HTML}{A4243B}
\definecolor{lightgray}{HTML}{C3BABA}
\definecolor{darkgray}{HTML}{9A8F97}

\title{\vspace{-30pt} \bf A multimodal foundation model \\ for tissue proteomics and morphology in cancer}

\date{}

\author{%
\small Benedikt von Querfurth$^{\orcidlink{0009-0001-0384-7923}\,1}$\footnotemark[1]\,, 
\small Lukas Klein$^{\orcidlink{0009-0000-3617-6276}\,1,2,3}$\footnotemark[1]\,, 
\small C\'edric Vincent-Cuaz$^{\orcidlink{0009-0004-5056-7977}\,4,5}$\,,\\ 
\small Eeshaan Jain$^{\orcidlink{0009-0002-6041-7193}\,1}$\,,
\small Yexiang Cheng$^{\orcidlink{0009-0005-2407-6653}\,1}$\,,
\small Johann Wenckstern$^{\orcidlink{0009-0007-1511-4659}\,1}$\,, 
\small Phil F. Cheng$^{\orcidlink{0000-0003-2940-006X}\,6}$\,, 
\small Petros Liakopoulos$^{\orcidlink{0009-0005-2015-6795}\,6}$\,,\\  
\small Olivier Michielin$^{\orcidlink{0000-0003-4926-6355}\,6}$\,,
\small Martina Haberecker$^{\orcidlink{}\,7}$\,, 
\small Andreas Wicki$^{\orcidlink{0000-0002-2924-8080}\,2,3}$\,, 
\small Pascal Frossard$^{\orcidlink{0000-0002-4010-714X}\,1,4}$\,, 
\small Charlotte Bunne$^{\orcidlink{0000-0003-1431-103X}\,1,8}$\footnotemark[3] \\[2ex]}

\begin{document}
\maketitle

\vspace{-30pt}
{\scriptsize \noindent
\footnotemark[1] These authors contributed equally. \\
\footnotemark[3] Correspondence to: \href{mailto:charlotte.bunne@epfl.ch}{Charlotte Bunne} (\url{charlotte.bunne@epfl.ch}). \\

\noindent $^{1}$ School of Computer and Communication Sciences, EPFL, Lausanne, Switzerland, \\
$^{2}$ Department of Medical Oncology and Hematology, University Hospital Zurich, Zurich, Switzerland, \\
$^{3}$ Faculty of Medicine, University of Zurich, Zurich, Switzerland,\\
$^{4}$ School of Engineering, EPFL, Lausanne, Switzerland, \\
$^{5}$ Department for BioMedical Research, University of Bern, Bern, Switzerland, \\
$^{6}$ Department of Oncology, Geneva University Hospitals, Geneva, Switzerland\\
$^{7}$ Department of Pathology and Molecular Pathology, University Hospital Zurich, Zurich, Switzerland, \\
$^{8}$ School of Life Sciences, EPFL, Lausanne, Switzerland. \\
}

\begin{abstract}
\looseness -1
Cancer diagnosis, prognosis, and treatment decisions increasingly rely on integrating tissue morphology with specific molecular readouts of the tumor and its microenvironment. H\&E histopathology provides scalable morphology, whereas multiplex spatial proteomics quantifies dozens of proteins \emph{in situ} with spatial context, but remains selectively deployed, often with different marker panels across cohorts. This creates a translational bottleneck: most patients lack high-depth molecular profiling, and existing models rarely integrate morphology and spatial molecular data in a single framework. We present VirTues-M2, a multimodal foundation model that jointly learns from H\&E and spatial proteomics using a novel vision transformer-based architecture. By encoding marker identity during pretraining, VirTues-M2 is robust to missing modalities, imperfect modality pairing or registration, and variable marker panels, enabling unified inference for tissue phenotyping, panoptic cell segmentation, virtual staining, annotation-free biomarker discovery, and reinforcement learning-based adaptive marker panel selection to reduce assay burden. 
As a concrete demonstration of translational utility, in a newly collected cohort of 1,053 treatment-naive non-small-cell lung cancer (NSCLC) patients with paired spatial proteomics and H\&E, VirTues-M2 discovers high-risk and low-risk tissue signatures that stratify overall survival \new{in both the discovery cohort and an independent validation cohort}, outperforming current state-of-the-art stratification schemes; all enabled by VirTues-M2’s ability to jointly analyze both modalities within a single model. 
\end{abstract}

\section{Introduction} \label{sec:1} 

Cancer diagnosis, prognosis, and treatment decisions increasingly integrate evidence from multiple, potentially complementary technologies, including tissue profiling, targeted molecular assays, and genomic testing, among others. 
Within this landscape, hematoxylin and eosin (H\&E) histopathology remains the most broadly available tissue readout in clinical practice, capturing morphological and architectural information at scale \citep{boehm2022harnessing}.
In parallel, highly multiplexed spatial proteomics (Method of the Year 2024 by Nature Methods \citep{editorial2024method}) extends H\&E by quantifying dozens of proteins \emph{in situ} while preserving spatial context. 
It enables molecularly specific readouts of cell identity, functional state, and microenvironment organization that complement morphology, including a more detailed characterization of the tumor immune microenvironment that impacts on immunotherapy response and resistance.
Despite rapid methodological progress and clear translational promise \citep{albrecht2024bridging,miglino2025feasibility,editorial2025making}, multiplex spatial proteomics is not yet part of routine clinical oncology due to cost, tissue requirements, operational complexity and the lack of a clear clinical framework for deployment. 
In addition, substantial variation in marker panels and acquisition protocols across studies and sites limits assay standardization and cross-site comparability, further hindering broad clinical deployment.
Consequently, most patients are evaluated with H\&E alone, and spatial proteomics is typically reserved for selected research cohorts, despite its complementary molecular specificity. 
This asymmetry creates a clear need for frameworks that can learn from cohorts with deep spatial readouts and transfer molecular and microenvironmental insight to routine H\&E under the heterogeneous and partially observed conditions of clinical practice.
Such a framework must address two coupled challenges:
First, comprehensive tissue characterization increasingly requires integrating complementary modalities: H\&E captures morphology and architecture, while spatial proteomics reveals molecular composition and cell state in their spatial context. 
Second, any clinically useful approach must translate insights from high-depth assays into tools that remain effective even when only routine histology is available.

\begin{figure}[t!]
    \centering
    \includegraphics[width=\linewidth,page = 1]{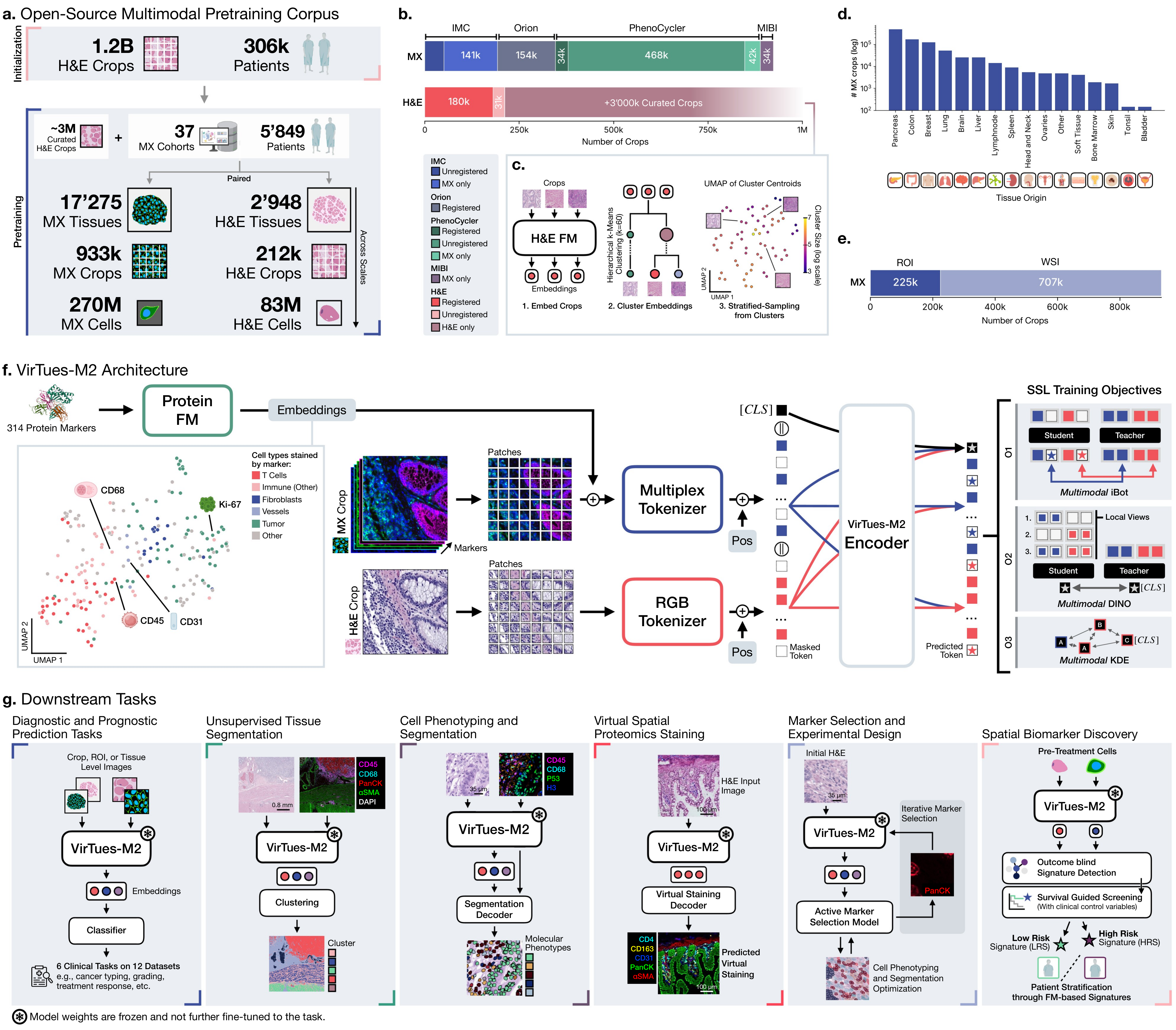}
    \caption{\textbf{Overview of the VirTues-M2 platform.} 
    \textbf{a}, Training data composition. Top: Hibou H\&E foundation model initialization pretraining data. Bottom: continual pretraining data showing multiplex (mx) with paired H\&E tissues and curated H\&E crops, with counts of tissues, crops, and cells.
    \textbf{b}, For continual pretraining, multiplex data distribution across technologies and registration status are shown as stacked bars. Bottom bars shows paired H\&E data with registration status as well as the additional 3 million curated crops as a fading bar.
    \textbf{c}, H\&E curation pipeline. From left to right: crops are embedded via an H\&E FM, clustered using hierarchical k-means, and stratified-sampled from clusters. UMAP visualization shows cluster centroids colored crop count per cluster.
    \textbf{d}, Bar chart showing crop distribution in log scale by tissue origin.
    \textbf{e}, Bar chart showing crop distribution by tissue format (WSI vs. \new{region-of-interest, ROI}).
    \textbf{f}, Architecture overview. Left: protein markers embedded via a protein FM, shown as UMAP colored by associated cell type. Center: multiplex and H\&E crops undergo modality-specific tokenization (blue and red pathways, respectively) with positional encodings before joint processing. Right: three SSL objectives shown, i.e., O1: multimodal iBOT (patch-level masked prediction, star symbols indicate predicted tokens), O2: multimodal DINO (crop-level \texttt{[CLS]} alignment), O3: multimodal KDE (letters indicate \texttt{[CLS]} tokens from different samples within the same batch).
    \textbf{g}, Downstream tasks evaluated with frozen encoder weights (indicated by snowflake): diagnostic and prognostic prediction, unsupervised tissue segmentation, cell phenotyping and segmentation, virtual spatial proteomics staining, marker selection for experimental design, and spatial biomarker discovery. 
    Created in BioRender.
}\label{fig:overview}
\end{figure}

A natural way to meet these challenges is to learn general-purpose tissue representations from large, unlabeled data that can transfer across cohorts, tasks, and input availability.
In this context, recent AI advances in self-supervised learning have led to the development of foundation models (FMs) pretrained on large corpora of unlabeled data \citep{bunne2024build, bommasani2021opportunities}.
These models learn general tissue representations that support diverse downstream tasks with minimal annotation, including prediction, stratification, and quantitative tissue phenotyping \citep{chen2024uni,vorontsov2024foundation,wenckstern2026virtual}. 
However, most existing FMs are built for a single modality (unimodal), either H\&E \citep{chen2024uni,vorontsov2024foundation,wang2024pathology} or specific multiplex assays \citep{wenckstern2026virtual,shaban2025foundation}, which limits their ability to learn how morphology relates to molecular cell states and microenvironment organization. 
Other non-foundational approaches \citep{li2026ai,valanarasu2025multimodal} use, due to previous unavailability of large multimodal patient cohorts, virtual staining to infer multiplex signals for a fixed panel of protein markers and then tailor the downstream pipeline to particular tasks, rather than training a generalist representation jointly across modalities.
This motivates a single multimodal FM that jointly learns from routine histology and spatial proteomics, so that specific molecular insights can inform standard analysis of slides while enabling unified inference when both modalities are available.

To this end, we introduce the multimodal Virtual Tissues platform (VirTues-M2), a novel FM for tissue biology that jointly learns from routine histopathology and multiplex spatial proteomics within a single architecture. 
We first address a primary bottleneck in the field concerning insufficient scale and diversity of multimodal training data by assembling the largest open-source multiplex spatial proteomics collection with partially paired H\&E images, spanning \new{5,849} patients across \new{37} studies, \new{17,275} multiplex tissues, acquired with multiple spatial proteomics technologies including cyclic approaches such as PhenoCycler (formerly CODEX) \citep{black2021codex} and one-shot high-plex approaches such as Orion \citep{lin2023high}, as well as imaging mass cytometry (IMC) \citep{giesen2014highly} \new{and multiplexed ion beam imaging (MIBI)} \citep{keren2019mibi} technologies, covering diverse cancer indications and tissue preparations (\autoref{fig:overview}a-e).
VirTues-M2 is enabled by a new vision transformer-based architecture that can embed heterogeneous tissue modalities into a unified representation space, while remaining robust to incomplete pairing, imperfect alignment, and substantial variability in spatial proteomics marker panels (\autoref{fig:overview}f). 
By incorporating protein marker identity directly into the model and pretraining across diverse cohorts and marker panels, VirTues-M2 harmonizes heterogeneous spatial proteomics assays into a common representation, enabling cross-study comparability while learning tissue features that connect morphology to molecular phenotype.

\new{These capabilities translate into practical advances that are difficult to achieve with unimodal models \citep{chen2024uni,vorontsov2024foundation,wang2024pathology,wenckstern2026virtual,shaban2025foundation} or late fusion schemes \citep{chen2025visual, vaidya2025molecular, jaume2024transcriptomics} (\autoref{fig:overview}g). VirTues-M2 supports a single workflow across H\&E-only, spatial proteomics-only, and joint multimodal inference. It improves \textit{one-shot} cell segmentation, typing, and downstream prediction by capturing both tissue architecture and molecular cell state. In addition, it enables cross-modal inference, including virtual staining from H\&E and prediction of unmeasured markers from partial panels, reducing assay burden when tissue, cost, or turnaround time constrain profiling. 
Coupled with a reinforcement learning (RL)-based marker-selection strategy, the model adaptively recommends which markers to measure starting from an H\&E slide, guiding clinicians and translational laboratories toward minimal, informative assays that approach full-panel performance. A systematic scaling law analysis shows continued gains with increasing pretraining data volume, while larger models do not consistently improve performance at the current data scale. The analysis further demonstrates that the design of the VirTues-M2 transformer enables learning from unpaired observations and paired but unregistered measurements, extending the usable training corpus beyond precisely registered multimodal data and informing priorities for future data generation.
Finally, VirTues-M2 enables annotation-free discovery of spatial biomarkers that capture complementary prognostic information in tissue morphology and protein expression within a single pretrained model. In a large treatment-naive NSCLC cohort with paired IMC and H\&E\citep{cords2024cancer}, combining these signatures improves patient stratification over either modality alone and established clinical predictors. The learned signature classifiers transfer without retraining to an independent NSCLC cohort, with the strongest performance arising from joint analysis of both modalities within VirTues-M2’s native multimodal representation.}

\section{Results} \label{sec:2}

\subsection*{A multimodal foundation model architecture for tissue biology} \label{sec:2.1}

Most tissue foundation models remain unimodal: H\&E models  capture relevant molecular variation only insofar as it is reflected in the staining \citep{vorontsov2024foundation,chen2024uni,nechaev2024hibou,shaban2025foundation}, while spatial proteomics models learn from measured protein expression without access to the complementary morphological information in H\&E \citep{wenckstern2026virtual,sorin2023single,kraus2024masked}. Neither learns the correspondence between these modalities during pretraining; a relationship so far mostly investigated through downstream analyses \citep{jaume2024hest}. Existing pipelines that consider both modalities impute a fixed marker panel from H\&E or fine-tune modality-specific encoders for individual endpoints \citep{li2026ai,valanarasu2025multimodal}. These approaches resemble early vision-language systems that combined separately pretrained components \citep{ramesh2021zero,liu2023visual,alayrac2022flamingo}, rather than learning shared representations through native multimodal pretraining and token-level fusion as done in state-of-the-art vision-language models \citep{deng2025emerging,team2024chameleon}. Tissue biology, however, presents an additional challenge: the molecular modality is a complementary spatial measurement rather than a natural language description. Integrating it with morphology requires accounting for its spatial structure and substantial heterogeneity while learning cross-modal relationships from predominantly unpaired data.

VirTues-M2 closes this gap by enabling joint multimodal pretraining across H\&E and multiplex spatial proteomics within a single architecture, so that cross-modal relationships are learned as part of the representation itself rather than added as a late-stage fusion step. To enable such joint multimodal pretraining, we assembled the largest \textit{open-source} multiplex (mx) dataset with partially paired H\&E images, focusing on cancer tissues (Methods). 
This corpus encompasses \new{5,849} patients across \new{37} cohorts, comprising \new{17,275} multiplex and \new{2,948} paired H\&E tissue samples (\autoref{fig:overview}a), including a newly collected large patient cohort from the University Hospital Zurich (USZ).
These tissues were processed into ${\sim}$\new{933,000} multiplex and \new{212,000} H\&E crops with varying levels of alignment: pairs registered pixel-wise, unregistered but paired samples (from the same tissue or patient), or unpaired H\&E or multiplex crops. 
This heterogeneous alignment structure reflects data availability within clinical workflows, where perfect registration between modalities is often infeasible or impractical. 
Further, the multiplex data spans \new{four} acquisition technologies, \new{more than thirteen} primary cancer types, and two tissue formats, i.e., WSI and region-of-interest (ROI), enabling generalization across heterogeneous experimental settings (\autoref{fig:overview}b-d). Also, to strengthen unimodal histopathology coverage, we augmented our multimodal pretraining corpus with three million additional unpaired H\&E crops, ensuring robust performance when only H\&E is available (\autoref{fig:overview}b). Because large H\&E collections contain many near-duplicate regions, we reduced redundancy by grouping crops by visual similarity and sampling across groups to prioritize diverse, informative morphologies, thereby broadening the histologic coverage seen during pretraining (\autoref{fig:overview}c) \citep{chen2025revisiting}. 

\begin{figure}[t!]
    \centering
    \includegraphics[width=\linewidth,page = 2]{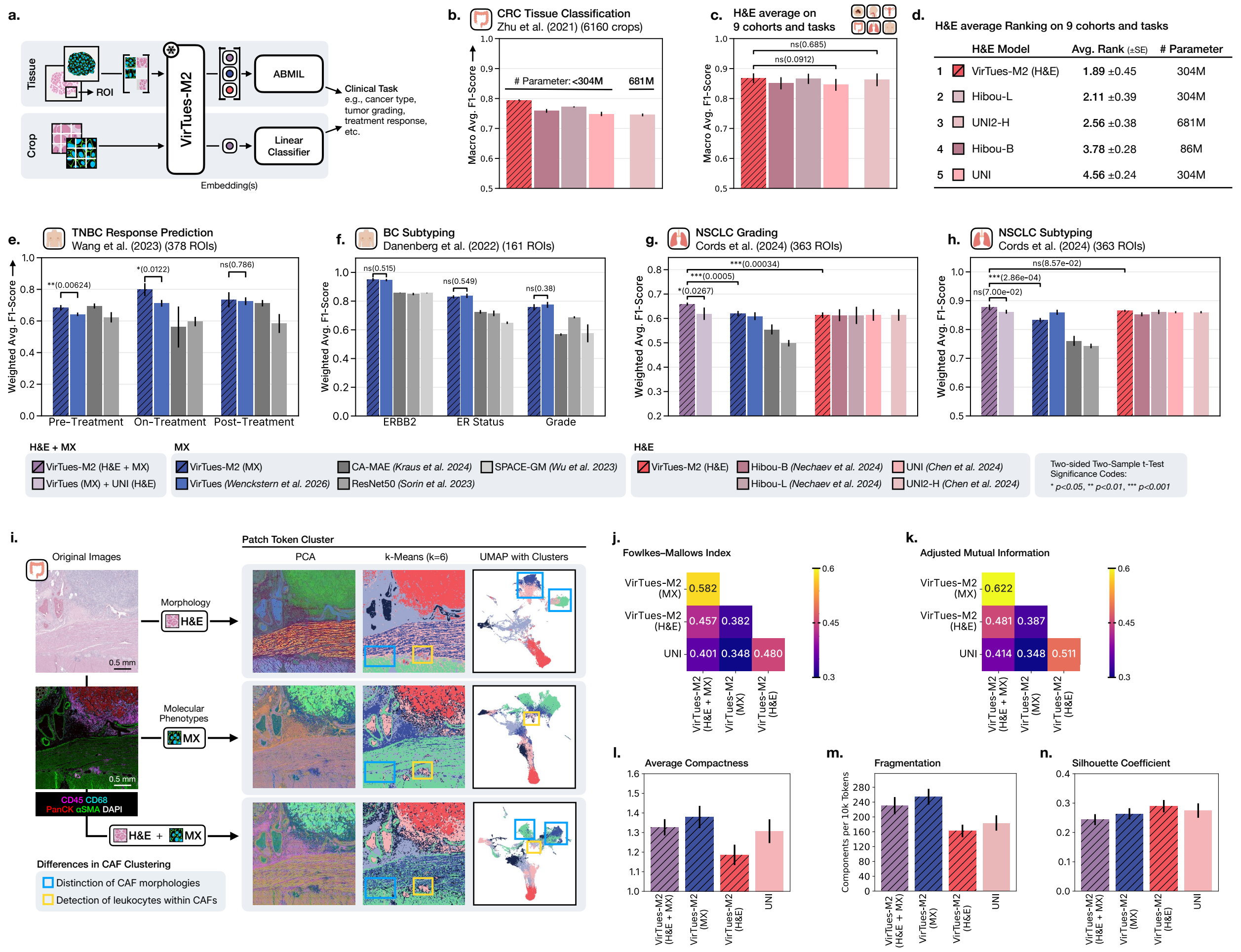}
    \caption{\textbf{VirTues-M2 embeddings enable accurate clinical predictions and capture complementary tissue features across modalities.} 
    \textbf{a}, Clinical prediction evaluation framework. After embedding each crop, crop-level tasks use a linear classifier on single embeddings and ROI-or tissue-level tasks aggregate embeddings via ABMIL.
    \textbf{b}, Macro-averaged F1 scores for \new{colorectal cancer (CRC) tissue classification across H\&E FMs on the \citet{zhu2021hard} cohort.} Models are grouped by amount of parameters, highlighting that UNI2-H (681 million parameters) is more than double in size than the other models.
    \textbf{c}, Macro-averaged F1 scores aggregated across nine H\&E benchmark datasets spanning diverse cancer types and prediction tasks. We observe that VirTues-M2 performs statistically similar ($p=0.685$ and $p=0.0912$) as the other unimodal models.
    \textbf{d}, Average ranking of H\&E FMs across nine benchmark datasets with the amount of parameters per model. Ordered by average ranking.
    \textbf{e}, F1-Score of multiplex (mx) FMs on Triple-negative breast cancer (TNBC) response prediction. P-values ($p=0.006$, $p=0.012$ and $p=0.786$) indicate significant difference on two out of three task between VirTues-M2 and VirTues. \new{VirTues-M2 scores an AUC of 0.760 ± 0.023 on the pre-treatment timepoint.}
    \textbf{f}, Macro-average F1 score across several diverse breast cancer (BC) subtyping tasks on the ROI level. P-values show no significant difference between VirTues-M2 and VirTues model performance.
    \textbf{g}, Weighted-average F1 score on a Non-Small Cell Lung Cancer (NSCLC) grading task comparing all multimodal and unimodal models. VirTues-M2 performs significant (two-sided two-sample t-test) better than all other unimodal or late-fusion multimodal versions.
    \textbf{h}, Weighted-average F1 score on a NSCLC subtyping task comparing all multimodal and unimodal models. Multimodal VirTues-M2 performs significant ($p=2.86e-04$) better than with multiplex data only.
    \textbf{i}, \new{PCA ($k=3$, mapped to RGB) and k-means ($k=6$) clustering of H\&E, multiplex, and multimodal patch embeddings from \citet{lin2023high}} visualized spatially and in UMAP space. H\&E embeddings capture morphological subtypes (e.g., blue box), multiplex embeddings identify molecular phenotypes (e.g., orange box), and multimodal embeddings integrate both feature types. Colors in the UMAP indicate the K-means clusters of the patch tokens.
    \textbf{j}, Fowlkes-Mallows index measuring cluster similarity (k-means with $k=6$). Brighter color indicates higher similarity.
    \textbf{k}, Adjusted Mutual Information measuring cluster similarity. Brighter color indicates higher similarity.
    \textbf{l}, Average cluster compactness across embedding types. Lower values indicate more compact clusters. H\&E clusters are spatially compact, multiplex clusters more dispersed, and multimodal clusters intermediate. 
    \new{\textbf{m}, Fragmentation score, the number of spatially connected components per 10k patch tokens; higher values indicate clusters scattered across many small regions rather than few contiguous ones. MX-based clusters are more fragmented than H\&E-based ones.}
    \new{\textbf{n}, Silhouette coefficient, measuring how well each patch token fits its own cluster relative to the nearest neighboring cluster ($-1$ to $1$). Values are comparable across all settings, indicating that the higher fragmentation of mx-based clusters does not reflect poorer cluster separation in embedding space.} 
    Error bars show 95\% bootstrapped ($n=1000$) CI. Statistical two-sample t-tests are two-sided.
    Created in BioRender.
}\label{fig:prediction_tasks}
\end{figure}

On the learning side, VirTues-M2 introduces a \emph{single-encoder}, \emph{early-fusion} architecture that jointly learns from multiplex spatial proteomics and H\&E histology (\autoref{fig:overview}f). In contrast to approaches that pretrain separate unimodal backbones and fuse embeddings only downstream, VirTues-M2 maps each modality to patch tokens and processes all tokens jointly throughout the network. This enables cross-modal relationships to be learned during pretraining and yields a single backbone for H\&E-only, multiplex-only, or joint inference (Methods). To initialize this backbone, we used Hibou, a large pretrained H\&E model that captures common morphological patterns in routine slides \citep{nechaev2024hibou}, followed by multimodal continual pretraining \new{(see \autoref{suppl:fig:scaling_laws}d and \autoref{fig:prediction_tasks}b for ablations on the size and choice of initialization model)}.
A key technical barrier to early fusion is the heterogeneity of multiplex assays. VirTues-M2 was pretrained on \new{314} distinct protein markers across \new{37} integrated clinical studies. Marker panels vary substantially across cohorts in both composition and channel order, requiring the model to identify which protein each channel measures to learn biologically meaningful representations. Existing approaches often assume fixed marker dictionaries \citep{shaban2025foundation,kraus2024masked,li2026ai} or enforce a specific ordering \citep{kraus2024masked}, limiting robustness to alternative panel configurations. VirTues-M2 addresses this with marker-aware multiplex tokenization (\autoref{suppl:fig:tokenizer}). Following \citet{wenckstern2026virtual}, we condition multiplex patch embeddings on marker identity using protein sequence embeddings from ESM-2 \citep{lin2023evolutionary}, enabling channels to be interpreted by biological meaning rather than index position. This design supports learning across heterogeneous panels and accommodates markers not observed during pretraining by embedding their protein sequences and incorporating them as additional channels.

The challenge of training VirTues-M2 encoder then requires tailored self-supervised objectives that learn both local, spatially resolved features and global, slide-level semantics under missing modalities and imperfect alignment. 
To this end, we combine complementary objectives at two scales adapted from DINOv2 \citep{oquab2023dinov2} (\autoref{fig:overview}f). 
At the patch level, we utilize a masked-patch prediction objective (multimodal iBOT \citep{zhou2021ibot}) to learn stable local representations that support spatial tasks. 
While at the crop level, we adopt a self-distillation objective that matches global representations across augmented views and modalities (multimodal DINO \citep{caron2021emerging}), supplemented by a kernel density estimation (KDE) objective to encourage well-structured global embeddings for patient-level prediction. 
To accommodate fully registered pairs, loosely matched pairs, and unimodal samples, we incorporate modality dropout so training remains effective when one modality is absent, while joint token processing in the shared encoder promotes integration when both modalities are available (Methods, \autoref{suppl:fig:arch}).
After training, we benchmark VirTues-M2 by freezing the pretrained backbone and training only task-specific heads across a suite of clinically and biologically motivated tasks, including diagnostic and prognostic prediction, panoptic cell segmentation, virtual staining, spatial biomarker discovery, and minimal marker set design (\autoref{fig:overview}g).

\subsection*{VirTues-M2 maintains competitive unimodal performance and gains from complementary multimodal input} \label{sec:2.2}

\new{Across a broad set of clinical prediction tasks, we evaluated VirTues-M2 with H\&E-only, multiplex-only, and joint inputs against established modality-specific foundation models and representation learning baselines. When a target property is fully determined by information available in one modality, adding another modality is not expected to improve prediction; competitive performance with specialized unimodal models is therefore the relevant objective. Where morphology and protein expression provide complementary information about the target, joint input is expected to offer additional benefit. Our evaluation examines both settings, assessing performance with individual modalities and the gains from their integration.}
Benchmarks span crop-, region of interest (ROI)-, and tissue-level settings to probe both local and global embedding quality (\autoref{fig:prediction_tasks}a). 
Crop-level tasks used linear probing on frozen embeddings, whereas ROI- and tissue-level tasks used attention-based multiple instance learning (ABMIL) \citep{ilse2018attention} to aggregate embeddings (Methods).

On H\&E, we benchmarked VirTues-M2 across nine \new{external} cohorts\new{, none of which appear in the pretraining corpus,} spanning diverse cancer types and clinical tasks \citep{barbano2021unitopatho,brancati2022bracs,nechaev2025spider,zhu2021hard,han2022wsss4luad, aresta2019bach}. 
\new{In colorectal cancer (CRC) tissue classification, distinguishing between four colorectal tissue categories, VirTues-M2 achieved the highest performance (macro-average F1 of $0.79$) compared with established H\&E foundation models including UNI \citep{chen2024uni} and Hibou \citep{nechaev2024hibou} across model sizes}, outperforming UNI2-H (macro-average F1 of $0.75$) despite fewer parameters (\autoref{fig:prediction_tasks}b). 
Across all nine H\&E benchmarks, VirTues-M2 ranked first overall and achieved comparable average performance to larger state-of-the-art models ($p=0.685$) (\autoref{fig:prediction_tasks}c,d; \autoref{suppl:fig:he_all}; \autoref{suppl.tab:pred_he_f1}; \autoref{suppl.tab:pred_he_acc}).

On multiplex spatial proteomics, we evaluated two breast cancer cohorts including triple-negative breast cancer (TNBC) and diverse breast cancers (BC) across six clinical prediction tasks (\autoref{fig:prediction_tasks}e,f; \autoref{suppl.tab:wang}; \autoref{suppl.tab:danenberg}) \citep{wang2023spatial,danenberg2022breast}. We compared VirTues-M2 against multiplex foundation models (VirTues \citep{wenckstern2026virtual} and the channel-agnostic masked autoencoder (MAE), CA-MAE \citep{kraus2024masked}) and supervised baselines (ResNet50 \citep{sorin2023single} and SPACE-GM \citep{wu2022space}). \new{VirTues-M2 matched the leading multiplex foundation model on most tasks, outperforming it on response prediction at both the pre- and on-treatment time points ($p=0.006$ and $p=0.012$)}.

To evaluate multimodal performance, we benchmarked models on an NSCLC cohort with paired but unregistered tissue cores, assessing grading (\autoref{fig:prediction_tasks}g) and subtyping (\autoref{fig:prediction_tasks}h; \autoref{suppl.tab:multimodal}). 
We compared VirTues-M2 receiving only an H\&E or multiplex input versus having access to both modalities. 
In addition, we include a late-fusion baseline that combines embeddings from independently pretrained H\&E and multiplex state-of-the-art unimodal foundation models (UNI + VirTues), reflecting common strategies for multimodal integration with disjoint encoders. 
In unimodal settings, H\&E and multiplex models achieved comparable performance for both tasks, and VirTues-M2 matched these baselines when restricted to a single modality. 
When provided with both modalities, VirTues-M2 outperformed its unimodal counterparts as well as the late-fusion baseline on NSCLC grading \new{($p=3.40e{-4}$ versus H\&E-only; $p=5.00e{-4}$ versus multiplex-only)} and outperformed late fusion \new{($p=0.0267$)}. 
On NSCLC subtyping, VirTues-M2 using both modalities outperformed VirTues-M2 using multiplex alone \new{($p=2.86e{-4}$)}.

\new{
Because comprehensive, well-registered spatial proteomics data remain costly and scarce, we systematically examined how VirTues-M2's performance depends on pretraining data volume, composition, and model capacity to determine where future gains are best invested:
The pretraining scaling law analysis showed that performance scales consistently with data volume (\autoref{suppl:fig:scaling_laws}a). However, balanced sampling of H\&E and multiplex images is important to overall downstream performance (\autoref{suppl:fig:scaling_laws}b). The analysis further confirms that the current model size is optimized to the training data available, i.e., \autoref{suppl:fig:scaling_laws}d shows that model capacity is capped by the amount of multiplex data available. 
Lastly, the multimodal scaling law analysis underlines VirTues-M2's strength of learning robust representations from a heterogeneity of multimodal data typical of clinical cohorts: The model exhibits continual performance increase even when training only on non-registered and unpaired H\&E-multiplex data instead of perfectly registered H\&E and multiplex images (see Supplementary Information for details).
}

\subsection*{Multimodal embeddings recover complementary morphologic and molecular tissue structure}

To characterize the information captured by VirTues-M2 representations across modalities, we performed unsupervised tissue segmentation using patch embeddings from H\&E, multiplex, or joint inputs. 
\new{We embedded all patches from a ROI and visualized the spatial organization of the learned features by projecting embeddings onto the first three principal components, mapped to RGB, and by clustering them with k-means ($k=6$) (\autoref{fig:prediction_tasks}i; \autoref{suppl:fig:kmeans_cluster}).}

H\&E-derived embeddings primarily separated morphologic compartments, including tumor regions, vasculature, and fibrotic or stromal areas (\autoref{fig:prediction_tasks}i, top), whereas multiplex-derived embeddings emphasized molecular defined single-cell phenotypes, conditioned on specifically characterized neighborhoods (\autoref{fig:prediction_tasks}i, middle). 
Several broad structures, such as the tumor compartment, were delineated by both modalities. 
In contrast, joint multimodal embeddings preserved the compartment-level organization captured by H\&E while simultaneously resolving the finer-grained, molecularly defined states highlighted by multiplex, effectively integrating complementary information from both views into a single representation (\autoref{fig:prediction_tasks}i, bottom).
In one example region (\autoref{fig:prediction_tasks}i, blue box), H\&E-based clustering separated two morphologically distinct stromal compartments consistent with cancer-associated fibroblast (CAF)-rich areas, whereas multiplex-based clustering highlighted an immune-enriched population positioned between these stromal regions (orange box). 
The multimodal embedding recovered both patterns in a single clustering solution: the morphologically distinct upper and lower CAF populations as separate clusters, as well as the intervening leukocyte population as a distinct cluster (leukocyte are also vastly populated around the tumor, see CD45 marker in the original multiplex).

To quantify correspondence between clustering solutions, we computed the Fowlkes-Mallows index \citep{fowlkes1983method} and adjusted mutual information \citep{vinh2009information} between clusters derived from VirTues-M2 embeddings (H\&E, multiplex, multimodal) and UNI H\&E embeddings (\autoref{fig:prediction_tasks}j,k).
UNI and VirTues-M2 (H\&E) clusterings were similar, H\&E- and multiplex-derived clusterings dissimilar, and multimodal clusters similar to both.
\new{
Spatially, H\&E-derived clusters were more compact (\autoref{fig:prediction_tasks}l) and formed fewer connected components (\autoref{fig:prediction_tasks}m) than multiplex-derived clusters.
Silhouette coefficients were comparable across all four settings (\autoref{fig:prediction_tasks}n), so the greater fragmentation of multiplex clusters does not reflect poorer separation in embedding space.
Multiplex embeddings therefore capture distributed single-cell phenotypes that intermix spatially, while H\&E embeddings resolve larger-scale tissue architecture.
Multimodal embeddings recover both scales, clustering in agreement with each unimodal solution rather than collapsing onto either.
}

\begin{figure}[t!]
    \centering
    \includegraphics[width=\linewidth,page = 3]{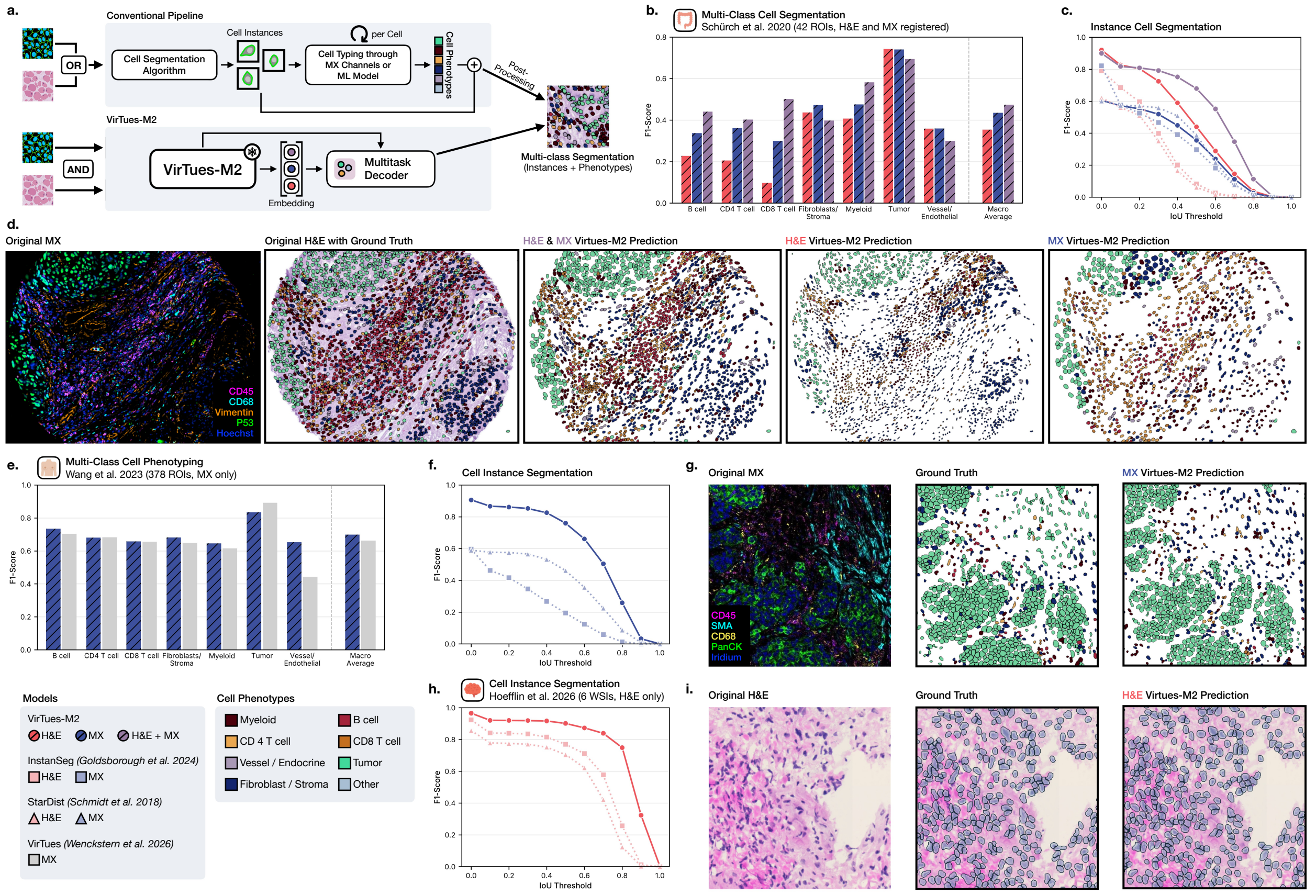}
    \caption{\textbf{Multimodal VirTues-M2 enables accurate cell instance segmentation and phenotyping in one end-to-end pipeline.}
    \textbf{a}, Comparison of conventional segmentation pipelines versus end-to-end VirTues-M2 approach with frozen encoder and trainable multitask decoder.
    \new{
    \textbf{b--d}, Benchmarking of multi-modal panoptic segmentation (H\&E, multiplex (mx), and multimodal embeddings): \textbf{b}, cell phenotyping performance across seven cell types; \textbf{c}, cell instance segmentation performance (F1-score) at varying IoU matching thresholds; \textbf{d}, qualitative comparison showing that H\&E provides accurate instance boundaries while mx enables correct phenotype assignment, with multimodal embeddings combining both strengths. Filled cell colors denote different cell phenotypes.
    \textbf{e--g}, Panoptic segmentation on multiplex images from the \citet{wang2023spatial} cohort: \textbf{e}, phenotyping performance against VirTues, which VirTues-M2 slightly outperforms on macro average while matching it on most phenotypes; \textbf{f}, segmentation performance against InstanSeg and StarDist, which VirTues-M2 outperforms; \textbf{g}, qualitative example compared to ground truth segmentations.
    \textbf{h,i}, Cell instance segmentation on H\&E images from the
    \citet{hoefflin2026spatial} brain tissue cohort: \textbf{h}, quantitative performance, where VirTues-M2 consistently outperforms its counterparts for brightfield cell segmentation across all IoU thresholds; \textbf{i}, qualitative example.
    }
}\label{fig:segmentation}
\end{figure}

\subsection*{VirTues-M2 enables \textit{one-shot} cell phenotyping and segmentation} \label{sec:2.3}

Cell instance segmentation and phenotyping are foundational steps in quantitative tissue analysis for multiplexed imaging, yet in clinical and research practice they are typically performed with multi-stage pipelines that separate boundary detection from phenotype assignment. 
In multiplex imaging, segmentation is commonly driven by generic image processing or nucleus-based heuristics, followed by marker-intensity quantification and gating to assign cell types (\autoref{fig:segmentation}a), resulting in the use of separate task-specific models \citep{pachitariu2025cellpose, marks2025cellsam}. These staged workflows require assay-specific tuning, propagate errors from segmentation to phenotyping, and make it difficult to exploit the complementarity between morphology and molecular specificity when both modalities are available.

VirTues-M2 enables \textit{one-shot} panoptic inference that predicts cell instances and phenotypes jointly within a single model, incorporating both morphology from H\&E as well as molecular information from multiplex data. 
\new{To evaluate this setting, we used VirTues-M2 as a frozen backbone within a fused CellViT \citep{horst2024cellvit} and InstanSeg-style \citep{goldsborough2024instanseg} architecture and fine-tuned only the decoder head to jointly segment cells and assign phenotypes (Methods, \autoref{suppl:fig:cellvit}).}

\new{
We evaluated cell phenotyping on the registered \citet{schurch2020coordinated} cohort, which was excluded from training all segmentation decoders, thereby testing transfer to a separate cohort without decoder retraining. Per-class F1 scores were obtained by convolving the predicted phenotyping mask with the ground truth cell instance segmentation mask (\autoref{fig:segmentation}b), which allowed us to benchmark the segmentation model across several modality configurations. The strongest macro-averaged performance came from VirTues-M2 with both modalities, suggesting an advantage over the corresponding unimodal variants.
For cell instance segmentation, we measure the F1-Score across different Intersection over Union (IoU) thresholds (\autoref{fig:segmentation}c).
VirTues-M2 alone not only outperforms stardist \citep{schmidt2018} and InstanSeg \citep{goldsborough2024instanseg} in the H\&E setting and matches their performance in the unimodal multiplex case, but performs best in a multimodal setting with an F1-score at an IoU at 0.5 of 0.650 versus 0.452 using only H\&E and 0.349 using only multiplex.}
Qualitative examples highlight why multimodality matters for one-shot panoptic inference (\autoref{fig:segmentation}d; \autoref{suppl:fig:seg_examples}). 
Multiplex-only inference recovered phenotype-specific signal but showed reduced accuracy in delineating individual cell boundaries or missing cells entirely.
H\&E-only inference delineated boundaries accurately but more frequently confused phenotypes. 
Joint multimodal inference combined accurate boundary delineation with improved phenotype assignment, yielding more coherent panoptic maps consistent with the quantitative results (\autoref{fig:segmentation}b-d).

\new{To assess the cross-cohort panoptic segmentation capabilities of our segmentation decoder, we further benchmarked cell phenotyping and instance segmentation on TNBC patient cohort imaged with IMC \citep{wang2023spatial}. Compared to the phenotype segmentation performance of VirTues \citep{wenckstern2026virtual}, VirTues-M2 matches the performance across all cell phenotypes (\autoref{fig:segmentation}e) and outperforms the instance segmentation baselines (\autoref{fig:segmentation}f).}
Similarly, using only H\&E as the input, the VirTues-M2 segmentation head outperforms stardist \citep{schmidt2018} and InstanSeg \citep{goldsborough2024instanseg} (\autoref{fig:segmentation}h) and is able to detect small cells that where not detected in the original authors annotations (\autoref{fig:segmentation}i).

\begin{figure}[t!]
    \centering
    \includegraphics[width=\linewidth,page = 4]{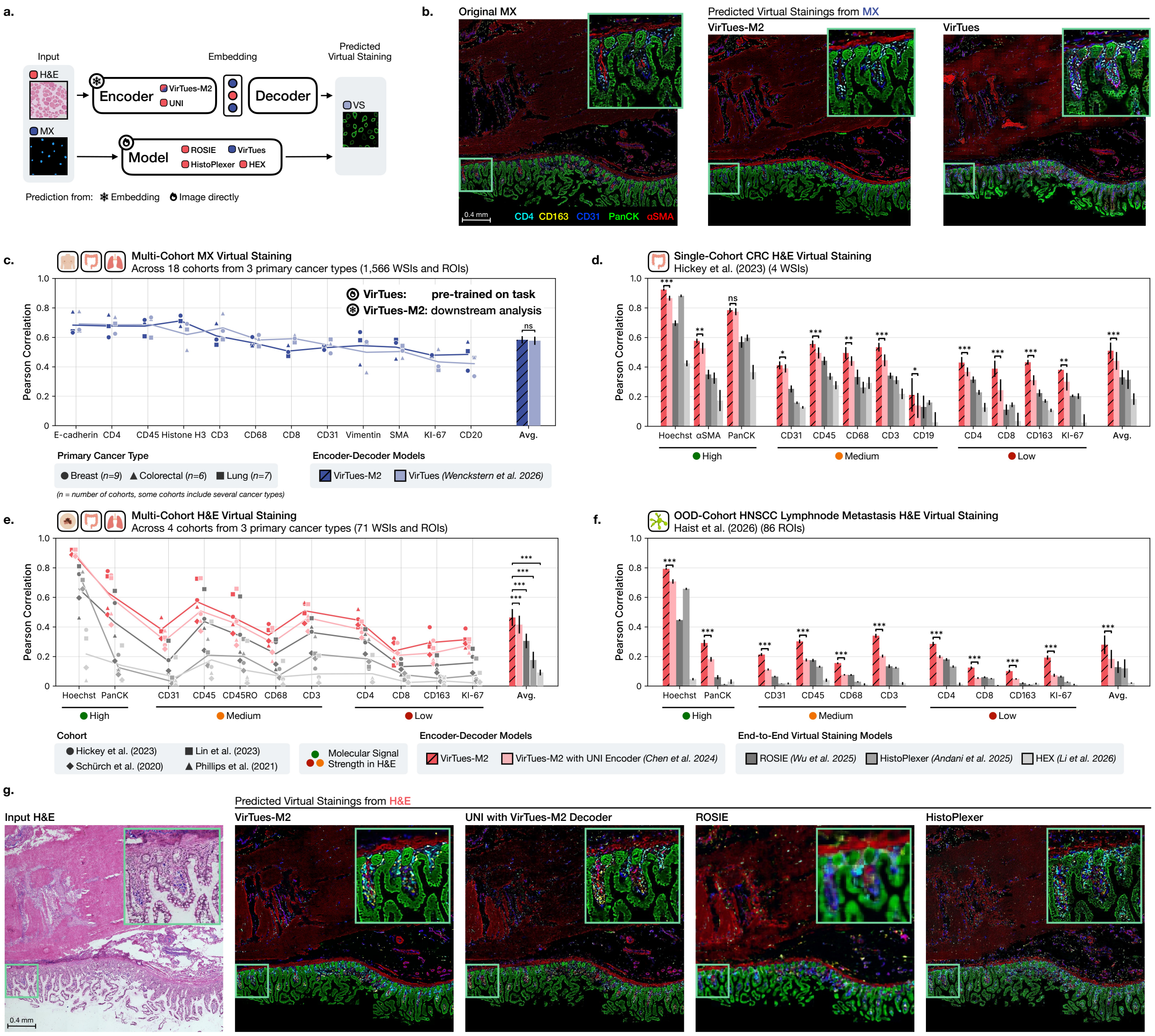}
    \caption{\textbf{VirTues-M2 enables accurate virtual staining from multiplex and H\&E inputs, outperforming supervised approaches in cross-cohort generalization.}
    \textbf{a}, Evaluation framework comparing frozen encoder-based prediction (VirTues-M2, UNI), in which only the decoder is trained on the task (snowflake symbol), against fully supervised models trained from scratch (\new{HEX}, ROSIE, HistoPlexer) and VirTues, whose encoder and decoder were jointly pretrained by masked autoencoding (MAE) on multiplex virtual staining (flame symbol).
    \textbf{b}, Virtual staining from complementary multiplex markers demonstrates high accuracy at the WSI level.
    \new{\textbf{c}, Multiplex virtual staining performance (Pearson correlation) across 18 cohorts from three primary cancer types (some cohorts include several cancer types). Overall average performance not significant different.}
    \new{\textbf{d}, H\&E-based virtual staining performance (Pearson correlation) on a single CRC cohort. Markers are grouped by the extent to which their expression is reflected in H\&E morphology, and thus their expected predictability, as assessed by a clinical expert.}
    \new{\textbf{e}, H\&E-based virtual staining performance (Pearson correlation) across four cohorts. VirTues-M2 significantly outperformed the supervised baselines on all markers and was matched by UNI only on Hoechst.}
    \new{\textbf{f}, Out-of-distribution (OOD) H\&E virtual staining performance (Pearson correlation) on an unseen cancer type (human head and neck squamous cell carcinoma (HNSCC)) and organ (lymph nodes). VirTues-M2 significantly outperforms all other baselines on all markers.}
    \textbf{g}, Qualitative comparison of H\&E-based virtual staining on tissue and ROI scale.
    Error bars show 95\% bootstrapped ($n=1000$) CI. Statistical two-sample t-tests are two-sided.
}\label{fig:inpainting}
\end{figure}

\subsection*{Virtual staining and marker completion from H\&E and multiplex panels} \label{sec:2.4}

Multiplex spatial proteomics enables molecularly specific tissue profiling but remains constrained by cost, turnaround time, and limited tissue availability, often necessitating reduced marker panels or selective deployment. 
VirTues-M2 addresses these constraints by enabling virtual staining and marker completion from a shared multimodal representation, allowing prediction of protein markers either from routinely available H\&E or from a complementary subset of multiplex channels without additional staining.
\new{We evaluated VirTues-M2's virtual staining abilities in two settings (\autoref{fig:inpainting}a): multiplex panel completion, which predicts a held-out target marker from a complementary subset of markers, simulating the setting of reduced panels, and H\&E virtual staining, which predicts marker expression from H\&E images alone.}
For multiplex panel completion, we compared against VirTues, a multiplex FM which includes the capability of marker reconstruction and can incorporate all 148 markers we evaluate.
For evaluation, we use three established metrics: Pearson correlation (PC),  structural similarity index
measure (SSIM) \citep{wang2004image}, and MSE as all of them measure different positive properties of a virtual staining (Methods).

\new{
We froze the VirTues-M2 encoder and trained only a decoder (\autoref{suppl:fig:virtual_staining_arch}) across 18 cohorts spanning three major cancer types, i.e., breast ($n=9$), CRC ($n=6$), and lung ($n=7$), some cohorts include several cancer types, as well as multiple technologies and organs \citep{allam2022spatially, cords2024cancer, hu2023single, monkman2024spatial, rigamonti2024integrating, eling2025multi, cords2023cancer, danenberg2022breast, fischer2023multiplex, jackson2020single, meyer2025stratification, schulz2018simultaneous, liu2022reproducible, greenwald2026temporal, lin2023multiplexed, lin2023high, schurch2020coordinated, hickey2023organization} (Methods).
Using a decoder that supports arbitrary marker sets and cohort-specific panels, VirTues-M2 matched the virtual staining quality of VirTues across all cohorts and markers, showing no significant difference in average Pearson correlation (\autoref{fig:inpainting}c) despite keeping the backbone frozen while VirTues was pretrained end-to-end for this task.
The multimodal representation therefore retains the information that multiplex inpainting requires without being explicitly optimized toward it.
Per-marker, cohort and metric results are reported in \autoref{suppl.tab:mx_17_cohorts_breast}, \autoref{suppl.tab:mx_17_cohorts_crc} and \autoref{suppl.tab:mx_17_cohorts_lung}.
Visual comparison of original and virtually stained whole-slide images showed close agreement at the tissue level, with preserved spatial localization of marker signal at higher magnification (\autoref{fig:inpainting}b; \autoref{suppl:fig:mx_inpainting}).
To assess transferability, we applied both models to an out-of-distribution (OOD) cohort of head and neck cancer metastases in lymph nodes \citep{haist2026lymph} (\autoref{suppl:fig:vs_mx_ood}). 
In contrast to the VirTues Decoder, the VirTues-M2 virtual staining head was not trained on head and neck cancer tissue. Nevertheless, both models achieve comparable overall performance on this tissue type, with marker-specific differences that may reflect variation in their virtual staining training data.
}

For virtual staining from H\&E, we compared VirTues-M2 against UNI, using the same decoder and just switching the frozen foundation model backbone, as well as three supervised models: HEX \citep{li2026ai}, ROSIE \citep{wu2025rosie}, and HistoPlexer \citep{andani2025histopathology} (\autoref{fig:inpainting}a).
\new{Because the supervised baselines require a fixed marker set and ordering, evaluation was restricted to the markers shared across training cohorts.
We first trained all models on a single cohort and evaluated them on a held-out test set from that same cohort (\autoref{fig:inpainting}d; per-marker and per-metric results in \autoref{suppl.tab:he_hickey2023organization_pearson}, \autoref{suppl.tab:he_hickey2023organization_spearman}, and \autoref{suppl.tab:he_hickey2023organization_ssim}), as it is common in related virtual staining work \citep{valanarasu2025multimodal, andani2025histopathology}. VirTues-M2 significantly outperformed all baselines on every marker and was matched only by the UNI backbone on PanCK. 
}

\new{The previous single-cohort protocol provides limited evidence of generalization to new clinical settings. We therefore additionally trained all models across four cohorts, requiring generalization across technologies, cancer types, and organs. In this multi-cohort setting, VirTues-M2 again outperformed all other models on every marker and was matched only by the UNI backbone on Hoechst (\autoref{fig:inpainting}e; per-marker, per-cohort, and per-metric results in \autoref{suppl.tab:he_all4_pearson}, \autoref{suppl.tab:he_all4_spearman}, and \autoref{suppl.tab:he_all4_ssim}; visual comparison in \autoref{fig:inpainting}g and \autoref{suppl:fig:he_inpainting}). Notably, both frozen foundation model backbones lost little performance relative to the single-cohort setting, whereas the supervised baselines degraded substantially, indicating that foundation model representations supply generalization that task-specific supervised training does not.}
\new{To test generalization to an independent cohort, we applied the models trained across four cohorts, without further adaptation, to HNSCC lymph-node metastases from \citet{haist2026lymph}. This cohort was excluded from both VirTues-M2 pretraining and virtual-staining decoder training, and its tissue and disease context were absent from the decoder-training cohorts (\autoref{fig:inpainting}f; \autoref{suppl.tab:he_haist2026lymph_pearson}- \ref{suppl.tab:he_haist2026lymph_ssim}; \autoref{suppl:fig:vs_he_ood_examples}).
While performance dropped as expected (except for the nuclear marker), VirTues-M2 still significantly outperformed all baselines across all markers, showing that its advantage persists even under strong distribution shifts.}

\subsection*{Minimal diagnostic tests and tissue-efficient marker panels} \label{sec:2.5}

\begin{figure}[t!]
    \centering
    \includegraphics[clip, width=\linewidth,page = 5]{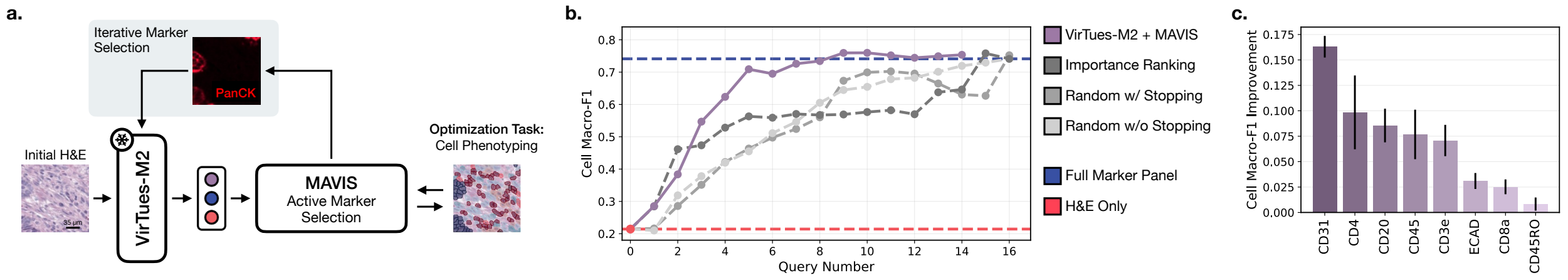}
    \caption{\textbf{Active marker selection with VirTues-M2 and MAVIS reduces required marker panels by half while maintaining full performance} 
    \textbf{a}, MAVIS pipeline leveraging multimodal VirTues-M2 embeddings to iteratively select optimal markers starting from an initial H\&E image for cell phenotyping.
    \textbf{b}, MAVIS with VirTues-M2 achieves equivalent performance to the full marker panel with only eight selected markers, outperforming random and importance-based selection strategies.
    \textbf{c}, Marker-wise improvement in cell macro-F1 for panoptic segmentation when each marker is added to the H\&E baseline, ranked by gain. 
    Error bars show 95\% bootstrapped ($n=1000$) CI.
}\label{fig:mavis}
\end{figure}

Although virtual staining and marker completion can reduce reliance on exhaustive panels, clinical and translational workflows still require principled choices about which markers to measure experimentally and which information can be recovered computationally. 
This choice is particularly consequential in spatial proteomics, where additional markers increase assay complexity and consume limited tissue\citep{jain2025test,jain2026measure}.
Therefore, we integrated VirTues-M2 with the reinforcement learning framework  MAVIS (Multi-modal Active VIew Selection) \citep{jain2025test}, that uses foundation model embeddings to sequentially select informative measurements in a tissue-dependent manner (\autoref{fig:mavis}a; Methods). 

We benchmarked this setting in a clinically realistic workflow that begins with an H\&E slide and adaptively queries a minimal set of markers, while targeting the performance achieved with a full multiplex panel. 
For evaluation, we used one-shot cell phenotyping as the evaluation task.
We compared this MAVIS-based approach with three baseline selection strategies: random selection with a fixed budget, random selection with an adaptive stopping rule, and importance ranking that adds markers based on their marginal utility estimated individually (\autoref{fig:mavis}b). 
Because the VirTues-M2 segmentation decoder requires a joint embedding that incorporates H\&E and multiplex inputs for multimodal segmentation, unimodal foundation models that cannot jointly encode both modalities were not directly applicable as baselines.

\new{
MAVIS coupled with VirTues-M2 matched full-panel performance using only eight selected markers, significantly reducing the required panel size (\autoref{fig:mavis}b) and outperforming all baseline strategies.
The selection trajectories were interpretable (\autoref{fig:mavis}c, \autoref{suppl.fig:mavis_markers}b): because cell instance segmentation was already supported by H\&E morphology, early queries prioritized markers whose cellular identities H\&E cannot resolve, T cell compartments (CD3e, CD4, CD8a), endothelium (CD31) and B cells (CD20), which are morphologically indistinguishable from surrounding stroma or from one another.
CD31 contributed the largest single gain, as endothelial cells cannot be assigned at the single-cell level in H\&E despite vascular structures being visible (\autoref{suppl.fig:mavis_markers}a), whereas later selections such as Pan-CK, SMA and Ki67 added little once lineage identity was established.
These results demonstrate that VirTues-M2 can be coupled to sequential decision-making to identify compact, task-relevant marker panels from an H\&E-first starting point.
}

\begin{figure}[t!]
    \centering
    \includegraphics[width=\linewidth,page = 7]{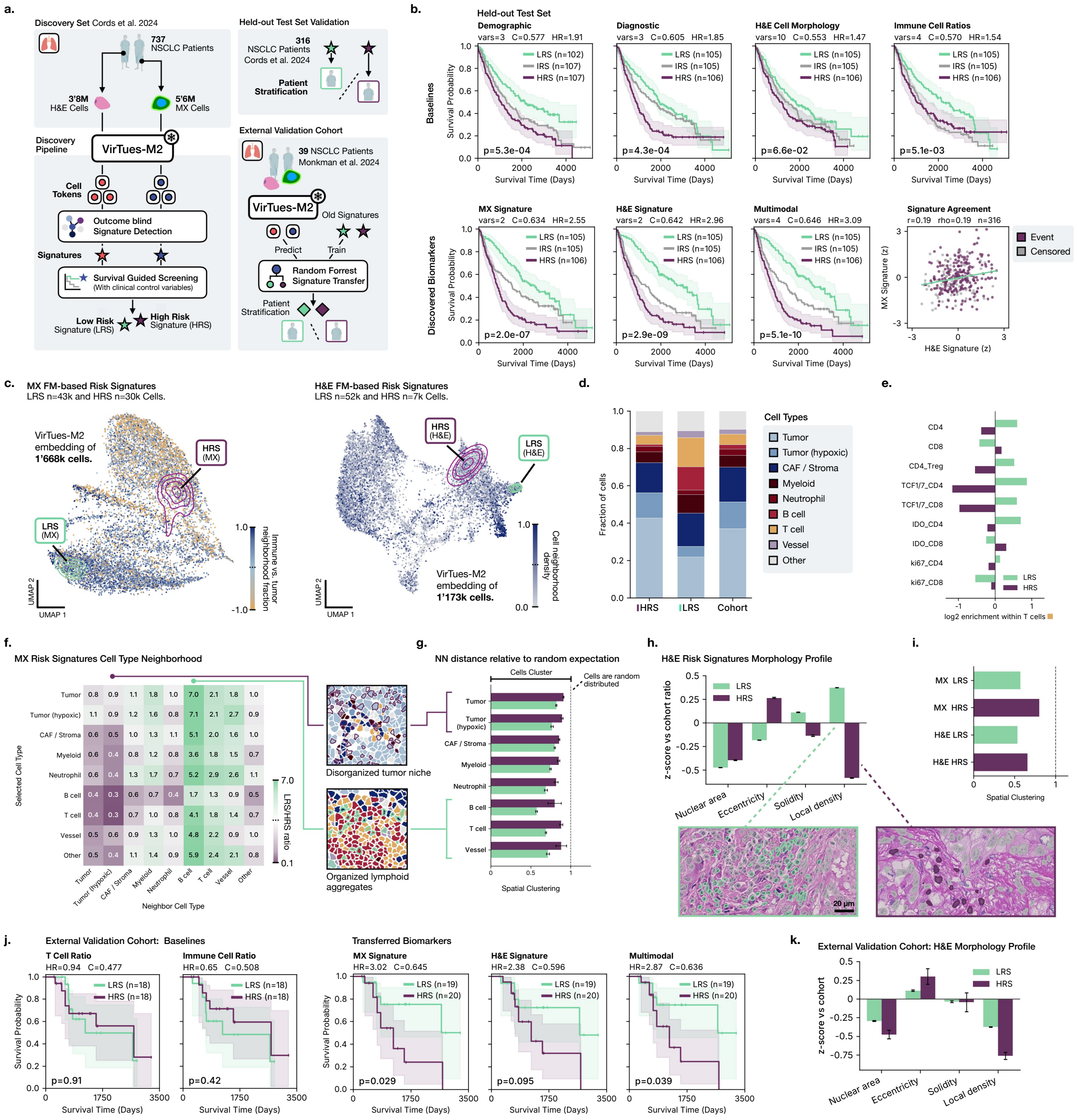}
    \caption{\textbf{Automated biomarker discovery using VirTues-M2 identifies multimodal risk signatures that transfer to held-out test set and validation cohort.}
    \textbf{a}, Annotation-free biomarker discovery pipeline using VirTues-M2 embeddings for signature detection via cross-patient cell clustering. The most predictive FM-based signatures, i.e. cross-patient cell clusters, for patient stratification in high and low survival are then selected as low-risk (LRS) and high-risk (HRS) signatures and evaluated on a held-out test set as well as a second NSCLC cohort.
    \new{
    \textbf{b}, Kaplan-Meier survival curves comparing patient stratification in LRS, HRS and intermediate-risk signature (IRS) using demographic, diagnostic, H\&E morphological and immune cell ratio baselines versus automatically discovered VirTues-M2 signatures from unimodal and combined modalities based on tertiles. The signatures from both modalities show low similarity measured by Pearson ($r$) and Spearman ($\rho$) correlation.
    \textbf{c}, UMAP visualization of multiplex-derived (mx) cell embeddings showing LRS (lime green) and HRS (purple) signatures occupying distinct regions across cell neighborhoods indicated by colors.
    \textbf{d}, Distribution of cell types within HRS, LRS and across the whole cohort.
    \textbf{e}, Log2 enrichment of T cell states within each signature, computed relative to the cohort-wide T cell composition. 
    \textbf{f}, Ratio of neighborhood composition between LRS and HRS: for each cell type (rows), the fraction of its 50 nearest neighbors belonging to each type (columns), divided between the two signatures. Values above 1 indicate more frequent contact in LRS.
    \textbf{g}, Spatial clustering per cell type, given as the mean nearest-neighbour distance between cells of the same type relative to random placement at equal density; values below 1 indicate aggregation.
    \textbf{h}, Nuclear morphology and local density of the H\&E signature cells, given as z-scores relative to the cohort mean, with representative crops of each signature (colored cells).
    \textbf{i}, Spatial clustering of each signature in both modalities.
    \textbf{j}, Kaplan-Meier survival curves on the validation cohort comparing T cell and immune cell ratio baselines to unimodal and combined transferred VirTues-M2 signatures. Patient stratification based on the median. 
    \textbf{k}, Nuclear morphology and local density of predicted H\&E signature cells in validation cohort, given as z-scores relative to the cohort mean.}
    }
    \label{fig:biomarker}
\end{figure}

\subsection*{Foundation model-derived risk signatures enhance patient stratification} \label{sec:2.6}

Current spatial biomarker discovery is largely driven by expert-defined phenotypes and hand-crafted features of tissue architecture \citep{wang2023spatial, cords2023cancer, meyer2025stratification}.
Beyond being labor-intensive, this ties discovered signatures to a specific marker panel and annotation scheme, so applying them to a new cohort requires the panel to be reproduced and the phenotypes re-annotated, which limits transfer to the cohort in which they were defined.
Foundation model embeddings capture cellular phenotypes and their local tissue context without requiring cell type or region annotations, enabling systematic discovery of outcome-associated signatures and their evaluation in new cohorts without re-annotation. To exploit these properties for biomarker discovery, we developed an automated, annotation-free pipeline that uses VirTues-M2 embeddings to identify survival-associated cell populations and locate their counterparts in independent cohorts (\autoref{fig:biomarker}a; Methods).

\new{
In a cohort of $1,053$ treatment-naive NSCLC patients with paired IMC \citep{cords2024cancer} and H\&E tissue cores, we used this pipeline to discover survival-associated cell populations. For each core, we extracted VirTues-M2 cell tokens from both modalities, capturing cellular phenotype and local microenvironmental context (\autoref{suppl:fig:cell_tokens} validates the inclusion of contextual information). We then performed high-resolution unsupervised Leiden clustering to define fine-grained cell states and computed each patient's relative abundance of cells per cluster. Clusters were evaluated by Cox regression on overall survival, adjusted for demographic and diagnostic covariates, and the most significant clusters with negative and positive coefficients were defined as a low-risk (LRS) and high-risk signature (HRS), respectively (Methods).
On the held-out test set ($n = 316$), both signatures combined into a single abundance-based score stratified patients using pre-specified tertiles without cutoff optimization (mx: $C = 0.634$, $HR = 2.55$; H\&E: $C = 0.642$, $HR = 2.96$), exceeding demographic ($C = 0.577$), diagnostic ($C = 0.605$), H\&E morphology ($C = 0.553$) and immune cell ratio ($C = 0.570$) baselines (\autoref{fig:biomarker}b). 
The two modality scores agreed only weakly ($r = 0.19$), and their combination improved stratification further ($C = 0.646$, $HR = 3.09$, $p=5.1e{-10}$), indicating complementary rather than redundant information.
}

\new{
To interpret the tissue programs captured by these signatures, we examined their cellular composition and spatial organization.
In the multiplex embedding, LRS occupied immune-rich and HRS tumor-dominated neighborhoods, while in H\&E, LRS occupied highly dense regions of tissue (\autoref{fig:biomarker}c; \autoref{suppl:fig:biomarker_discovery}), indicating that both signatures are defined by microenvironmental context rather than by cell identity alone.
Consistent with this, LRS was enriched for B and T cells and depleted of tumor relative to the cohort (\autoref{fig:biomarker}d), and within the T cell compartment specifically enriched for stem-like TCF1/7$^+$ CD4 and CD8 populations (\autoref{fig:biomarker}e). In LRS niches, every cell type contacted B and T cells $1.3$- to $7.1$-fold more often than in HRS (\autoref{fig:biomarker}f; signature-specific neighborhood compositions in \autoref{suppl:fig:biomarker_cell_neighbors}), and all cell types were more tightly aggregated (nearest-neighbor distance $0.53$-$0.79$ of random expectation), which is consistent with organized lymphoid aggregates (\autoref{fig:biomarker}g). HRS niches showed the reverse: contact was shifted toward tumor and hypoxic tumor cells, and all cell types approached random spatial mixing ($0.78$-$0.91$), consistent with more disorganized tumor tissue.
The H\&E signatures recovered a similar relationship from morphology alone (LRS nuclei rounder and densely packed, HRS elongated and sparse), with cells of both modalities more spatially aggregated in LRS (\autoref{fig:biomarker}h,i), indicating that tissue organization is where the two signatures converge.
}

\new{
Finally, we tested whether the biomarkers transfer to an independent cohort.
We annotated LRS and HRS signatures using random forest classifiers trained on the \citet{cords2024cancer} discovery split to recover signature membership from cell tokens alone, applied unchanged to the cell tokens of $39$ NSCLC patients from \citet{monkman2024spatial}.
Patients were again stratified by predicted population abundance, using a median rather than a tertile split given the small cohort size.
The mx and multimodal signatures separated survival in the expected direction (mx: $HR = 3.02$, $C = 0.645$, $p = 0.029$; multimodal: $HR = 2.87$, $C = 0.636$, $p = 0.039$), and the H\&E signature showed the same direction of effect without reaching significance at $\alpha = 0.05$ ($HR = 2.38$, $C = 0.596$, $p = 0.095$).
By contrast, the annotated T cell and immune cell ratios provided by the original study did not separate survival at all ($p = 0.91$ and $p = 0.42$; \autoref{fig:biomarker}j).
The transferred H\&E populations also preserved the LRS-versus-HRS morphological contrast observed in discovery (\autoref{fig:biomarker}i), indicating that their relative phenotype is retained across cohorts.}
These results show that VirTues-M2 reveals complementary prognostic information in tissue morphology and protein expression through annotation-free discovery of interpretable cellular signatures, while supporting transfer to independent cohorts.
\section{Discussion} \label{sec:3} 

VirTues-M2 brings routine histology and multiplex spatial proteomics into a single foundation model, enabling morphology and molecular phenotype to inform tissue analysis through native multimodal pretraining. Alongside the model, we assembled the largest \textit{open-source} multimodal tissue dataset to date, spanning 5,849 patients, 37 studies, four acquisition technologies, and 314 protein markers. Learning across incomplete pairing and heterogeneous panels allows diverse clinical studies to contribute to a shared encoder that supports H\&E-only, multiplex-only, and joint analysis without task-specific backbone fine-tuning. Multimodal learning need not improve every endpoint: its value lies in retaining strong performance where one modality captures the relevant information and improving prediction where the modalities contribute complementary evidence.
 
\looseness -1 Most prominently, the learned multimodal representations of VirTues-M2 open new opportunities for discovering clinically relevant tissue biology beyond predefined cell types and expert-selected features. In NSCLC, VirTues-M2-based annotation-free discovery identifies signatures that provide stronger patient stratification than current clinical schemes and established morphological and immune-based predictors. Joint analysis further improves stratification over either modality alone, suggesting that tissue morphology and protein expression reveal complementary aspects of prognostic biology. Their integration offers a way to investigate how molecular phenotypes and tissue organization together relate to disease outcome. Importantly, VirTues-M2 also enables recovering related prognostic signatures from H\&E alone, extending the reach of multimodal biomarker discovery: molecularly characterized cohorts can reveal tissue patterns that can then be studied at the scale of routine histology, including in patients for whom multiplex measurements are unavailable.

\looseness -1 Combining modalities can also change how tissue is characterized and which measurements are required. A single \textit{one-shot} model integrates morphological boundaries and molecular specificity for joint cell segmentation and typing, while virtual staining and panel completion extend analysis to partially measured samples. Coupled with an RL-based marker-selection strategy \citep{jain2025test}, VirTues-M2 guides subsequent measurements and approaches full-panel performance with fewer markers. These capabilities suggest a workflow in which models help extract information from existing measurements and select the additional measurements needed, potentially reducing cost, acquisition time, and tissue consumption.

\looseness -1 To understand which investments would most advance multimodal tissue modeling, we systematically varied model capacity, pretraining data volume, and cross-modal pairing. These results highlight the joint importance of data scale, biological diversity, and cross-modal pairing. VirTues-M2 benefits from precisely registered measurements, while its transformer design also enables learning from unpaired observations and paired but unregistered data, allowing a broader range of clinical cohorts to contribute to pretraining. Such analyses can guide investment in future data generation: where should we expand data diversity, and where will additional multimodal measurements most advance biological discovery and model capabilities? Coordinated community efforts to collect and openly share these data could establish a common foundation for tissue modeling, allowing each cohort to advance biological discovery and clinical applications beyond its original purpose.

\looseness -1 Several steps remain before these capabilities can inform clinical practice. Prospective multi-site evaluation is needed to establish robustness and assess institutional and demographic confounding. Biomarker transfer requires larger independent cohorts: While this study expands the available multimodal data with a newly assembled cohort of >1,000 NSCLC patients with paired H\&E and spatial proteomics for biomarker discovery, the independent survival validation was limited in patient numbers and requires confirmation in larger external cohorts. Molecular predictions remain constrained by the measured input and require calibrated uncertainty and explicit assessment of failure modes. Adaptive marker acquisition likewise needs experimental validation to establish its practical benefits.

\looseness -1 VirTues-M2 establishes native multimodal learning as a foundation for connecting tissue morphology with molecular phenotype across heterogeneous clinical datasets. Its significance extends beyond combining measurements: learning their relationships enables complementary tissue biology to inform biomarker discovery and allows related prognostic patterns to be investigated through routine H\&E. This provides a concrete direction for digital pathology 2.0, where molecularly characterized cohorts contribute knowledge that can be applied beyond the patients directly profiled. With larger paired datasets and prospective validation, the opportunity is to bring the depth of molecular tissue analysis to the scale at which pathology is practiced.

{\footnotesize
\bibliographystyle{unsrtnat}
\bibliography{references}
} 

\section*{Methods} \label{sec:4}

\subsection*{Data acquisition.} \label{sec:4.1}
\textbf{Multiplex data collection.}
For pretraining of VirTues-M2, a total of \new{37} spatial proteomics cohorts were collected either from open-source data portals or were requested from the authors of the respective studies; \new{no private or in-house data was used.}
The collection covers data from \new{four technologies, namely IMC, Orion, PhenoCycler (previously CODEX) and MIBI, encompasses 314 distinct protein markers and includes around 270 million cells from 17,275 different multiplex tissues from 5,849 patients originating from more than 13 organs as shown in \autoref{suppl.tab:dataset_overview}.}
While IMC is an ablative imaging method making subsequent same-slide H\&E tissue staining impossible, the immunofluorescence-based technologies PhenoCycler and Orion both allow imaging of morphological and proteomic information on the same tissue section.
As the typical resolution of IMC is the coarsest among the covered technologies, all images were resized to match the IMC datasets' $1\mu m/px$ resolution.
Afterwards, for registered images, a tissue segmentation model \citep{jaume2024hest} is used to segment the tissue from the background based on the H\&E image.
For segmenting unpaired or unaligned multiplex images, we use the log transformed maximum signal across the channels and apply an adaptive threshold \citep{otsu1975threshold} to segment the background from the tissue.
We preprocess the spatial proteomic images by creating crops of $256\times256$ pixels using the tissue mask to restrict them to tissue regions of the image. 
\new{As our dataset corpus comprises a large amount of tissue cores consisting of pre-determined regions of interests, we argue that classical grid tiling as commonly used for whole slide histopathology images is not suitable. To gain the highest coverage of these regions of interest and discard as little tissue as possible, we formulate the generation of crops as a greedy optimization problem in which the coverage of the tissue should be maximized with the least amount of crops using a target coverage of 95\% of the tissue space.}
Post crop generation, the 99th percentile is calculated channel-wise for each mx cohort, only considering parts of the image containing tissue. 
For normalization of the images, after loading a crop, we clip the marker intensities between zero and the 99th percentile of the corresponding channels, followed by rescaling the marker expressions to the range of 0 to 1 and applying gaussian blurring with a kernel size of $3\times3$ and a unit variance to reduce noise in the image \citep{sequeira2024penguin}.
The total of \new{314} unique proteins of all antibody panels used in the various cohorts were manually matched with UniProt identifier, based on which ESM-2 \citep{lin2023evolutionary} embeddings of the canonical UniProt amino acid sequence are generated and saved to disk.

\paragraph{H\&E data collection.}
For the histopathology data collection, we additionally curate a unimodal dataset using 14,558 whole slide images from 5,538 patients from the TCGA \citep{tomczak2015review} and GTEx \citep{lonsdale2013genotype} databases as shown in \autoref{suppl.tab:dataset_overview_he}, selecting WSIs of the same tissue types as included in our mx database.
Matching the resolution of the spatial proteomics data collection, all WSIs are resized to a resolution of $1\mu m/px$ and non-overlapping crops from tissue regions are generated resulting in 29M unique crops.
Following \citet{vo2024automatic} and \citet{chen2025revisiting}, we embed all crops using the histopathology FM UNI~\citep{chen2024uni} and perform a hierarchical clustering and sampling to reduce visual redundancy and promote morphological diversity. Hierarchical clustering is applied jointly to all H\&E crops from both multiplex and H\&E-only cohorts, comprising a total of 29.5 million crops. The clustering tree depth is set to four, with the number of clusters at each level (from bottom to top) fixed to $\{29,500;7,690;480;60\}$, consistent with established data curation guidelines. From the resulting hierarchy, we select a subset of 2.95M crops from the H\&E-only data to complement the mx data. Sampling is performed at the bottom-cluster level by selecting crops whose distances to the cluster centroid are approximately uniformly distributed, ensuring balanced coverage of common and rare morphologies.
Normalization of the H\&E images is done using the same standardization values as the VirTues-M2 initialization model, Hibou-L.

\new{All multiplex cohorts are split at a 80-20 train-test ratio at the patient level for later model evaluation, preventing data leakage from patient-level batch effects. 
Data from TCGA or GTEx is not used for model evaluation purposes. Thus, all curated crops extracted from these databases are used during pretraining.}

\subsection*{VirTues-M2 architecture.} \label{sec:4.2}
The VirTues-M2 architecture is designed to accommodate the heterogeneous nature of spatial proteomics images while also processing conventional histopathology images.
In contrast to the fixed RGB channels of brightfield imaging, the channels of a spatial proteomics image each correspond to a distinct protein marker, and both the number and identity of these markers vary substantially across acquisitions depending on the imaging technology, disease of interest and organ.
Traditional vision architectures are neither able to handle such channel diversity nor designed to process different imaging modalities or to bridge across modalities in a latent space \citep{bao2024channel}.
Furthermore, commonly employed (pre-)training strategies for vision FMs, such as DINOv2~\citep{oquab2023dinov2}, are not designed to account for multiple modalities in their training objectives or data sampling.
Tackling these unique challenges, both from an architectural and training point of view, while efficiently leveraging available data, is a core contribution of the proposed VirTues-M2 model.
To \new{integrate cross-modal interactions} of the pixel-wise aligned multimodal images and to enhance model scalability, VirTues-M2 is built using an early-fusion approach with modality-wise tokenizers and a vision transformer backbone, as shown in \autoref{fig:overview}.

\paragraph{Multiplex tokenization.}
The multiplex tokenization of VirTues-M2 builds upon the tokenization and marker attention layers introduced by VirTues \citep{wenckstern2026virtual}: Given a normalized multiplex image of size $W \times H \times C$ pixels, each channel $c \in \mathcal{C}$ is tiled into a grid of $N = \frac{W}{d_p} \times \frac{H}{d_p}$  patches of size $d_p \times d_p = 14 \times 14$ pixels. Each patch is flattened and passed through the same linear layer with output dimensionality $d_{model}$. In parallel, for each channel the corresponding protein is embedded using the Protein Language Model ESM-2~\citep{lin2023evolutionary} and projected to the dimension $d_{model}$ via a learnable linear projection. These projected protein embeddings are then fused via addition with the corresponding tokens of the mx image. Next, we group the projected image patches by their respective grid position $(i,j)$, resulting in $N$ sets $\{x_{c} | c \in \mathcal{C}\}_{ij}$ and append an additional learnable patch multiplex summary token to each of these sets. Lastly, we apply two layers of marker attention~\citep{wenckstern2026virtual} to each of these sets and retain from the outputs only the now encoded $N$ patch multiplex summary tokens. This yields the final multiplex tokens $x_{\texttt{mx}} \in \mathbb{R}^{Nxd_{\texttt{model}}}$. Note that this procedure transforms the channel-dependent input size, into a channel-independent number of tokens, alleviating the computation overhead of the subsequent vision transformer (see \autoref{suppl:fig:tokenizer}). The detailed hyperparameters of the multiplex tokenizer can be found in \autoref{suppl.tab:dinov2_hparams}.

\paragraph{H\&E tokenization.}
For tokenization of the H\&E images, we use the traditional RGB image tokenization\citep{dosovitskiyimage}: Each H\&E image is divided into patches of size $d_p \times d_p \times 3$. These patches are then flattened and passed through a linear layer of output dimensionality $d_{\texttt{model}}$, resulting in $x_{\texttt{HE}} \in \mathbb{R}^{N \times d_{\texttt{model}}}$. This ensures an evenly balanced number of tokens for both modalities.

\paragraph{Unimodal processing.}
In settings where only single modalities are available, we only tokenize the present modality and impute the missing modality tokens using the learnable modality-specific masking tokens $\Box_{\texttt{mx}} \in \mathbb{R}^{d_{\texttt{model}}}$ or $\Box_{\texttt{HE}} \in \mathbb{R}^{d_{\texttt{model}}}$, respectively.
This allows us not only to rely on paired multimodal images but also to utilize unimodal cohorts during both training and inference, extending the model's utility.

\paragraph{VirTues-M2 encoder.}
The encoder of VirTues-M2 is based on the vision transformer architecture, namely a ViT-Large~\citep{dosovitskiyimage} with $d_{\texttt{model}} = 1024$, SwiGLU as the activation function for the feed-forward layers and the attention heads, and 4 learnable register tokens $x_{\texttt{REG}} \in \mathbb{R}^{4\times d_{\texttt{model}}}$ \citep{darcet2024vision}.
More detailed model hyperparameters are shown in \autoref{suppl.tab:dinov2_hparams}. 
Shared learnable positional embeddings are added to both modalities' tokens $x_\texttt{mx}$ and $x_\texttt{HE}$ to encode the spatial relationships between tokens.  
The input of the VirTues-M2 encoder is the concatenation of a learnable classification token $x_{\texttt{CLS}} \in \mathbb{R}^{1\times d_{\texttt{model}}}$, the register tokens and modality-specific tokens \\
\begin{equation}
x_{\texttt{VirTues-M2}} = (x_\texttt{CLS},x_\texttt{REG},x_\texttt{mx},x_\texttt{HE})
\end{equation}
resulting in $(2N + 5)$ input tokens independent of the present combination of modalities and spatial proteomics channels.
During the forward pass through the vision transformer all tokens attend to each other, yielding contextualized output embeddings $\tilde{x}_{\texttt{VirTues-M2}} = (\tilde{x}_\texttt{CLS}, \tilde{x}_\texttt{REG}, \tilde{x}_\texttt{mx}, \tilde{x}_\texttt{HE})$. The classification token embedding $\tilde{x}_\texttt{CLS}$ is meant to capture global information across modalities, whereas $\tilde{x}_\texttt{mx}$ and $\tilde{x}_\texttt{HE}$ capture modality-specific information while being informed by the other modality.

During pretraining, the encoder is additionally equipped with a shared multi-layer perceptron head, which is applied to the classification token embedding and modality-specific tokens, mapping these to learnable prototypes for loss computations.

\subsection*{Multimodal pretraining.} \label{sec:4.3}
To enable the model to learn both unimodal and cross-modal representations, we build on the self-supervised student-teacher pretraining framework of DINOv2 \citep{oquab2023dinov2} and extend it to: a) allow uni- and multimodal training; b) capture global cross-modality information in a single token $x_\texttt{CLS}$; c) embed the diversity of channels captured in the spatial proteomics images; d) facilitate later in-painting of a full modality.
To this end, we modify the iBOT-based patch-level objective \citep{zhou2021ibot} as well the DINO-based crop-level objective \citep{caron2021emerging}, replace the default KoLeo regularizer \citep{sablayrolles2018spreading}, adapt the data augmentation for multimodal and multiplexed inputs and adjust the data sampling to varying cohort sizes.

\paragraph{Multimodal patch-level objective.}
For the patch-level training objective, we adapt the masking approach by iBOT \citep{zhou2021ibot} and DINOv2 \citep{oquab2023dinov2}:
Given an image, two global views of shape $H \times W = 224 \times 224$ are selected, tokenized and encoded using the teacher instance of VirTues-M2. For each view we pass the modality-wise embeddings $\tilde{x}^t_\texttt{mx}$ and $\tilde{x}^t_\texttt{HE}$ through the teacher's prototype head and apply a softmax, resulting in prototypes ${p}^t_\texttt{mx}$ and ${p}^t_\texttt{HE}$.
To account for modality-wise distribution shifts of these prototypes and increase training stability, we then apply Sinkhorn-Knopp centering to the prototypes of each modality separately. The same two views are tokenized by the student model, but then masked by randomly replacing a fraction $r$ of the modality-wise tokens with the masking tokens $\Box_{\texttt{mx}}$ or $\Box_{\texttt{HE}}$, respectively, before being passed through the student's encoder and prototype head.
The resulting student prototypes ${p}^s_\texttt{mx}$ and ${p}^s_\texttt{HE}$ are softmax-transformed without centering.
Finally we compute for all tokens that were masked in the student the iBOT loss
\begin{align}
    \mathcal{L_\texttt{iBOT}}(p^s, p^t) &= -\sum_i p_i^t \log p_i^s.
\end{align}
Note that this loss computation is performed separately for each modality to allow for balancing of the modalities.

To account for different modality availabilities of data samples and to promote multimodal learning, we employ varying masking strategies during training:
For a unimodal data sample, the missing modality is imputed using the corresponding masking token for both the teacher's and student's input and is not used any further during training.
Furthermore, for the available modality a proportion $r$, uniformly sampled between 10\% and 50\%, of the present patch tokens is masked for the student's input with probability 50\%.
In the case of a multimodal data sample, we independently apply two different masking strategies for the two global views.
We design the first view to promote full modality-to-modality interactions by increasingly masking one randomly selected modality. \new{In preparation for the H\&E-to-multiplex virtual staining task, we slightly increase the probability to mask the multiplex image to a probability of 60\%. Conversely, H\&E is masked with a probability of 40\%.}
Furthermore, a linear scheduler controlling the sampling range of masking ratio $r$ is used, increasing from 30-50\% to 90-100\% during the first quarter of the pretraining. We opted for an iteratively increasing level of masking to promote the full latent in-painting of modalities during pretraining, creating richer latent cross-modal features on which the VirTues-M2 decoder can operate on. 
The second view is meant to foster cross-modal understanding of VirTues-M2. 
Specifically, we apply the same masking strategy as in the unimodal case, except for a higher masking ratio range of 60-90\%, across all patch tokens, incentivizing the model to attend to patch tokens both across modalities and spatially.

\paragraph{Multimodal crop-level objective.}
Learning a global representation that captures information across modalities, as well as within single modalities, is the main objective of the crop-level DINO-based training objective.
Besides the two global teacher views used for the patch-level objective, eight local views of \new{$98\times98$} pixels are generated from the same image.
After embedding each global and local view using the student, we extract the \texttt{CLS} tokens $\tilde{x}_\texttt{CLS}^s$ from the student model and gather the teacher embeddings of the global views $\tilde{x}_\texttt{CLS}^t$.
Similarly to the patch-level objective, the \texttt{CLS} tokens are passed through the shared prototype MLP head, a Sinkhorn-Knopp centering is applied to the teacher tokens, and a softmax for all \texttt{CLS} tokens, resulting in the teacher tokens $c^t$ and student tokens $c^s$.
These are then passed to the DINO loss function
\begin{equation}
    \mathcal{L}_\texttt{DINO}(c^s, c^t) = -\sum c^t * \log c^s.
\end{equation}

For multimodal data samples, we randomly drop a modality from the local views with 70\% probability, replacing it with masking tokens, to capture relevant information across modalities and to refine unimodal encoding capabilities within the crop-level representations.

\paragraph{Regularizer.}
We follow recent histopathology FMs \citep{karasikov2025training, zimmermann2024virchow2} replacing the default KoLeo regularizer of DINOv2 by a kernel density estimator one \\
\begin{equation}
    H_{\texttt{KDE}}(f) = -\frac{1}{n} \sum_{i=1}^{n} \log \sum_{j=1}^{n}k_{\texttt{vMF}}(z_i, z_j)\
\end{equation}

with the von Mises-Fisher kernel $k_{\texttt{vMF}}$ defined as\\
\begin{equation}
    k_{\texttt{vMF}}(x,y) = \exp(\kappa x^T y),
\end{equation}

using a scaling constant $\kappa = 0.1$.
This regularizer is used to scatter learned representations avoiding dimensional collapse, while ensuring more stable training especially in cases where very similar samples are expected during training, such as similar histopathological crops and is applied to the \texttt{CLS} tokens $\tilde{x}_\texttt{CLS}^{s, \texttt{Global}}$ representing the global crops passed through the student before processing by the MLP head.

The overall objective optimized during pretraining of VirTues-M2 is

\label{eq:mm_dino_loss}
\begin{align}
    \mathcal{L}_{\texttt{VirTues-M2}}(p^s_\ast,p^t_\ast, c^s, c^t) = &\frac{1}{2}\mathcal{L}_\texttt{iBOT}^\texttt{mx}(p^s_\texttt{mx}, p^t_\texttt{mx}) + \frac{1}{2}\mathcal{L}_\texttt{iBOT}^\texttt{HE}(p^s_\texttt{HE}, p^t_\texttt{HE}) \\ 
    &+ \mathcal{L}_\texttt{DINO}(c^s, c^t) - 0.1H_\texttt{KDE}(\tilde{x}_\texttt{CLS}^{s, \texttt{Global}}). \
\end{align}

\paragraph{Data augmentation.}
For both modalities, we first generate two global views and multiple local views for each field of view using randomly resized cropping and geometric transformations. Global crops are obtained by loading a 256x256 image crop and cropping \& resizing them to a fixed resolution (e.g., $224\times224$), whereas local crops focus on smaller scales and are resized to a lower resolution (e.g., $98\times98$). Random rotations and random symmetries composed of flips and 90° rotations are applied to the crops. The same geometric transformations are applied jointly to all channels and modalities to preserve spatial correspondence between stains or markers. Multiplex images are capped at 30 channels, images exceeding this are randomly subsampled to 30, which keeps GPU memory usage more uniform, prevents sudden memory spikes and hinders the model to overfit on certain antibody panels per cohort.

The following channel augmentations are applied for both global and local views. For multiplex imaging, the subsequent photometric augmentations are designed to mimic realistic variability in staining intensity, channel balance and image quality. 
A randomly selected subset comprising 75–90\% of the original marker set is used for each crop, ensuring that different crops within the same image use different marker subsets. 
A random brightness scaling and a random channel shift are applied with high probability, such that the overall signal intensity and the relative contributions of individual channels vary between views~\citep{kim2025self}. In addition, Gaussian blurring is randomly applied to both global and local crops, with higher blur probabilities for local crops, to simulate defocus, motion blur and other optical degradations that frequently occur in microscopy.

For H\&E brightfield images, a color jitter transform perturbs the brightness, contrast, saturation and hue of the RGB channels, thereby modeling differences in scanner calibration, illumination and staining protocol between slides and laboratories. Furthermore, a stain-space transformation is applied in the HED (hematoxylin-eosin-DAB) color space, where small random biases and scaling factors perturb the concentrations of each of the three biomarkers, emulating realistic fluctuations of H\&E staining~\citep{tellez2018whole}.
To avoid a too unrealistic color perturbation, only one of the two color augmentations is applied uniformly at random with a probability of 80\%.
A random conversion to near-grayscale is introduced with a probability of 20\%, removing most chromatic information in some views. As in the multiplex case, Gaussian blur is also applied to H\&E crops with view-dependent probabilities. Finally, all crops are normalized using modality-specific normalization.

\paragraph{Data sampling.}
To ensure balanced learning across modalities and imaging technologies, we randomly sample a H\&E or a multiplex sample with equal probability.
As the majority of our H\&E crops stems from the hierarchical clustering based curation, we do not employ any further sample balancing here.
Given the diversity and imbalance of different multiplex cohorts (\autoref{fig:overview}, \autoref{suppl.tab:dataset_overview}), we weight the sampling of a crop for every multiplex cohort $D$, with $|D|$ samples, using the following formula \\

\begin{equation}
    w_{D} = (\frac{1}{|D|})^\alpha
    \label{eq:data_sampling}
\end{equation} \\
\new{with $\alpha = 0.25$ to ensure a more balanced sampling across cohorts. Additionally, we oversample multimodal crop samples by a factor of 2. Lastly, given the large amount of crops induced by the \texttt{shapira2022understanding}~\citep{shapira2022understanding} dataset, we down-weight sampling of this dataset by a factor of 3.5. }
\\

\paragraph{Optimization.}
To support the large batch sizes required by recent histopathology FMs \citep{zimmermann2024virchow2, chen2024uni}, VirTues-M2 is trained on 16 compute nodes, each equipped with four GH200 GPUs of 96 GB each. On a single GPU, the maximum feasible batch size is 26 samples, thus scaling to 64 GH200 GPUs enabled training with an effective global batch size of 1,664. As proposed by the DINOv2 \citep{oquab2023dinov2} authors, Fully Sharded Data Parallel (FSDP) is used to split the model replicas required by the optimizer across GPUs to more evenly shard the memory footprint used for the student, teacher, AdamW's \citep{loshchilov2018decoupled} first and second moments.
We initialize both the student and teacher models with the weights of the histopathology foundation model, Hibou-L \citep{nechaev2024hibou}.
Furthermore, xformers \citep{xFormers2022} is used to more efficiently utilize the GPU hardware and speed up the attention layers.
All other pretraining hyperparameters used for the final model training are displayed in \autoref{suppl.tab:dinov2_hparams}.

\subsection*{Diagnostic and prognostic prediction.} \label{sec:4.4}
\paragraph{Baselines.}
We compare VirTues-M2 against state-of-the-art models proposed for spatial proteomics: ResNet-50 \citep{sorin2023single}, CA-MAE~\citep{kraus2024masked}, SPACE-GM~\citep{wu2022space} and VirTues~\citep{wenckstern2026virtual}. 
For the ResNet-based approaches, we utilize a ResNet-50 model pretrained on the ImageNet-1K~\citep{deng2009imagenet} dataset as proposed by \citet{sorin2023single}. Channel-wise embeddings are projected onto their top 16 principal components, and crop-level representations are constructed by concatenating these components across all channels, yielding a feature dimensionality comparable to that of VirTues-M2.
As this convolutional architecture produces only crop- or niche-level embeddings and does not provide pixel- or patch-level representations, ResNet-50 is evaluated exclusively on crop-level downstream tasks.
\new{For VirTues, we use the \texttt{virtues-sp32} model weights from HuggingFace \citep{wenckstern2026virtual} and follow the proposed preprocessing method and image normalization.}

Unimodal histopathology benchmarking is conducted using the available pretrained weights of Hibou-B and Hibou-L \citep{nechaev2024hibou}, as well as UNI and UNI-2h \citep{chen2024uni}. Together, these models span base-, large-, and huge-scale vision transformer backbones. Following common practice \citep{chen2025revisiting}, all input images from both ROI and crop-level cohorts were rescaled to 
$224 \times 224$ pixels. This preprocessing step enables consistent evaluation across different optical magnifications, ranging from approximately 5$\times$ (e.g., BRACS) to 20$\times$ (e.g., the SPIDER benchmarks). For all evaluations involving VirTues-M2, we used the teacher model checkpoint obtained after \new{20,000} training steps, as performance plateaued beyond this point (see \autoref{suppl:fig:scaling_laws}e,f for the training dynamics).

\paragraph{Tissue-level representation.}
Representations spanning more than single crops of size $224\times224$ px are obtained by generating a non-overlapping grid of $K$ crops and gathering the high-level embeddings $\tilde{x}_\texttt{CLS}^i$ using the frozen VirTues-M2 encoder as bags $\mathcal{B} = \{\tilde{x}_\texttt{CLS}^1, ..., \tilde{x}_\texttt{CLS}^K\}$. Then,
an attention-based multiple-instance aggregation classifier \citep{ilse2018attention} is trained to find an optimal convex combination over crop embeddings to represent $\mathcal{B}$

\begin{equation}
    x_{\mathcal{B}} = \sum_{k = 1}^K a_{k} {\tilde{x}_\texttt{CLS}^k}
\end{equation}
with
\begin{equation}
    a_{k} = \frac{\exp{w^T(\tanh{(V{{\tilde{x}_\texttt{CLS}^k}})}\odot \sigma(U{\tilde{x}_\texttt{CLS}^k})})}{\sum_{k'}\exp{w^T(\tanh{(V{\tilde{x}_\texttt{CLS}^{k'}})}\odot \sigma(U{\tilde{x}_\texttt{CLS}^{k'}})})},
\end{equation} \\

where $U, V\in \mathbb{R}^{d_{\mathrm{hidden}} \times d_{\mathrm{model}}}$ and $w\in \mathbb{R}^{d_{\mathrm{hidden}}}$ are learnable parameters, $\sigma$ the sigmoid function and $\odot$ the element-wise multiplication. \new{Since the clinical tasks shown in \autoref{fig:prediction_tasks}f are determined at the cell or pixel level, we additionally add the multiplex patch tokens $p_\texttt{mx}$ to the bag of instances ensuring a fair comparison to VirTues and upscale the image to match the content per patch token to VirTues for all three tasks evaluated on the \citep{danenberg2022breast} cohort.}

The gated ABMIL model described above is trained in a supervised manner, using the cross entropy loss, the Adam optimizer~\citep{kingma2014adam} with learning rate of $10^{-4}$, a single branch of ABMIL comprising eight attention heads with a hidden dimension of $d_\mathrm{hidden} = 256$.
The model is trained for a maximum of 100 epochs with an early stopping of 10 epochs monitoring the loss on a 20\% hold-out subset of the train set. 
In a multi-head attention setting, each head operates separately using its own learnable weight vector $w~\in\mathbb{R}^{d_{\mathrm{hidden}}}$ and the resulting head-specific representations are concatenated.
For evaluating VirTues, we stick to the authors implementation and use a maximum of 500 randomly selected instances (patch summary tokens) per bag.

\subsection*{Cell phenotyping and segmentation.} \label{sec:4.5}

\paragraph{VirTues-M2 segmentation decoder.} \new{Similarly to the VirTues cell segmentation head \citep{wenckstern2026virtual}, the cell segmentation decoder used to predict panoptic cell segmentation, i.e., cell boundaries paired with the cells' phenotype, is based on the CellViT~\citep{horst2024cellvit} and the InstanSeg \citep{goldsborough2024instanseg} architecture.}
CellViT's decoder follows the hierarchical decoding idea of UNets~\citep{ronneberger2015u}, but combines it with a transformer encoder.
Applied to VirTues-M2, \new{three nearly identical models are composed for the different modality settings that VirTues-M2 can handle to showcase the models' multi-modal utility. All three segmentation decoders retrieve embeddings from different layers of the VirTues-M2 encoder.} 
Specifically \new{in the multimodal case}, the patch token embeddings generated by the 6th, 12th, 18th and last transformer blocks are concatenated along their embedding dimensions, forming tokens $x_\texttt{Seg-Decoder}^L = (\tilde{x}^L_\texttt{mx}, \tilde{x}^L_{\texttt{HE}})  \in \mathbb{R}^{N\times 2 d_{\texttt{model}}}$ with $L \in \{0,6,12,18,24\}$ \new{whereas the unimodal decoders retrieve their corresponding patch tokens directly without concatenation.
These tokens are then passed separately to the instance segmentation branch and cell phenotyping decoding branch.}
Beginning with $x_\texttt{Seg-Decoder}^{24}$, the tokens are processed by a deconvolution layer after which they are concatenated with $x_\texttt{Seg-Decoder}^{18}$. Using a sequence of convolutional and deconvolutional layers, the embeddings from the 12th, 6th and tokenizer layers are concatenated, showing the U-Net inspired processing streamline.
\new{As a result, each of the unimodal segmentation branches comprises 36M parameters whereas the multimodal counterparts consist of 49M parameters.}

\new{The instance segmentation head predicts a five-channel feature map which is converted into cell instance masks following the instance segmentation prediction and post-processing procedure of InstanSeg \citep{goldsborough2024instanseg}.
The cell phenotyping branch outputs one channel per phenotype class, including a \textit{background} and an \textit{unknown} class, and an argmax over channels yields the predicted phenotype at each pixel.
Furthermore, the \textit{unknown} class is set to $-\infty$ during inference excluding it from the predicted phenotypes consequentially forcing the model to decide for one cell phenotype.}

\paragraph{Cell segmentation training.}
\new{
From the pretraining corpus we identified eleven patient cohorts for which cell instance segmentation and cell phenotypes were available. To train and evaluate a single decoder across all datasets, we unified the cell phenotypes by mapping the detailed cell types to the categories \textit{B}, \textit{CD8}, \textit{CD4}, \textit{Myeloid}, \textit{Fibroblasts / Stroma}, \textit{Vessel / Endothelial} and \textit{Tumor} cells, as detailed in \autoref{tab:cell_type_mapping_cords2023cancer}-\autoref{tab:cell_type_mapping_lin2023highsubset}.
Additionally, we generated cell instance segmentations for the multimodal slides of the \citet{hickey2023organization} dataset using InstanSeg on the H\&E images. For training of the multiplex and multimodal decoder, the same train-test split as the pretraining one were used for the patient cohorts from \citet{cords2023cancer, cords2024cancer, danenberg2022breast, hoch2022multiplexed, meyer2025stratification, moldoveanu2022spatially, eling2025multi, wang2023spatial, rigamonti2024integrating, hoefflin2026spatial, hickey2023organization, lin2023high}. 
For the unimodal H\&E module, training is restricted to datasets which are accompanied by histopathology images, namely \citet{cords2024cancer, hickey2023organization, hoefflin2026spatial, lin2023high, rigamonti2024integrating}. 
In all modality settings, the dataset by \citet{schurch2020coordinated} is used as a full hold-out evaluation dataset.}

\new{The segmentation decoder is trained on top of the frozen pretrained VirTues-M2 backbone with the instance segmentation and cell typing
branches optimized jointly under separate objectives, as done in VirTues \citep{wenckstern2026virtual}. For instance segmentation
we adopt the InstanSeg loss unchanged \citep{goldsborough2024instanseg}. For cell typing, we combine a focal Tversky loss \citep{horst2024cellvit} with a standard cross-entropy loss in equal proportion,
$\mathcal{L}_{\mathrm{cell\text{-}typing}}
= \tfrac{1}{2}\mathcal{L}_{\mathrm{FT}}
+ \tfrac{1}{2}\mathcal{L}_{\mathrm{CE}}$, where
\begin{align*}
\mathcal{L}_{\mathrm{FT}}
=
\sum_{k=1}^{K} w_k
\left(
1 -
\frac{\mathrm{TP}_k + \varepsilon}
{\mathrm{TP}_k + \alpha \mathrm{FN}_k + \beta \mathrm{FP}_k + \varepsilon}
\right)^{\gamma}.
\end{align*}
Soft true positives, false negatives and false positives are computed from the
predicted and ground-truth pixel-wise class probabilities $\hat{p}$ and $p$ as
$\mathrm{TP}_k = \sum_{i=1}^{N} p_{ik}\hat{p}_{ik}$,
$\mathrm{FN}_k = \sum_{i=1}^{N} p_{ik}(1-\hat{p}_{ik})$ and
$\mathrm{FP}_k = \sum_{i=1}^{N} (1-p_{ik})\hat{p}_{ik}$. We set $\alpha = 0.7$,
$\beta = 0.3$ and $\gamma = \frac{4}{3}$, and down-weight the \textit{background} and \textit{unknown} classes with $w_k = 0.05$ ($w_k = 1$ otherwise).
Each of the models is trained for 200 epochs using a learning rate of 0.001, an effective batch size of 64, AdamW as the optimizer and the same $256 \times 256$ crops generated for pretraining randomly cropped to 224x224.}

\subsection*{Virtual staining.} \label{sec:4.6}
\paragraph{VirTues-M2 decoder.}
We developed a virtual staining decoder that combines foundation model representations with ESM-2 protein embeddings \citep{lin2023evolutionary} to condition the decoder explicitly on marker identity (\autoref{suppl:fig:virtual_staining_arch}).
This design enables prediction of arbitrary marker combinations rather than restricting the model to a fixed panel.

\new{
The architecture comprises two main components.
A frozen VirTues-M2 encoder tokenizes the H\&E or multiplex input ($224 \times 224$) into $16 \times 16$ patch tokens.
These are passed to a decoder consisting of two parts: a semantic decoder, inspired by the VirTues decoder, which translates the tokens into marker-specific semantic features, and an upsampling head, which restores the spatial detail required for subcellular-resolution prediction.
To retain information at multiple levels of abstraction, linear skip projection layers extract hidden representations after encoder blocks 6, 12, and 18 and route them, together with the final patch tokens, to both the semantic decoder and the upsampling head.
}

\new{
In the decoder, the ESM-2 embedding of each target marker is added to the encoder tokens, after which two full-attention blocks and a layer normalization produce decoded multiplex tokens of shape $p \times 16 \times 16 \times d$, where $p$ is the number of requested protein marker.
The upsampling head applies a $1 \times 1$ convolution ($d = 1024 \rightarrow 128$) followed by four PixelShuffle convolution blocks \citep{shi2016real} that progressively increase spatial resolution from $16 \times 16$ to $256 \times 256$, and a resize operation to $224 \times 224$.
Before the final $1 \times 1$ convolution ($d = 128 \rightarrow 1$), we add the output of an input guide layer that operates directly on the H\&E image, sharpening the spatial detail of the predicted multiplex channels.}
As the training objective, we use the mean squared error (MSE) between the decoded crop $\tilde{x}^p$ and the normalized ground truth image crop $y^p$

\begin{equation}
    \mathcal{L} = \norm{\tilde{x}^p - y^p}^{2}_2.
    \label{eq:vs_mse}
\end{equation}

We train two separate virtual staining models. 
The first one is trained in an mx-to-mx encoding-decoding setting to evaluate virtual staining from one set of markers to a distinct set of proteins.
During training, we randomly select 70\% of markers as input to predict the other 30\%.
During inference we then predict each marker of a multiplex based on the other present marker.
The second model is trained for the H\&E-to-mx virtual staining task.
For fair comparison with supervised baselines that require fixed marker configurations, we restrict evaluation to eleven markers shared across all four cohorts.
Given the heterogeneous cohort sizes, we leverage the same stratified sampling technique as we used during the pretraining to ensure balanced representation of all cohorts.

\paragraph{Baselines.}
\new{Among existing mx-to-mx methods, only VirTues can accommodate the heterogeneous marker panels across 18 cohorts comprising 148 markers.} 
To ensure fair comparison, we retrain VirTues on these cohorts via masked autoencoding as described by \citet{wenckstern2026virtual}. 
However, MAE pretraining explicitly optimizes both the encoder and the decoder for virtual staining prediction, whereas for VirTues-M2 the encoder is frozen and only the decoder is trained on the task. 
For all baselines, we weight the data sampling according to \autoref{eq:data_sampling}.

For H\&E-to-mx virtual staining, we compare against one different foundation model backbone and three end-to-end supervised approaches. 
To compare the quality of the embeddings, we implement the same decoder as used for VirTues-M2, but substitute the encoder with UNI.
This ensures a fair comparison, as the model and embedding dimensions of the encoder match those of VirTues-M2.

\new{
We use HEX \citep{li2026ai}, ROSIE \citep{wu2025rosie} and HistoPlexer \citep{andani2025histopathology} as fully supervised baselines. 
For HEX, we use the official provided implementation, which combines a MUSK backbone \citep{xiang2025vision} with a regression head that predicts a single mean expression score per marker for each crop.}
ROSIE employs a ConvNext architecture that takes $128 \times 128$-pixel H\&E patches as input and predicts mean mx marker intensities for the central $8 \times 8$ pixels. 
Contiguous virtual stainings are generated by sliding-window inference, iteratively stitching patch-level intensity predictions across whole tissues or crop sections.
Thus, to stain a $256\times256$ crop, ROSIE requires 1,024 forward passes, which can be batched.

HistoPlexer employs a conditional Generative Adversarial Network (cGAN) architecture with four complementary loss components: an adversarial loss for realistic image generation, an L1 loss for pixel-level accuracy, a Gaussian pyramid loss to handle alignment imperfections across serial sections, and a patch-wise contrastive loss for embedding-space consistency between corresponding patches.
The complexity of the combined loss functions led to training instability across our heterogeneous cohorts, necessitating methodological additions to the original implementation, including label smoothing, spectral normalization for the discriminator, and distinct gradient scaling for the generator and discriminator.

All virtual staining hyperparameters and model selection are determined on a held-out 15\% validation set.

\paragraph{Evaluation metrics.}
For evaluation, we use three established metrics: MSE, SSIM (structural similarity index measure), and PC (Pearson correlation), as they measure different positive properties of a virtual staining. 
MSE, which also serves as our training loss, is defined in \autoref{eq:vs_mse}. 

The PC measures the linear relationship between predicted and ground-truth marker intensities

\begin{equation}
    \mathrm{PC} = \frac{\sum_{i}(\tilde{x}_{i}^p - \bar{\tilde{x}}^p)(y_i^p - \bar{y}^p)}{\sqrt{\sum_{i}(\tilde{x}_{i}^p - \bar{\tilde{x}}^p)^2}\sqrt{\sum_{i}(y_i^p - \bar{y}^p)^2}},
    \label{eq:vs_pc}
\end{equation}

where $\bar{\tilde{x}}^p$ and $\bar{y}^p$ denote the mean values of the predicted and ground-truth images, respectively, and $i$ indexes over all pixels.

The SSIM evaluates perceptual similarity by comparing luminance, contrast, and structure
\begin{equation}
    \mathrm{SSIM} = \frac{(2\mu_{\tilde{x}}\mu_{y} + c_1)(2\sigma_{\tilde{x}y} + c_2)}{(\mu_{\tilde{x}}^2 + \mu_{y}^2 + c_1)(\sigma_{\tilde{x}}^2 + \sigma_{y}^2 + c_2)},
    \label{eq:vs_ssim}
\end{equation}
where $\mu_{\tilde{x}}$ and $\mu_{y}$ are the local means, $\sigma_{\tilde{x}}^2$ and $\sigma_{y}^2$ are the local variances, and $\sigma_{\tilde{x}y}$ is the local covariance of $\tilde{x}^p$ and $y^p$. The constants $c_1$ and $c_2$ stabilize division when the denominators are small. 

These three metrics provide complementary assessments: MSE quantifies pixel-level error, PC evaluates intensity correlation independent of absolute scaling, and SSIM captures perceptual similarity including structural information.

\subsection*{Minimal diagnostic tests with VirTues-M2 and RL.} \label{sec:4.7}
Multimodal Active VIew Selection (MAVIS)\citep{jain2025test} is a RL framework that dynamically selects diagnostic tests at inference time by unifying modality selection (choosing between diagnostic platforms) and feature selection (selecting specific measurements within a platform) into a single view selection task. 
The problem is formulated as a Markov decision process where the state comprises FM embeddings, classifier logits, and a mask of acquired views, while actions include available views, a \texttt{[stop]} token to terminate acquisition, and a \texttt{[jump]} token to transition between modalities. 
The reward function combines instantaneous rewards reflecting diagnostic improvement at each selection step with episodic rewards for correct final predictions, encouraging greedy selection of maximally informative views while penalizing unnecessary acquisitions. 
The policy is trained using Proximal Policy Optimization with invalid action masking to prevent redundant acquisitions.

The original MAVIS leverages two pretrained FMs as feature extractors without fine-tuning: UNI2-H for H\&E images and VirTues for multiplex.
However, we adapted MAVIS to work only with one FM, i.e., VirTues-M2, also improving computational efficiency.
In our multimodal setting, MAVIS learns a hierarchical decision process that always starts with the H\&E tissue and all multiplex marker have the same acquisition cost.

\subsection*{Spatial biomarker discovery framework.} \label{sec:4.8}
We filter the \citet{cords2024cancer} cohort based on patients with survival data and at least $200$ segmented cells within each modality, resulting in a total of $1,053$ patients. 
For all ROIs, single-cell embeddings are generated from both modalities using the frozen VirTues-M2 encoder based on the provided cell segmentation masks. 
\new{Rather than using only local cell information, the transformer architecture encodes both neighborhood context and local patch information into the embedding space of the patch to which the segmented cell belongs, enabling the discovery of cell states defined by their spatial microenvironment.
We validate this property in \autoref{suppl:fig:cell_tokens}.}

\new{
To identify fine-grained cell states analogous to specific cell subtypes or neighborhoods, we perform unsupervised Leiden clustering iteratively in high resolutions ($r \in \{4.0, 4.2, 4.4\}$) on the discovery set ($n=737$ with $505$ events). 
These resolutions substantially exceed the default value of $r = 1$ to capture subtle cell subpopulations that may be obscured at coarser clustering granularities.} 
We retain only clusters containing at least five cells per patient to ensure robust proportion estimates, and remove clusters exhibiting high multicollinearity to ensure independent signatures.
For each Leiden resolution, we compute for every patient a vector representing the relative proportion of cells belonging to each cluster, yielding a matrix $X_{prop} \in \mathbb{R}^{(n \times c)}$ where $n$ represents the number of patients and $c$ the number of clusters identified at that resolution. 

\new{
The predictive strength of each cluster proportion was screened on the discovery set by Cox regression on overall survival, adjusted for demographic and diagnostic covariates to ensure independence from these confounders, with hyperparameters selected by five-fold cross-validation split by patient. 
Across all resolutions, p-values were corrected within each modality using the Benjamini-Hochberg false discovery rate, and we selected per modality the most significant cluster with a positive coefficient (high-risk signature, HRS) and the most significant with a negative coefficient (low-risk signature, LRS). 
To evaluate the selected signatures on the held-out test set ($n = 316$ with $217$ events), we combine them per patient and modality into a single risk signature feature $s$, given by the log-ratio of the high-risk to low-risk cluster proportion,}

\new{
\begin{equation} \label{eq:risk_sign}
    s = \log(X_{prop}^{\text{HRS}} + \epsilon) - \log(X_{prop}^{\text{LRS}} + \epsilon), \quad \epsilon = 10^{-6},
\end{equation}
}
\new{which is invariant to the total number of cells sampled per core.}

\new{
Baselines comprised demographic variables (age, sex, smoking status), diagnostic variables (stage, tumour size, grade), H\&E cell morphology (mean and standard deviation of nuclear area, eccentricity, solidity, and major and minor axis length), and immune cell ratios (CD4 and CD8 fractions, their log-ratio, and the neutrophil-to-lymphocyte ratio). 
To compare baselines and signatures without cutoff optimization, patients were stratified into low-, intermediate- and high-risk groups by tertiles of the fitted risk score.
}

\new{
For signature transfer, we trained a random forest classifier via five-fold cross validation on the discovery split cell tokens by sampling balanced from the LRS, HRS, and ''Other'' clusters for each modality ($n_{trees}=300$, $\text{max\_depth}=16$, $n_{h\_bins} = 64$).
On both cohorts we standardize the cell tokens based on the mean and standard deviation of the \citet{cords2024cancer} discovery cohort to minimize distribution shifts between them.
The trained classifiers were then applied to the \citet{monkman2024spatial} validation cohort ($n = 39$ patients, $16$ events), where no cluster labels are available.
Since cluster proportions cannot be counted directly, we estimate the LRS and HRS abundance of a patient as the mean predicted class probability across all of their cells, and compute $s$ as in \autoref{eq:risk_sign}.
Patients were stratified by the median of $s$ rather than by tertiles, given the smaller cohort size.
As demographic and diagnostic variables are not available for this cohort, we use the expert cell type annotations provided with the cohort as baselines, computing the T cell (CD4, Effector CD4, Treg, CD8) and immune cell (T cells, B cells, Monocytes, Macrophages, Granulocytes, Mast Cells) fraction per patient and stratifying each at its median.
}

\section*{Data availability}
All datasets used in this study are openly available from the original studies as referenced in \autoref{suppl.tab:dataset_overview} with the exception of the imaging data from \citet{rigamonti2024integrating, wang2023spatial, shapira2022understanding}, the H\&E data from \citet{monkman2024spatial}, the cell type annotations from \citet{lin2023high} and data from the TuPro consortium which are available from the respective authors or consortium upon request. The newly collected H\&E data accompanying the IMC image data from \citet{cords2024cancer} will be released upon publication of this work. \new{All open-source datasets will be released in a standardized data format through the \texttt{spora} framework \citep{querfurth2026spora} to facilitate reproducibility.}
Sequences of the protein markers were obtained from UniProtKB (\url{https://www.uniprot.org/}).

\section*{Code availability}
All code was implemented in Python using PyTorch as the primary deep learning package. The source code for the model and Supplementary Information for VirTues-M2 is available at \url{https://github.com/bunnelab/virtues-m2}. 

\section*{Acknowledgements}
We thank Gabriele Gut for discussions concerning the manuscript. We further acknowledge support from John W. Hickey, Garry Nolan, Federica Marchesi, Marika Viatore, James Monkman, Lasse Meyer, Christian Sch\"urch, Jia-Ren Lin, Raza Ali, and Magdalena Matusiak with accessing and analyzing the datasets.
This work was supported by the Swiss AI Initiative through a Swiss National Supercomputing Centre (CSCS) allocation under the Swiss AI Large Grants \texttt{a146} and \texttt{ab007} (“Virtual Patient Platform”), an AI2050 Early Career Fellowship, and by the ISREC Foundation.
This manuscript used data acquired from the database (https://hpap.pmacs.upenn.edu/) of the Human Pancreas Analysis Program (HPAP; RRID:SCR\_016202; PMID: 31127054; PMID: 36206763). HPAP is part of a Human Islet Research Network (RRID:SCR\_014393) consortium (UC4-DK112217, U01-DK123594, UC4-DK112232, and U01-DK123716).
Icons were created using resources from Flaticon.com.

\section*{Author contributions}
B.v.Q., L.K., C.V.C., O.M., A.W., P.F., and C.B. conceived the study; B.v.Q., L.K., C.V.C., E.J., J.W., and C.B. devised the model architecture and its pretraining; B.v.Q., L.K., E.J., Y.C., J.W., P.F.C., P.L., and M.H. curated the datasets; B.v.Q., L.K., C.V.C., E.J., Y.C., J.W., P.F., and C.B. developed the evaluation framework and downstream tasks; B.v.Q., L.K., C.V.C., and E.J. performed the experiments; B.v.Q., L.K., C.V.C., P.F.C., and C.B. wrote the manuscript; O.M., A.W., P.F., and C.B. funded the study. All authors approved the final version of the manuscript.

\section*{Competing interests}
The authors declare no competing interests.
C.B. serves on the scientific advisory board of Recursion Pharmaceuticals. No patent applications have been filed on this work.

\newpage
\setcounter{page}{1}

\setcounter{section}{0}
\renewcommand{\thesection}{\Alph{section}}
\setcounter{figure}{0}
\renewcommand{\figurename}{Figure}
\renewcommand{\thefigure}{S\arabic{figure}}%
\setcounter{table}{0}
\renewcommand{\tablename}{Table}
\renewcommand{\thetable}{S\arabic{table}}%

\appendix
\section*{\LARGE Supplementary Material}

\section{VirTues-M2 architecture}

\begin{figure}[H]
    \centering
    \includegraphics[width=\linewidth,page = 1]{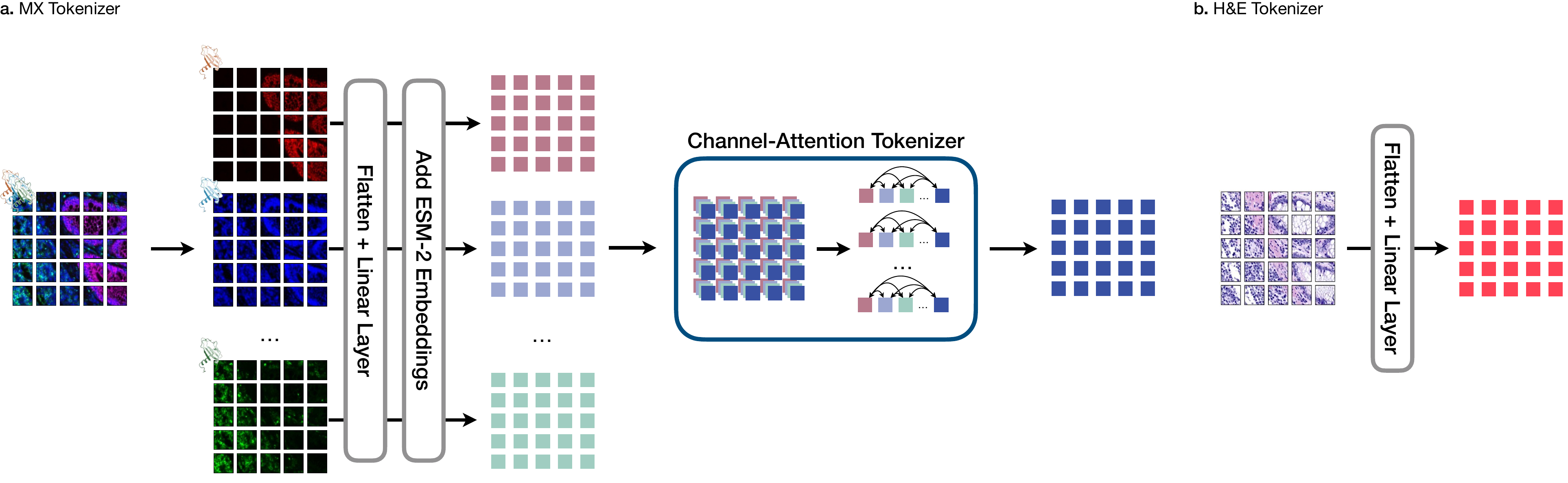}
    \caption{\textbf{Schematic of the VirTues-M2 tokenizers.} \textbf{a}, as ESM-2 \citep{lin2023evolutionary} embeddings are being used as an inductive bias, the tokenizer is not limited to any set of proteins used during training time and thus can easily be extended as well as completely zero-shot new channels during inference time.
    \textbf{b}, schematic of the VirTues-M2 H\&E tokenizer using a traditional RGB as in \citet{dosovitskiyimage}.
    }\label{suppl:fig:tokenizer}
\end{figure}

\begin{figure}[H]
    \centering
    \includegraphics[width=\linewidth,page = 2]{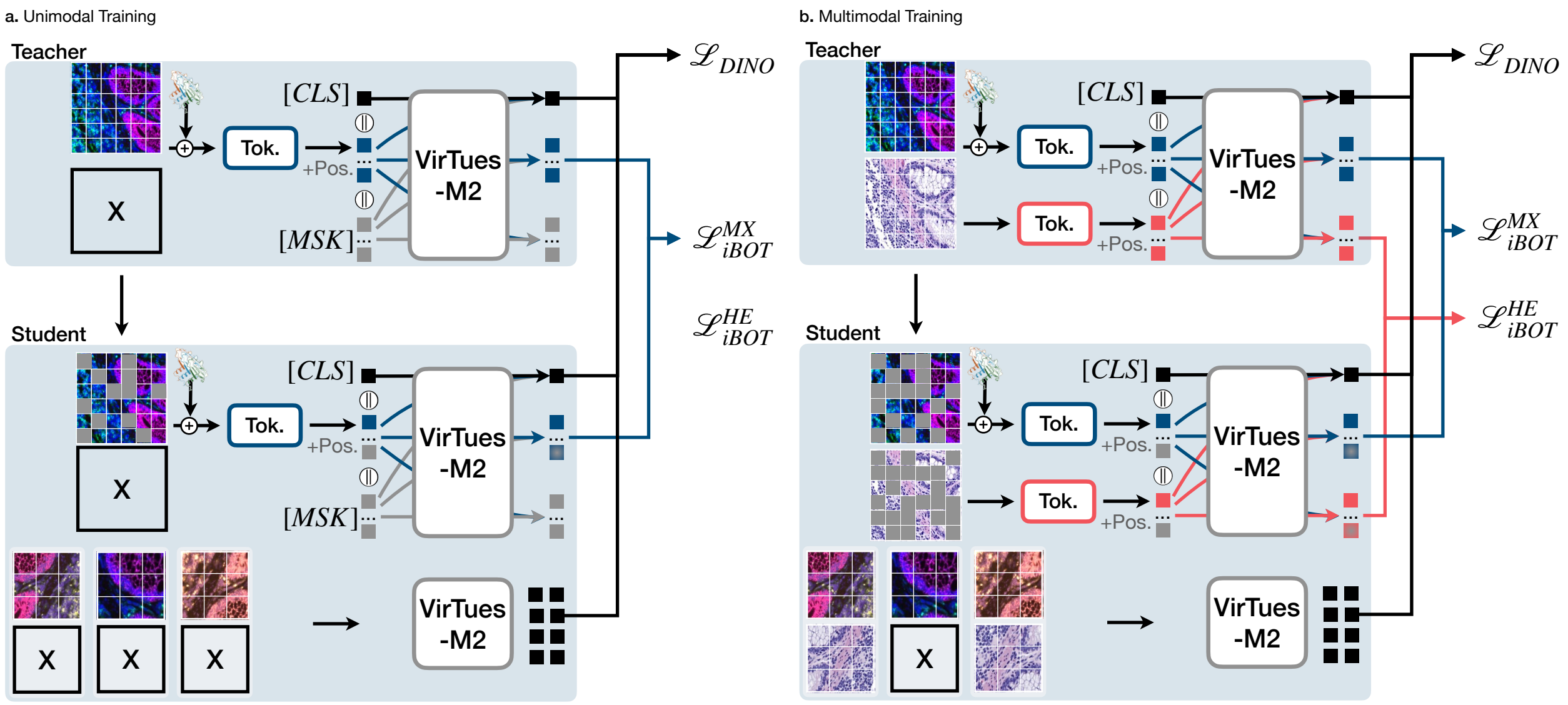}
    \caption{\textbf{Unimodal and multimodal training pipeline.} 
    Training architecture for unimodal (\textbf{a}) and multimodal (\textbf{b}) inputs. Both configurations use a teacher-student framework with shared VirTues-M2 encoders, computing DINO loss on \texttt{[CLS]} tokens and iBOT losses on masked patch tokens ($\mathcal{L}_\texttt{iBOT}^\texttt{MX}$, $\mathcal{L}_\texttt{iBOT}^\texttt{HE}$). Multimodal inputs incorporate modality-specific tokenization and positional encodings, enabling joint representation learning across channels.
}\label{suppl:fig:arch}
\end{figure}

\section{Diagnostic and prognostic prediction tasks metrics}

\begin{figure}[H]
    \centering
    \includegraphics[width=\linewidth,page = 3]{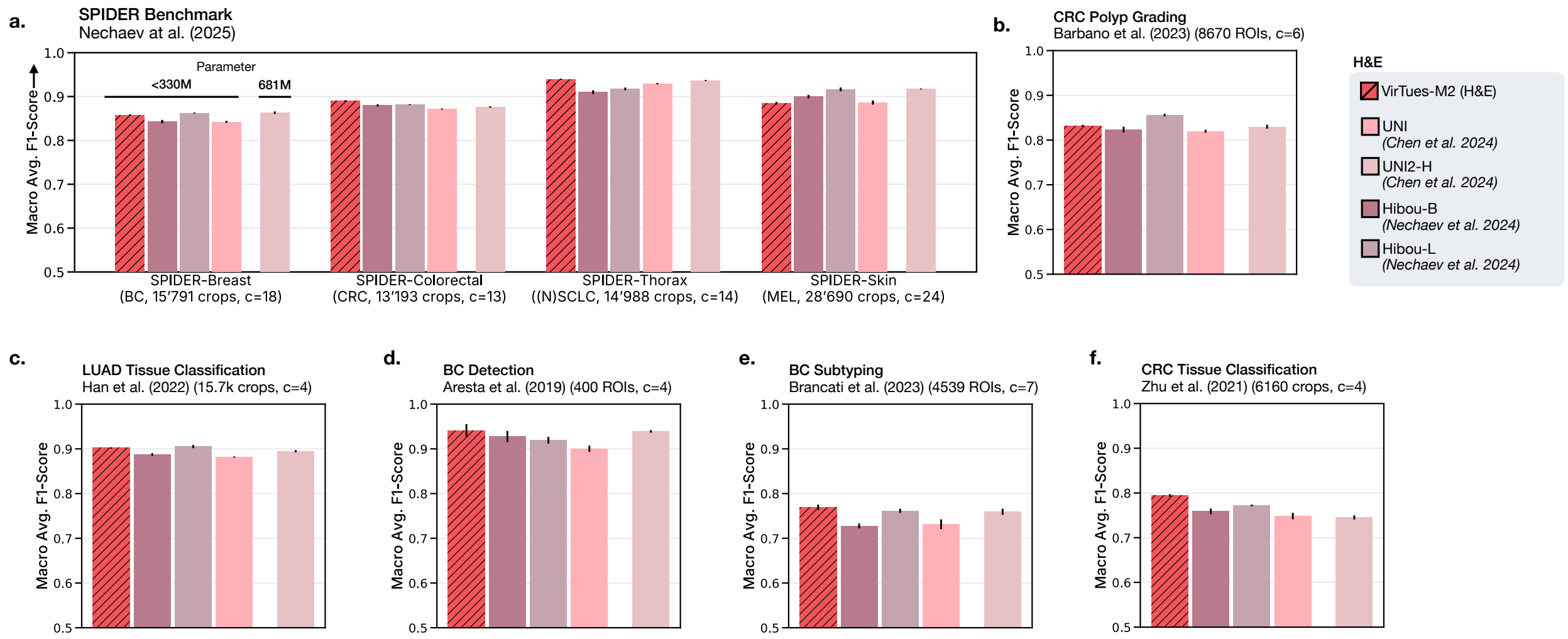}
    \caption{\textbf{Performance on all H\&E evaluation datasets.} Macro-averaged F1 scores for all 9 (\textbf{a}-\textbf{f}) H\&E datasets based on 5-fold cross validation.
}\label{suppl:fig:he_all}
\end{figure}

\begin{table}[H]
    \centering
    \caption{Macro F1 scores across 9 H\&E datasets. VirTues-M2 achieves the highest average score.}
    \label{suppl.tab:pred_he_f1}
    \begin{tabular}{llllll}
        \toprule
         & VirTues-M2 & Hibou-B & Hibou-L & UNI & UNI2-H \\
        Dataset &  &  &  &  &  \\
        \midrule
        BACH \citep{aresta2019bach} & 0.9408 & 0.9275 & 0.9191 & 0.9002 & 0.9391 \\
        BRACS \citep{brancati2022bracs} & 0.7693 & 0.7276 & 0.7614 & 0.7313 & 0.7593 \\
        SPIDER-breast \citep{nechaev2025spider} & 0.858 & 0.843 & 0.8625 & 0.8425 & 0.8633 \\
        SPIDER-colorectal \citep{nechaev2025spider} & 0.8896 & 0.8798 & 0.8815 & 0.8716 & 0.8762 \\
        SPIDER-skin \citep{nechaev2025spider} & 0.8847 & 0.8999 & 0.9164 & 0.8864 & 0.9173 \\
        SPIDER-thorax \citep{nechaev2025spider} & 0.9395 & 0.9100 & 0.9175 & 0.9294 & 0.9365 \\
        Chaoyang \citep{zhu2021hard} & 0.7944 & 0.7588 & 0.7720 & 0.7484 & 0.7452 \\
        Unitopatho \citep{barbano2021unitopatho} & 0.8318 & 0.8234 & 0.8559 & 0.8196 & 0.8297 \\
        WSSS4LUAD \citep{han2022wsss4luad} & 0.9020 & 0.8874 & 0.9051 & 0.8817 & 0.8946 \\
        \midrule
        Average & 0.8678 & 0.8508 & 0.8657 & 0.8457 & 0.8624 \\
        \bottomrule
    \end{tabular}
\end{table}

\begin{table}[H]
    \centering
    \caption{Balanced accuracy scores across 9 H\&E datasets.}
    \label{suppl.tab:pred_he_acc}
    \begin{tabular}{llllll}
        \toprule
         & VirTues-M2 & Hibou-B & Hibou-L & UNI & UNI2-H \\
        Dataset &  &  &  &  &  \\
        \midrule
        BACH \citep{aresta2019bach} & 0.9400 & 0.9275 & 0.9200 & 0.9000 & 0.9400 \\
        BRACS \citep{brancati2022bracs} & 0.7802 & 0.7407 & 0.7720 & 0.7438 & 0.7716 \\
        SPIDER-breast \citep{nechaev2025spider} & 0.8881 & 0.8750 & 0.8920 & 0.8744 & 0.8918 \\
        SPIDER-colorectal \citep{nechaev2025spider} & 0.8904 & 0.8808 & 0.8831 & 0.8732 & 0.8770 \\
        SPIDER-skin \citep{nechaev2025spider} & 0.8818 & 0.8996 & 0.9209 & 0.8888 & 0.9242 \\
        SPIDER-thorax \citep{nechaev2025spider} & 0.9460 & 0.9182 & 0.9273 & 0.9378 & 0.9439 \\
        Chaoyang \citep{zhu2021hard} & 0.8449 & 0.8121 & 0.8281 & 0.8057 & 0.8079 \\
        Unitopatho \citep{barbano2021unitopatho} & 0.8564 & 0.839 & 0.8731 & 0.8309 & 0.8463 \\
        WSSS4LUAD \citep{han2022wsss4luad} & 0.9228 & 0.9104 & 0.9252 & 0.9072 & 0.9152 \\
        \midrule
        Average & 0.8812 & 0.8670 & 0.8824 & 0.8624 & 0.8798 \\
        \bottomrule
    \end{tabular}
\end{table}

\begin{table}[H]
    \centering
    \caption{Further results on clinical tasks of \citet{wang2023spatial}.}
    \label{suppl.tab:wang}
    \begin{subtable}[t]{0.46\textwidth}
        \centering
        \caption{Macro average F1 Scores}
        \label{suppl.tab:wang_f1}
        \begin{tabular}{llll}
            \toprule
            Model & pCR-Base & pCR-On & pCR-Post \\
            Architecture &  &  &  \\
            \midrule
            VirTues-M2 (MX) & 0.6829 & 0.7987 & 0.6768 \\
            VirTues & 0.6446 & 0.7128 & 0.6616 \\
            CA-MAE & 0.6908 & 0.5643 & 0.6552 \\
            ResNet & 0.6228 & 0.5979 & 0.5625 \\
            \bottomrule
        \end{tabular}
    \end{subtable}
    \hfill
    \begin{subtable}[t]{0.46\textwidth}
        \centering
        \caption{Precision Scores}
        \label{suppl.tab:wang_precision}
        \begin{tabular}{llll}
            \toprule
            Model & pCR-Base & pCR-On & pCR-Post \\
            Architecture &  &  &  \\
            \midrule
            VirTues-M2 (MX) & 0.6872 & 0.8442 & 0.7162 \\
            VirTues & 0.6686 & 0.7600 & 0.6865 \\
            CA-MAE & 0.6948 & 0.5728 & 0.6577 \\
            ResNet & 0.6291 & 0.6465 & 0.5673 \\
            \bottomrule
        \end{tabular}
    \end{subtable}
    
    \vspace{1em}
    
    \begin{subtable}[t]{0.46\textwidth}
        \centering
        \caption{Recall Scores}
        \label{suppl.tab:wang_recall}
        \begin{tabular}{llll}
            \toprule
            Model & pCR-Base & pCR-On & pCR-Post \\
            Architecture &  &  &  \\
            \midrule
            VirTues-M2 (MX) & 0.6879 & 0.8079 & 0.6685 \\
            VirTues & 0.6599 & 0.7235 & 0.6542 \\
            CA-MAE & 0.6899 & 0.6097 & 0.6545 \\
            ResNet & 0.6286 & 0.6176 & 0.5689 \\
            \bottomrule
        \end{tabular}
    \end{subtable}
    \hfill
    \begin{subtable}[t]{0.46\textwidth}
        \centering
        \caption{Accuracy Scores}
        \label{suppl.tab:wang_acc}
        \begin{tabular}{llll}
            \toprule
            Model & pCR-Base & pCR-On & pCR-Post \\
            Architecture &  &  &  \\
            \midrule
            VirTues-M2 (MX) & 0.6833 & 0.8044 & 0.7405 \\
            VirTues & 0.6467 & 0.7200 & 0.7351 \\
            CA-MAE & 0.6967 & 0.6044 & 0.7135 \\
            ResNet & 0.6233 & 0.6133 & 0.5800 \\
            \bottomrule
        \end{tabular}
    \end{subtable}
\end{table}

\begin{table}[H]
    \centering
    \footnotesize
    \caption{Further results on clinical tasks of \citet{danenberg2022breast}.}
    \label{suppl.tab:danenberg}
    \begin{subtable}[t]{0.40\textwidth}
        \centering
        \caption{F1 Scores}
        \label{suppl.tab:danenberg_f1}
        \begin{tabular}{lllll}
            \toprule
            Model & ERBB2 & ERStatus & Grade \\
            Architecture &  &  &  \\
            \midrule
            VirTues-M2 (MX) & 0.8694 & 0.7768 & 0.7507 \\
            VirTues & 0.8577 & 0.7847 & 0.7684 \\
            SPACE-GM & 0.4744 & 0.4570 & 0.5078 \\
            CA-MAE & 0.4747 & 0.6210 & 0.5231 \\
            ResNet & 0.4936 & 0.6266 & 0.6582 \\
            \bottomrule
        \end{tabular}
    \end{subtable}
    \hfill
    \begin{subtable}[t]{0.50\textwidth}
        \centering
        \caption{Precision Scores}
        \label{suppl.tab:danenberg_precision}
        \begin{tabular}{lllll}
            \toprule
            Model & ERBB2 & ERStatus & Grade \\
            Architecture &  &  &  \\
            \midrule
            VirTues-M2 (MX) & 0.9024 & 0.7969 & 0.7607 \\
            VirTues & 0.8823 & 0.8160 & 0.7755 \\
            SPACE-GM & 0.4512 & 0.6339 & 0.4989 \\
            CA-MAE & 0.4518 & 0.6991 & 0.5320 \\
            ResNet & 0.5162 & 0.6482 & 0.6746 \\
            \bottomrule
        \end{tabular}
    \end{subtable}
    
    \vspace{1em}
    
    \begin{subtable}[t]{0.40\textwidth}
        \centering
        \caption{Recall Scores}
        \label{suppl.tab:danenberg_recall}
        \begin{tabular}{lllll}
            \toprule
            Model & ERBB2 & ERStatus & Grade \\
            Architecture &  &  &  \\
            \midrule
            VirTues-M2 (MX) & 0.8555 & 0.7645 & 0.7501 \\
            VirTues & 0.8472 & 0.7721 & 0.7667 \\
            SPACE-GM & 0.5000 & 0.5128 & 0.5529 \\
            CA-MAE & 0.5000 & 0.6117 & 0.5275 \\
            ResNet & 0.5017 & 0.6209 & 0.6555 \\
            \bottomrule
        \end{tabular}
    \end{subtable}
    \hfill
    \begin{subtable}[t]{0.50\textwidth}
        \centering
        \caption{Accuracy Scores}
        \label{suppl.tab:danenberg_acc}
        \begin{tabular}{lllll}
            \toprule
            Model & ERBB2 & ERStatus & Grade \\
            Architecture &  &  &  \\
            \midrule
            VirTues-M2 (MX) & 0.9514 & 0.8344 & 0.7595 \\
            VirTues & 0.9471 & 0.8433 & 0.7759 \\
            SPACE-GM & 0.9024 & 0.7457 & 0.6430 \\
            CA-MAE & 0.9035 & 0.7581 & 0.5838 \\
            ResNet & 0.8772 & 0.7271 & 0.6955 \\
            \bottomrule
        \end{tabular}
    \end{subtable}
\end{table}

\begin{table}[H]
    \centering
    \caption{Further results on clinical tasks of \citet{cords2024cancer}.}
    \label{suppl.tab:multimodal}
    \begin{subtable}[t]{0.48\textwidth}
        \centering
        \caption{Weighted F1 Scores}
        \label{suppl.tab:multimodal_f1}
        \begin{tabular}{lll}
            \toprule
            Model & Cancer Type & Grade \\
            Architecture &  &  \\
            \midrule
            VirTues-M2 (H\&E + MX) & 0.8812 & 0.6508 \\
            VirTues + UNI & 0.8617 & 0.6192 \\
            VirTues-M2 (MX) & 0.8275 & 0.6178 \\
            VirTues & 0.8597 & 0.6092 \\
            CA-MAE & 0.7605 & 0.5548 \\
            ResNet & 0.7428 & 0.4983 \\
            VirTues-M2 (H\&E) & 0.8657 & 0.6121 \\
            UNI & 0.8600 & 0.6140 \\
            Hibou-B & 0.8526 & 0.6131 \\
            Hibou-L & 0.8611 & 0.6125 \\
            UNI2-H & 0.8600 & 0.6140 \\
            \bottomrule
        \end{tabular}
    \end{subtable}
    \hfill
    \begin{subtable}[t]{0.48\textwidth}
        \centering
        \caption{Precision Scores}
        \label{suppl.tab:multimodal_precision}
        \begin{tabular}{lll}
            \toprule
            Model & Cancer Type & Grade \\
            Architecture &  &  \\
            \midrule
            VirTues-M2 (H\&E + MX) & 0.8726 & 0.6190 \\
            VirTues + UNI & 0.8550 & 0.5647 \\
            VirTues-M2 (MX) & 0.8175 & 0.6373 \\
            VirTues & 0.8500 & 0.6634 \\
            CA-MAE & 0.7517 & 0.4637 \\
            ResNet & 0.7294 & 0.3729 \\
            VirTues-M2 (H\&E) & 0.8576 & 0.5174 \\
            UNI & 0.8518 & 0.5724 \\
            Hibou-B & 0.8457 & 0.5475 \\
            Hibou-L & 0.8557 & 0.5946 \\
            UNI2-H & 0.8518 & 0.5724 \\
            \bottomrule
        \end{tabular}
    \end{subtable}
    
    \vspace{1em}
    
    \begin{subtable}[t]{0.48\textwidth}
        \centering
        \caption{Recall Scores}
        \label{suppl.tab:multimodal_recall}
        \begin{tabular}{lll}
            \toprule
            Model & Cancer Type & Grade \\
            Architecture &  &  \\
            \midrule
            VirTues-M2 (H\&E + MX) & 0.8783 & 0.5377 \\
            VirTues + UNI & 0.8509 & 0.5142 \\
            VirTues-M2 (MX) & 0.8169 & 0.4921 \\
            VirTues & 0.8553 & 0.5073 \\
            CA-MAE & 0.7378 & 0.4381 \\
            ResNet & 0.7226 & 0.3856 \\
            VirTues-M2 (H\&E) & 0.8580 & 0.4840 \\
            UNI & 0.8522 & 0.4952 \\
            Hibou-B & 0.8422 & 0.5028 \\
            Hibou-L & 0.8498 & 0.4892 \\
            UNI2-H & 0.8522 & 0.4952 \\
            \bottomrule
        \end{tabular}
    \end{subtable}
    \hfill
    \begin{subtable}[t]{0.48\textwidth}
        \centering
        \caption{Accuracy Scores}
        \label{suppl.tab:multimodal_acc}
        \begin{tabular}{lll}
            \toprule
            Model & Cancer Type & Grade \\
            Architecture &  &  \\
            \midrule
            VirTues-M2 (H\&E + MX) & 0.8807 & 0.6642 \\
            VirTues + UNI & 0.8621 & 0.6435 \\
            VirTues-M2 (MX) & 0.8276 & 0.6418 \\
            VirTues & 0.8592 & 0.6245 \\
            CA-MAE & 0.7638 & 0.5889 \\
            ResNet & 0.7448 & 0.5231 \\
            VirTues-M2 (H\&E) & 0.8657 & 0.6346 \\
            UNI & 0.8599 & 0.6369 \\
            Hibou-B & 0.8530 & 0.6341 \\
            Hibou-L & 0.8617 & 0.6374 \\
            UNI2-H & 0.8599 & 0.6369 \\
            \bottomrule
        \end{tabular}
    \end{subtable}
\end{table}
\clearpage

\section{\new{Pretraining Scaling Laws}}
\begin{figure}[t!]
    \centering
    \includegraphics[width=\linewidth,page = 14]{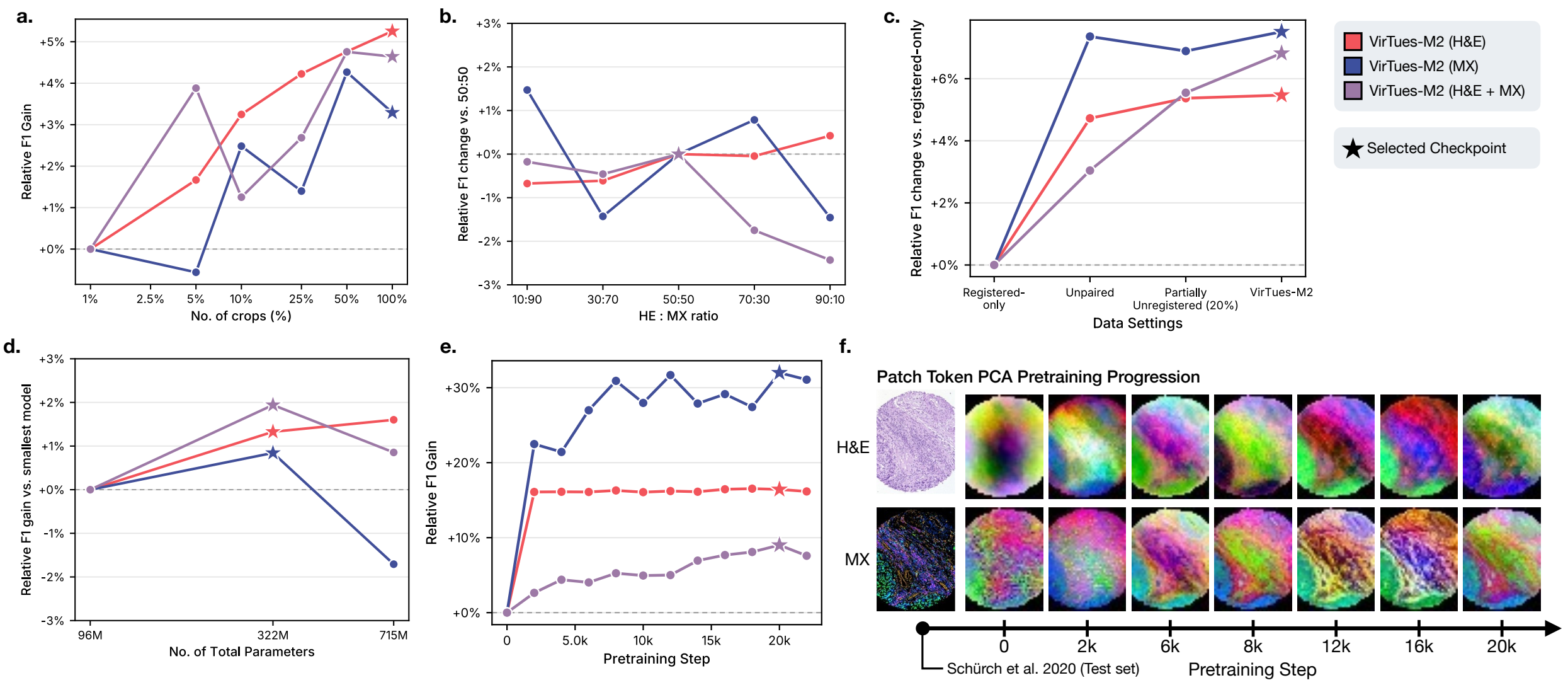}
    \caption{
    \new{
    \textbf{VirTues-M2 scaling ablation across model and data size, data settings and progression during pretraining.}
    Relative macro average f1 gain shown for \textbf{a} different dataset corpus sizes,
    \textbf{b}, varying H\&E-to-multiplex sampling ratios,
    \textbf{c}, four different data pairing settings,
    \textbf{d}, different model backbone sizes,
    \textbf{e}, progressive pretraining steps.
    \textbf{f}, PCAs visualization of patch-level token representations at different pretraining steps showing the progression of modality representations.
    }
}\label{suppl:fig:scaling_laws}
\end{figure}

\new{Given the methodological pretraining advancements developed for multimodal training of VirTues-M2, we run different ablation experiments to understand the trainings' scaling behavior, robustness under different data settings, and performance attributions during varying modality and dataset compositions.
For all ablations, we use the same effective batch size, 20,000 pretraining steps, dataset composition, training hyperparameters, and model backbone, unless explicitly stated otherwise.
For the unimodal H\&E performance, we average over the macro average F1 scores of the nine H\&E benchmarks shown in \autoref{suppl:fig:he_all}, for multiplex performance, we report the average macro F1 score for Pre-, On- and Post-Treatment on the response prediction on the TNBC patient cohort \citet{wang2023spatial}, the ERBB2 and ER Status prediction on the METABRIC cohort \citet{danenberg2022breast} and on the cancer grading and typing prediction task on the lung cancer cohort from \citep{cords2024cancer}. In the multimodal task, we use cancer grading and typing from \citet{cords2024cancer} as evaluation tasks.
Due to the variation in clinical data collected with each cohort, we could not use the same downstream tasks across all modalities. Thus, we present in \autoref{suppl:fig:scaling_laws} the \emph{relative} performance gain within each modality, making gains comparable between modalities.
}

\new{
\textbf{Dataset size.} 
The growing availability of histopathological data was one of the main success factors for H\&E foundation models \citep{chen2025revisiting, chen2024uni}. However, not only the amount of crops but also their diversity play an important role when training foundation models \citep{chen2025revisiting, simeoni2025dinov3}. Therefore, to understand the crop diversity in our pretraining corpus, in a first experiment we randomly subsample the number of crops from the overall pretraining corpus for both the multiplex as well as the H\&E data corpus, using 1\%, 5\%, 10\%, 25\% and 50\% of all available crops. For H\&E, we observe a clear scaling trend, with a relative Macro F1 score gain of 3.25\% when scaling from 1\% to 10\% of the data and further scaling to 5.25\% when further increasing the data by a factor of ten (\autoref{suppl:fig:scaling_laws}a.).
Similarly, for multiplex and multimodal performance, VirTues-M2 improves its performance by 2.48\% and 1.25\% for 10\% of the data and by 3.29\% and 4.64\% for 100\% of the data, respectively.
}

\new{\textbf{H\&E-to-MX ratio.}}
\new{In a second experiment we vary the ratio of H\&E-to-multiplex pretraining data. This experiment is meant to understand the effect of balanced modality sampling and to study catastrophic forgetting of the histopathology weight initialization. While the H\&E performance barely changes, multiplex performance is best using a 10:90 H\&E-to-multiplex ratio. Most notably, multimodal performance benefits most from a balanced sampling underlining the decision of a balanced 50:50 modality sampling for VirTues-M2.
}

\new{\textbf{Registration setting.}}
\new{
As reflected in our dataset collection (\autoref{suppl.tab:dataset_overview}) the modality pairings vary across datasets, ranging from strictly unimodal over paired but unaligned to pixel-wise alignment, emphasizing the importance of the multimodal pretraining scheme proposed which accounts for them. 
A third ablation experiment is meant to ablate the effect of the registration settings.
As shown in \autoref{suppl:fig:scaling_laws}, starting from only sampling from the five registered datasets, we see an immediate performance increase, most expressed for the unimodal settings, when using crops from all datasets but sampling them in an unpaired manner. Further multimodal performance improvements are shown when adding registered data. To simulate consecutive cut data alignments, the two modalities we randomly zoom into different parts of all sampled paired $256 \times 256$ crops with a probability of 20\% and only apply the contrastive DINO loss capturing global image features.
In contrast to the strong multimodal improvement, only a slight increase for the H\&E and a slight decrease for the multiplex setting is observed.
Without sampling artificially unregistered samples and without using unaligned crops, VirTues-M2 shows the best performance across all three modality settings.
}

\new{\textbf{Model size.}}
\new{Following conventional vision and histopathology foundation models \citep{nechaev2024hibou, chen2024uni, oquab2023dinov2, simeoni2025dinov3} we ablate the scaling laws of the model size in a fourth experiment. Using a vision transformer base (ViT-B) comprising 96 million parameters and initialized with Hibou-B weights \citet{nechaev2024hibou} performance increase across all modalities are shown when scaling to a model spanning more than triple the size (VirTues-M2 with a vision transformer large).
However, when scaling to a vision transformer huge, containing 715 million parameters and initialized with UNI2-h \citet{chen2024uni}), only the H\&E performance increases whereas multiplex performance drops under the one of the ViT-B. While we controlled for the effective batch size in all three experiments, we hypothesize that either a larger data corpus is needed to benefit from scaling further than a ViT-L or more hyperparameter tuning is needed.
}

\new{\textbf{Pretraining steps.}}
\new{In a fifth an last ablation experiment, we analyze the scaling laws over the time of pretraining. For this, in \autoref{suppl:fig:scaling_laws} e,f we show the models' performance every 2,000 pretraining steps alongside PCA visualizations of the patch-level token representations. After a steep performance increase of the unimodal settings, the H\&E perform is plateauing whereas multiplex performance is increasing until 20,000 pretraining steps. 
The multimodal performance is more steadily growing until 20,000 steps.
Across all settings, the best performance is exhibited at 20,000 steps emphasizing the selection of this checkpoint.
With regards to input tokens, each pretraining step corresponds to 1,664 training samples. Each sample comprises 2 global crops (2 x 517 tokens) and 8 local crops (8 x 103 tokens) that are processed by the student model. Therefore, every step contains 3.1M tokens on which gradients are computed - excluding the tokenizers. After 20,000 steps, the model consumed approximately 61.8 billion tokens. Assuming an average of 20 channels per multiplex sample and the presence of multiplex in at least 50\% of the samples, this number grows to a lower bound of approximately 850B input (channel patches) tokens which are seen by VirTues-M2 during pretraining.
}

\section{Patch clustering}

\begin{figure}[H]
    \centering
    \includegraphics[width=\linewidth,page = 4]{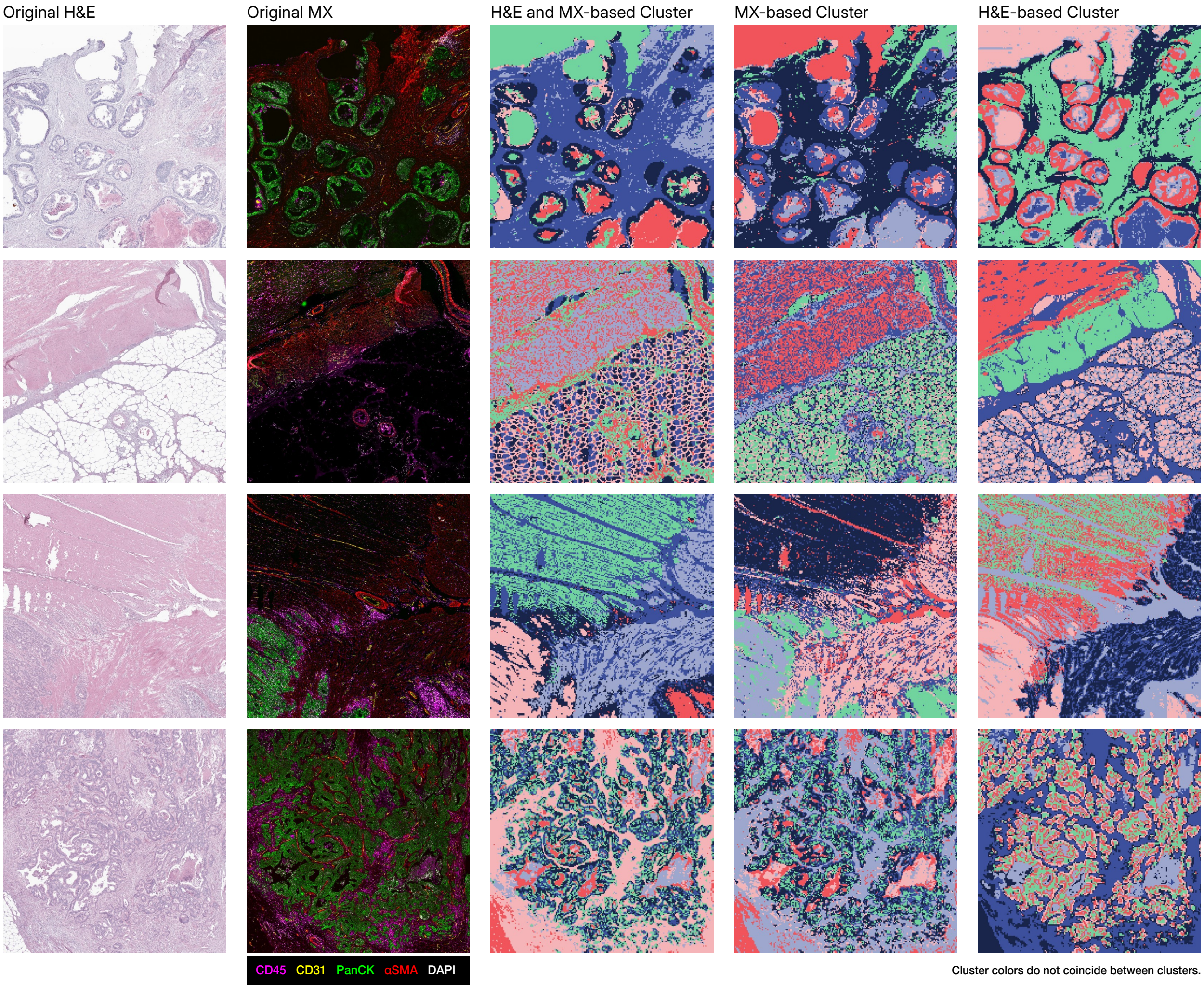}
    \caption{Further exemplary unsupervised clustering maps based on H\&E, multiplex and multimodal embeddings of VirTues-M2.
}\label{suppl:fig:kmeans_cluster}
\end{figure}

\section{Segmentation}

\begin{figure}[H]
    \centering
    \includegraphics[width=0.7\linewidth,page = 5]{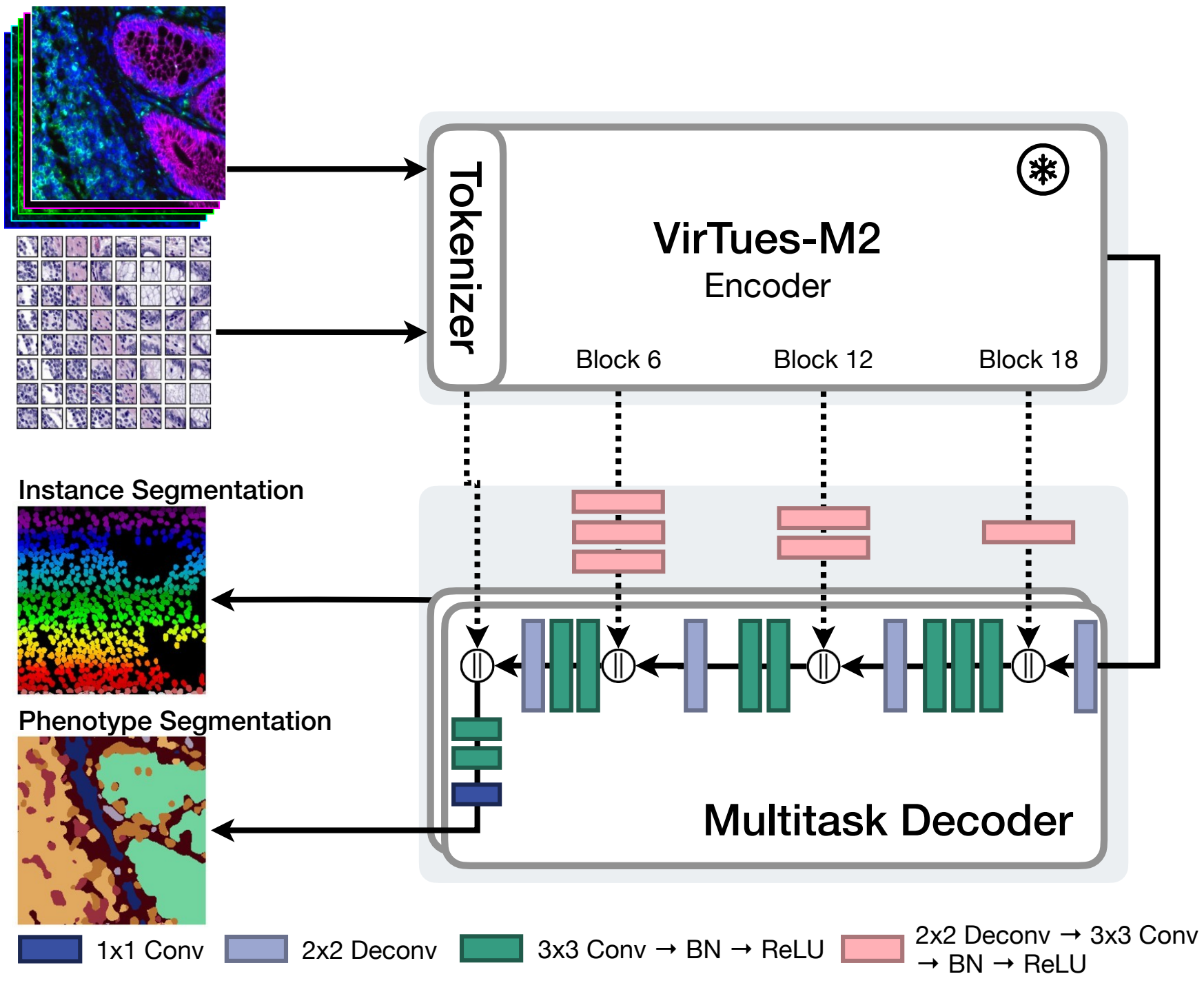}
    \caption{\textbf{VirTues-M2 integration into CellViT for panoptic segmentation.} Panoptic segmentation module for cell instance segmentation and cell classification based on \new{a mixture of the CellViT \citep{horst2024cellvit} and InstanSeg \citep{goldsborough2024instanseg} segmentation architectures.} Figure inspired by \citet{horst2024cellvit}.
}\label{suppl:fig:cellvit}
\end{figure}

\begin{figure}[H]
    \centering
    \includegraphics[width=\linewidth,page = 7]{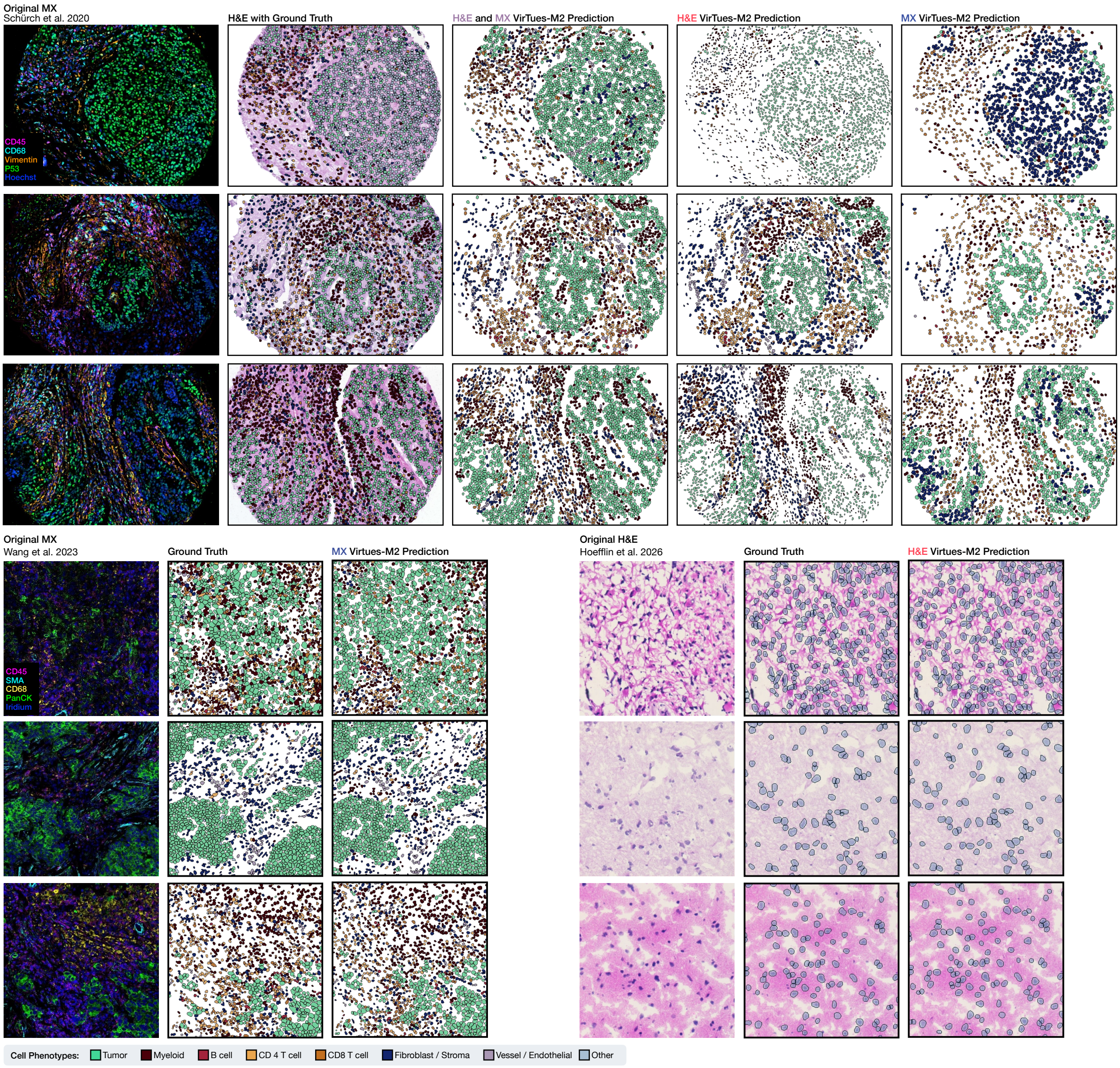}
    \caption{\new{Further exemplary panoptic segmentation maps based on multimodal, multiplex and H\&E inputs.}
}\label{suppl:fig:seg_examples}
\end{figure}

%
%

\begin{table*}[p]
\centering
\scriptsize
\setlength{\tabcolsep}{4pt}
\renewcommand{\arraystretch}{0.95}

\begin{minipage}[t]{0.48\textwidth}
\centering

\captionof{table}{\new{Cell Type Mapping for cords2023cancer}}
\label{tab:cell_type_mapping_cords2023cancer}
\begin{tabular}{l l}
\toprule
\textbf{High-Level Class} & \textbf{Original Cell Phenotypes} \\ \midrule
\multirow{1}{*}{B cell} & Bcell \\
\midrule\multirow{5}{*}{CD4 T cell} & CD4 Treg \\
& CD4 \\
& ki67 CD4 \\
& CD4 CXCL13 \\
& IDO CD4 \\
\midrule\multirow{5}{*}{CD8 T cell} & CD8 Treg \\
& CD8 \\
& ki67 CD8 \\
& CD8 CXCL13 \\
& IDO CD8 \\
\midrule\multirow{12}{*}{Fibroblasts / Stroma} & Vimentin high \\
& SMA Fibro \\
& CD10 CAF \\
& FN Cdh11 mCAF \\
& Fibroblast \\
& vCAF \\
& CXCL13 CAF \\
& CD73 CAF \\
& CCL21 CAF \\
& IDO CAF \\
& CD34 CAF \\
& CA9 CD10 CAF \\
\midrule\multirow{3}{*}{Myeloid} & Myeloid \\
& Macrophage \\
& Neutrophil \\
\midrule\multirow{3}{*}{Tumor} & hypoxia \\
& normal \\
& hypoxic \\
\midrule\multirow{2}{*}{Unknown} & other \\
& T cell \\
\midrule\multirow{3}{*}{Vessel / Endo.} & Blood \\
& Lymphatic \\
& HEV \\
\bottomrule
\end{tabular}

\vspace{1.5em}

\captionof{table}{\new{Cell Type Mapping for cords2024cancer}}
\label{tab:cell_type_mapping_cords2024cancer}
\begin{tabular}{l l}
\toprule
\textbf{High-Level Class} & \textbf{Original Cell Phenotypes} \\ \midrule
\multirow{1}{*}{B cell} & Bcell \\
\midrule\multirow{1}{*}{CD4 T cell} & CD4 \\
\midrule\multirow{1}{*}{CD8 T cell} & CD8 \\
\midrule\multirow{11}{*}{Fibroblasts / Stroma} & Collagen CAF \\
& hypoxic CAF \\
& mCAF \\
& SMA CAF \\
& tpCAF \\
& iCAF \\
& vCAF \\
& dCAF \\
& hypoxic tpCAF \\
& IDO CAF \\
& PDPN CAF \\
\midrule\multirow{2}{*}{Myeloid} & Myeloid \\
& Neutrophil \\
\midrule\multirow{2}{*}{Tumor} & normal \\
& hypoxic \\
\midrule\multirow{1}{*}{Unknown} & Other \\
\midrule\multirow{3}{*}{Vessel / Endo.} & HEV \\
& Blood \\
& Lymphatic \\
\bottomrule
\end{tabular}

\end{minipage}
\hfill
\begin{minipage}[t]{0.48\textwidth}
\centering

\captionof{table}{\new{Cell Type Mapping for danenberg2022breast}}
\label{tab:cell_type_mapping_danenberg2022breast}
\begin{tabular}{l l}
\toprule
\textbf{High-Level Class} & \textbf{Original Cell Phenotypes} \\ \midrule
\multirow{1}{*}{B cell} & B Cell or Plasma Cell \\
\midrule\multirow{1}{*}{CD4 T cell} & CD4+ T Cell \\
\midrule\multirow{1}{*}{CD8 T cell} & CD8+ T Cell \\
\midrule\multirow{1}{*}{Fibroblasts / Stroma} & Stromal Cell \\
\midrule\multirow{3}{*}{Myeloid} & Macrophage \\
& Granulocyte \\
& Antigen-Presenting Cell \\
\midrule\multirow{2}{*}{Tumor} & ER-Negative Epithelial Cell \\
& ER-Positive Epithelial Cell \\
\midrule\multirow{3}{*}{Unknown} & None \\
& Natural Killer Cell \\
& Proliferating Cell \\
\midrule\multirow{1}{*}{Vessel / Endo.} & Endothelial Cell \\
\bottomrule
\end{tabular}

\vspace{1.5em}

\captionof{table}{\new{Cell Type Mapping for hoch2022multiplexed}}
\label{tab:cell_type_mapping_hoch2022multiplexed}
\begin{tabular}{l l}
\toprule
\textbf{High-Level Class} & \textbf{Original Cell Phenotypes} \\ \midrule
\multirow{1}{*}{B cell} & CD38 \\
\midrule\multirow{1}{*}{CD4 T cell} & CD8- T cell \\
\midrule\multirow{1}{*}{CD8 T cell} & CD8+ T cell \\
\midrule\multirow{1}{*}{Fibroblasts / Stroma} & Stroma \\
\midrule\multirow{3}{*}{Myeloid} & Macrophage \\
& Neutrophil \\
& HLA-DR \\
\midrule\multirow{1}{*}{Tumor} & Tumor \\
\midrule\multirow{1}{*}{Unknown} & Other \\
\midrule\multirow{1}{*}{Vessel / Endo.} & Vasculature \\
\bottomrule
\end{tabular}

\vspace{1.5em}

\captionof{table}{\new{Cell Type Mapping for meyer2025stratification}}
\label{tab:cell_type_mapping_meyer2025stratification}
\begin{tabular}{l l}
\toprule
\textbf{High-Level Class} & \textbf{Original Cell Phenotypes} \\ \midrule
\multirow{1}{*}{B cell} & B cell \\
\midrule\multirow{1}{*}{CD4 T cell} & CD4 T cell \\
\midrule\multirow{1}{*}{CD8 T cell} & CD8 T cell \\
\midrule\multirow{1}{*}{Fibroblasts / Stroma} & Stromal \\
\midrule\multirow{3}{*}{Myeloid} & HLA DR \\
& Monocyte \\
& Neutrophil \\
\midrule\multirow{1}{*}{Tumor} & tumor \\
\midrule\multirow{1}{*}{Unknown} & Immune low \\
\midrule\multirow{1}{*}{Vessel / Endo.} & Endothelial \\
\bottomrule
\end{tabular}

\vspace{1.5em}

\captionof{table}{\new{Cell Type Mapping for moldoveanu2022spatially}}
\label{tab:cell_type_mapping_moldoveanu2022spatially}
\begin{tabular}{l l}
\toprule
\textbf{High-Level Class} & \textbf{Original Cell Phenotypes} \\ \midrule
\multirow{1}{*}{B cell} & B \\
\midrule\multirow{3}{*}{CD4 T cell} & Th.ae \\
& Treg \\
& Th.naive \\
\midrule\multirow{2}{*}{CD8 T cell} & Tc.ae \\
& Tc.naive \\
\midrule\multirow{1}{*}{Myeloid} & macro.mono \\
\midrule\multirow{1}{*}{Tumor} & melano \\
\midrule\multirow{1}{*}{Unknown} & others \\
\midrule\multirow{1}{*}{Vessel / Endo.} & CD31 \\
\bottomrule
\end{tabular}

\end{minipage}
\end{table*}

\begin{table*}[p]
\centering
\scriptsize
\setlength{\tabcolsep}{4pt}
\renewcommand{\arraystretch}{0.95}

\begin{minipage}[t]{0.48\textwidth}
\centering

\captionof{table}{\new{Cell Type Mapping for schulz2024immucan}}
\label{tab:cell_type_mapping_schulz2024immucan}
\begin{tabular}{l l}
\toprule
\textbf{High-Level Class} & \textbf{Original Cell Phenotypes} \\ \midrule
\multirow{2}{*}{B cell} & plasma \\
& B \\
\midrule\multirow{2}{*}{CD4 T cell} & CD4 \\
& Treg \\
\midrule\multirow{1}{*}{CD8 T cell} & CD8 \\
\midrule\multirow{1}{*}{Fibroblasts / Stroma} & Mural \\
\midrule\multirow{5}{*}{Myeloid} & MacCD163 \\
& DC \\
& HLADR \\
& Neutrophil \\
& pDC \\
\midrule\multirow{1}{*}{Tumor} & Tumor \\
\midrule\multirow{3}{*}{Unknown} & unlabelled \\
& NK \\
& BnT \\
\bottomrule
\end{tabular}

\vspace{1.5em}

\captionof{table}{\new{Cell Type Mapping for wang2023spatial}}
\label{tab:cell_type_mapping_wang2023spatial}
\begin{tabular}{l l}
\toprule
\textbf{High-Level Class} & \textbf{Original Cell Phenotypes} \\ \midrule
\multirow{2}{*}{B cell} & CD79a$^+$Plasma \\
& CD20$^+$B \\
\midrule\multirow{3}{*}{CD4 T cell} & CD4$^+$TCF1$^+$T \\
& Treg \\
& CD4$^+$PD1$^+$T \\
\midrule\multirow{4}{*}{CD8 T cell} & CD8$^+$T \\
& CD8$^+$TCF1$^+$T \\
& CD8$^+$PD1$^+$T {Ex} \\
& CD8$^+$GZMB$^+$T \\
\midrule\multirow{3}{*}{Fibroblasts / Stroma} & Fibroblasts \\
& PDPN$^+$Stromal \\
& Myofibroblasts \\
\midrule\multirow{5}{*}{Myeloid} & DCs \\
& M2 Mac \\
& PD-L1$^+$APCs \\
& Neutrophils \\
& PD-L1$^+$IDO$^+$APCs \\
\midrule\multirow{18}{*}{Tumor} & CA9$^+$ \\
& CK8/18$^{med}$ \\
& panCK$^{med}$ \\
& CD15$^{+}$ \\
& PD-L1$^{+}$GZMB$^{+}$ \\
& pH2AX$^{+}$DSB \\
& Vimentin$^{+}$EMT \\
& CK$^{hi}$GATA3$^{+}$ \\
& CK$^{lo}$GATA3$^{+}$ \\
& Helios$^{+}$ \\
& Basal \\
& Apoptosis \\
& PD-L1$^{+}$IDO$^{+}$ \\
& AR$^{+}$LAR \\
& TCF1$^{+}$ \\
& MHC I\&II$^{hi}$ \\
& CD56$^{+}$NE \\
& CA9$^{+}$Hypoxia \\
\midrule\multirow{1}{*}{Unknown} & CD56$^+$NK \\
\midrule\multirow{1}{*}{Vessel / Endo.} & Endothelial \\
\bottomrule
\end{tabular}

\end{minipage}
\hfill
\begin{minipage}[t]{0.48\textwidth}
\centering

\captionof{table}{\new{Cell Type Mapping for rigamonti2024integrating}}
\label{tab:cell_type_mapping_rigamonti2024integrating}
\begin{tabular}{l l}
\toprule
\textbf{High-Level Class} & \textbf{Original Cell Phenotypes} \\ \midrule
\multirow{1}{*}{B cell} & Bcells \\
\midrule\multirow{2}{*}{CD4 T cell} & CD4+ \\
& Treg \\
\midrule\multirow{1}{*}{CD8 T cell} & CD8+ \\
\midrule\multirow{2}{*}{Fibroblasts / Stroma} & aSMA+ \\
& Vim \\
\midrule\multirow{1}{*}{Myeloid} & Mf \\
\midrule\multirow{2}{*}{Tumor} & Tumor cells \\
& Ki67+ \\
\midrule\multirow{1}{*}{Unknown} & other \\
\midrule\multirow{1}{*}{Vessel / Endo.} & vessels \\
\bottomrule
\end{tabular}

\vspace{1.5em}

\captionof{table}{\new{Cell Type Mapping for hoefflin2026spatial}}
\label{tab:cell_type_mapping_hoefflin2026spatial}
\begin{tabular}{l l}
\toprule
\textbf{High-Level Class} & \textbf{Original Cell Phenotypes} \\ \midrule
\multirow{1}{*}{B cell} & Bcell \\
\midrule\multirow{1}{*}{CD4 T cell} & TcellCD4 \\
\midrule\multirow{1}{*}{CD8 T cell} & TcellCD8 \\
\midrule\multirow{2}{*}{Myeloid} & Mac \\
& Neutro \\
\midrule\multirow{6}{*}{Tumor} & cAC1 \\
& cOPC \\
& cUndiff \\
& cMESHyp \\
& cAC2 \\
& cMES \\
\midrule\multirow{5}{*}{Unknown} & Neuron \\
& Oligo \\
& Astro \\
& low \\
& excluded \\
\midrule\multirow{1}{*}{Vessel / Endo.} & Vasc \\
\bottomrule
\end{tabular}

\vspace{1.5em}

\captionof{table}{\new{Cell Type Mapping for lin2023highsubset}}
\label{tab:cell_type_mapping_lin2023highsubset}
\begin{tabular}{l l}
\toprule
\textbf{High-Level Class} & \textbf{Original Cell Phenotypes} \\ \midrule
\multirow{1}{*}{B cell} & B cell \\
\midrule\multirow{2}{*}{CD4 T cell} & Th \\
& Treg \\
\midrule\multirow{3}{*}{CD8 T cell} & Memory T \\
& CTL \\
& Exhausted CD8 T cell \\
\midrule\multirow{2}{*}{Fibroblasts / Stroma} & Vascular smooth muscle \\
& Stroma \\
\midrule\multirow{2}{*}{Myeloid} & Macrophage \\
& M2-like \\
\midrule\multirow{2}{*}{Tumor} & Tumor \\
& PD-L1+ tumor \\
\midrule\multirow{2}{*}{Unknown} & Other \\
& T Cell \\
\midrule\multirow{1}{*}{Vessel / Endo.} & Endothelial \\
\bottomrule
\end{tabular}

\end{minipage}
\end{table*}
\clearpage

\newpage
\section{Minimal diagnostic test and marker panels}

\begin{figure}[H]
    \centering
    \includegraphics[width=\linewidth,page = 6]{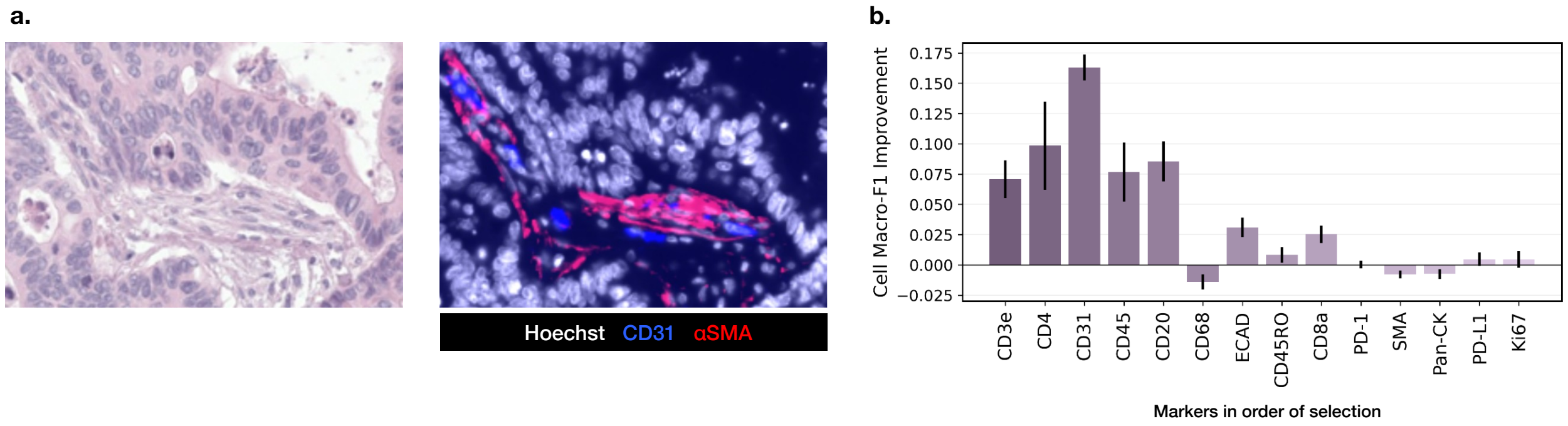}
    \caption{
    \textbf{Marker selection with MAVIS.}
    \textbf{a.} H\&E (left) and corresponding multiplex image (right) of a colorectal tumour region. Vascular structures are visible in H\&E, but CD31 (endothelium) and $\alpha$SMA (mural cells) are required to resolve the vessel wall into its cellular components.
    \textbf{b.} Improvement in cell macro-F1 for panoptic segmentation contributed by each marker, shown in the order MAVIS selected them.
    }\label{suppl.fig:mavis_markers}
\end{figure}
\section{Virtual staining} 

\begin{figure}[H]
    \centering
    \includegraphics[width=0.9\linewidth,page = 6]{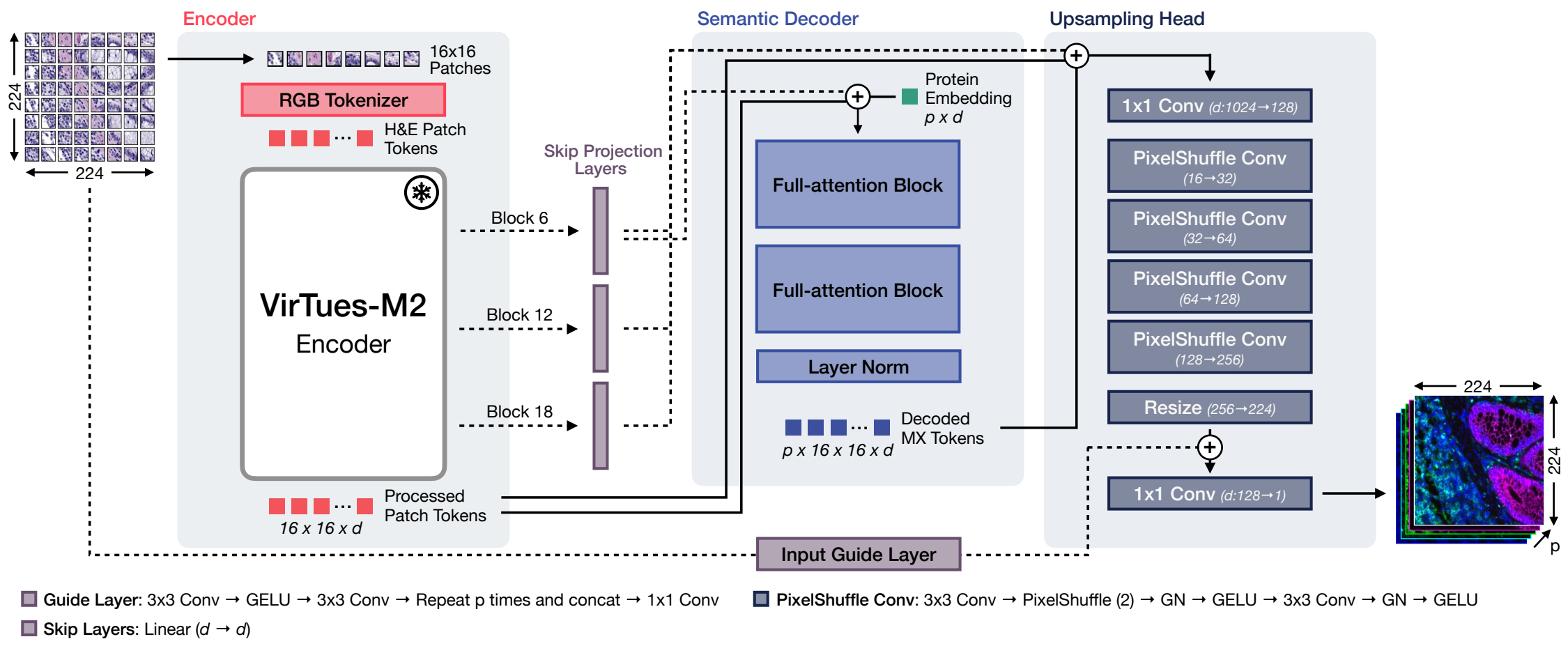}
    \caption{\new{Virtual staining architecture consisting of VirTues-M2 as the encoder and a specifically designed decoder} (Methods).}\label{suppl:fig:virtual_staining_arch}
\end{figure}

\begin{table}[H]
\centering
\caption{\new{Multiplex virtual staining performance on Breast cohorts (9 datasets, 102 markers).
M2 denotes VirTues-M2 and VT denotes VirTues. Metrics are averaged over tiles within a
dataset, then over datasets with equal weight; $n$ is the number of datasets in which the
marker is measured. Bold marks the better model.}}
\scriptsize
\setlength{\tabcolsep}{2.5pt}
\begin{tabular}[t]{@{}l r@{\hskip 5pt} r r@{\hskip 5pt} r r@{\hskip 5pt} r r@{}}
\toprule
Protein & $n$ & \multicolumn{2}{c}{Pearson $r$} & \multicolumn{2}{c}{Spearman $r$} & \multicolumn{2}{c}{SSIM} \\
\cmidrule(lr){3-4}\cmidrule(lr){5-6}\cmidrule(lr){7-8}
 &  & M2 & VT & M2 & VT & M2 & VT \\
\midrule
Cytokeratin 8/18 & 6 & \textbf{0.657} & 0.606 & \textbf{0.554} & 0.522 & \textbf{0.484} & -0.051 \\
CD45RO & 3 & \textbf{0.702} & 0.689 & \textbf{0.668} & 0.642 & \textbf{0.433} & 0.217 \\
Histone H3 & 9 & \textbf{0.710} & 0.653 & \textbf{0.698} & 0.663 & \textbf{0.497} & 0.198 \\
Estrogen Receptor Alpha & 3 & \textbf{0.707} & 0.678 & \textbf{0.706} & 0.705 & \textbf{0.500} & 0.124 \\
CD14 & 2 & 0.665 & \textbf{0.676} & 0.626 & \textbf{0.633} & \textbf{0.429} & 0.323 \\
Cytokeratin 19 & 2 & \textbf{0.694} & 0.529 & \textbf{0.619} & 0.512 & \textbf{0.495} & 0.078 \\
CXCL10\_mRNA & 1 & \textbf{0.691} & 0.526 & \textbf{0.690} & 0.588 & \textbf{0.450} & 0.200 \\
E-cadherin & 6 & 0.642 & \textbf{0.645} & 0.582 & \textbf{0.616} & \textbf{0.461} & 0.099 \\
CD4 & 7 & 0.676 & \textbf{0.693} & 0.632 & \textbf{0.646} & \textbf{0.416} & 0.264 \\
CD45 & 6 & \textbf{0.607} & 0.598 & \textbf{0.533} & 0.527 & \textbf{0.392} & 0.090 \\
EpCAM & 1 & \textbf{0.677} & 0.665 & \textbf{0.623} & 0.623 & \textbf{0.378} & 0.145 \\
beta-tubulin & 1 & \textbf{0.673} & 0.638 & \textbf{0.675} & 0.657 & \textbf{0.451} & 0.180 \\
B2M & 2 & \textbf{0.678} & 0.657 & \textbf{0.676} & 0.669 & \textbf{0.441} & 0.128 \\
p-mTOR & 3 & \textbf{0.666} & 0.635 & 0.597 & \textbf{0.621} & \textbf{0.457} & 0.140 \\
MMP9 & 1 & \textbf{0.597} & 0.593 & \textbf{0.613} & 0.612 & \textbf{0.338} & 0.257 \\
c-Myc & 3 & \textbf{0.665} & 0.633 & \textbf{0.613} & 0.609 & \textbf{0.426} & 0.067 \\
EGFR & 3 & \textbf{0.580} & 0.553 & 0.509 & \textbf{0.534} & \textbf{0.358} & 0.210 \\
TIM3 & 1 & 0.532 & \textbf{0.535} & \textbf{0.465} & 0.452 & \textbf{0.201} & 0.162 \\
HER2 & 5 & \textbf{0.658} & 0.633 & 0.600 & \textbf{0.609} & \textbf{0.443} & 0.217 \\
CD16 & 2 & 0.595 & \textbf{0.666} & 0.467 & \textbf{0.488} & \textbf{0.378} & 0.161 \\
cCasp3$+$cPARP & 2 & \textbf{0.655} & 0.639 & \textbf{0.591} & 0.587 & \textbf{0.390} & 0.060 \\
GATA3 & 3 & \textbf{0.661} & 0.654 & 0.525 & \textbf{0.551} & \textbf{0.475} & 0.211 \\
CAV1 & 1 & 0.597 & \textbf{0.640} & \textbf{0.594} & 0.530 & \textbf{0.438} & 0.299 \\
RORgammaT & 1 & \textbf{0.481} & 0.449 & \textbf{0.478} & 0.452 & \textbf{0.194} & 0.111 \\
HLA-ABC & 3 & \textbf{0.670} & 0.665 & 0.678 & \textbf{0.681} & \textbf{0.436} & 0.177 \\
Cadherin-11 & 1 & 0.678 & \textbf{0.689} & 0.536 & \textbf{0.542} & 0.393 & \textbf{0.430} \\
b-Catenin & 1 & 0.515 & \textbf{0.516} & 0.371 & \textbf{0.388} & \textbf{0.300} & 0.032 \\
CD146 & 1 & 0.588 & \textbf{0.614} & \textbf{0.463} & 0.438 & 0.323 & \textbf{0.455} \\
Cytokeratin 7 & 3 & \textbf{0.609} & 0.516 & 0.386 & \textbf{0.418} & \textbf{0.520} & 0.042 \\
CD31$+$vWF & 3 & 0.584 & \textbf{0.589} & \textbf{0.559} & 0.555 & \textbf{0.374} & 0.347 \\
CD3 & 9 & 0.549 & \textbf{0.587} & 0.457 & \textbf{0.483} & \textbf{0.350} & 0.200 \\
Collagen Ty1 & 1 & \textbf{0.616} & 0.505 & \textbf{0.581} & 0.421 & \textbf{0.249} & 0.071 \\
CXCL12 & 1 & \textbf{0.648} & 0.574 & \textbf{0.507} & 0.404 & \textbf{0.461} & -0.131 \\
FAP & 2 & 0.541 & \textbf{0.597} & 0.537 & \textbf{0.546} & \textbf{0.315} & -0.019 \\
Androgen Receptor AR & 2 & \textbf{0.594} & 0.573 & 0.472 & \textbf{0.527} & \textbf{0.399} & 0.168 \\
Keratin 14 (KRT14) & 4 & \textbf{0.352} & 0.290 & 0.146 & \textbf{0.228} & \textbf{0.311} & 0.094 \\
Twist & 1 & \textbf{0.590} & 0.575 & 0.483 & \textbf{0.491} & \textbf{0.347} & 0.172 \\
CD44 & 4 & \textbf{0.672} & 0.625 & \textbf{0.722} & 0.680 & \textbf{0.473} & 0.050 \\
Progesterone Receptor A/B & 3 & \textbf{0.587} & 0.539 & \textbf{0.569} & 0.557 & \textbf{0.367} & 0.164 \\
CD45RA & 2 & \textbf{0.626} & 0.517 & \textbf{0.631} & 0.552 & \textbf{0.405} & 0.152 \\
CD40 & 1 & \textbf{0.585} & 0.556 & \textbf{0.552} & 0.527 & \textbf{0.367} & 0.256 \\
HLA-DR & 7 & 0.564 & \textbf{0.614} & 0.568 & \textbf{0.594} & \textbf{0.426} & 0.112 \\
CD10 & 1 & \textbf{0.498} & 0.483 & \textbf{0.564} & 0.560 & \textbf{0.360} & 0.253 \\
Bcl-2 & 2 & \textbf{0.541} & 0.488 & 0.402 & \textbf{0.457} & \textbf{0.386} & 0.164 \\
CD11c & 7 & 0.536 & \textbf{0.549} & 0.467 & \textbf{0.486} & \textbf{0.343} & 0.275 \\
CD27 & 1 & 0.572 & \textbf{0.596} & 0.560 & \textbf{0.567} & \textbf{0.347} & 0.316 \\
CD163 & 3 & 0.551 & \textbf{0.564} & \textbf{0.405} & 0.399 & \textbf{0.398} & 0.233 \\
CD68 & 9 & \textbf{0.543} & 0.539 & 0.481 & \textbf{0.483} & \textbf{0.421} & 0.262 \\
panCK & 4 & \textbf{0.652} & 0.608 & \textbf{0.568} & 0.550 & \textbf{0.537} & 0.044 \\
S6 & 4 & \textbf{0.559} & 0.512 & 0.532 & \textbf{0.546} & \textbf{0.439} & 0.144 \\
Fibronectin & 4 & \textbf{0.554} & 0.376 & \textbf{0.597} & 0.434 & \textbf{0.303} & 0.084 \\
\bottomrule
\end{tabular}\hspace{0.9em}%
\begin{tabular}[t]{@{}l r@{\hskip 5pt} r r@{\hskip 5pt} r r@{\hskip 5pt} r r@{}}
\toprule
Protein & $n$ & \multicolumn{2}{c}{Pearson $r$} & \multicolumn{2}{c}{Spearman $r$} & \multicolumn{2}{c}{SSIM} \\
\cmidrule(lr){3-4}\cmidrule(lr){5-6}\cmidrule(lr){7-8}
 &  & M2 & VT & M2 & VT & M2 & VT \\
\midrule
CD33 & 1 & \textbf{0.553} & 0.551 & \textbf{0.625} & 0.623 & \textbf{0.384} & 0.330 \\
LYVE-1 & 1 & \textbf{0.456} & 0.347 & \textbf{0.432} & 0.345 & \textbf{0.262} & 0.214 \\
p53 & 3 & \textbf{0.526} & 0.508 & 0.435 & \textbf{0.465} & \textbf{0.329} & 0.032 \\
CD11b & 1 & \textbf{0.588} & 0.583 & 0.456 & \textbf{0.466} & \textbf{0.354} & 0.263 \\
Vimentin & 7 & \textbf{0.580} & 0.564 & \textbf{0.589} & 0.581 & \textbf{0.434} & -0.021 \\
CD73 & 1 & \textbf{0.473} & 0.394 & \textbf{0.434} & 0.397 & \textbf{0.300} & 0.140 \\
FSP1 & 1 & \textbf{0.539} & 0.526 & \textbf{0.441} & 0.426 & \textbf{0.343} & -0.006 \\
SMA & 8 & \textbf{0.583} & 0.542 & \textbf{0.530} & 0.469 & \textbf{0.448} & -0.042 \\
Granzyme B & 1 & \textbf{0.522} & 0.507 & \textbf{0.588} & 0.563 & \textbf{0.365} & 0.351 \\
CD206 & 1 & 0.545 & \textbf{0.587} & \textbf{0.469} & 0.466 & \textbf{0.425} & 0.376 \\
PD-L1 & 2 & \textbf{0.423} & 0.395 & \textbf{0.454} & 0.429 & \textbf{0.258} & 0.134 \\
CD31 & 4 & 0.536 & \textbf{0.586} & \textbf{0.506} & 0.481 & \textbf{0.361} & 0.326 \\
Slug & 1 & \textbf{0.526} & 0.499 & \textbf{0.444} & 0.434 & \textbf{0.350} & 0.230 \\
CD7 & 1 & 0.516 & \textbf{0.532} & 0.498 & \textbf{0.538} & \textbf{0.371} & 0.236 \\
Sox9 & 1 & 0.525 & \textbf{0.537} & 0.515 & \textbf{0.527} & \textbf{0.409} & 0.193 \\
PDGFRB & 3 & 0.521 & \textbf{0.559} & \textbf{0.515} & 0.495 & \textbf{0.314} & 0.193 \\
VISTA & 1 & \textbf{0.509} & 0.437 & \textbf{0.558} & 0.476 & \textbf{0.370} & 0.179 \\
FOXP3 & 4 & \textbf{0.483} & 0.427 & \textbf{0.441} & 0.401 & \textbf{0.289} & 0.156 \\
TCF1 & 2 & 0.463 & \textbf{0.484} & 0.461 & \textbf{0.463} & \textbf{0.271} & 0.230 \\
GLUT1 & 1 & \textbf{0.511} & 0.497 & \textbf{0.562} & 0.537 & \textbf{0.281} & 0.060 \\
CD8 & 7 & 0.473 & \textbf{0.565} & 0.408 & \textbf{0.428} & \textbf{0.358} & 0.300 \\
CD57 & 2 & \textbf{0.401} & 0.328 & \textbf{0.456} & 0.372 & \textbf{0.269} & 0.163 \\
TBET & 1 & \textbf{0.325} & 0.240 & \textbf{0.338} & 0.231 & \textbf{0.147} & 0.132 \\
Carbonic Anhydrase IX & 5 & \textbf{0.558} & 0.524 & \textbf{0.559} & 0.544 & \textbf{0.367} & 0.217 \\
PD-1 & 3 & \textbf{0.364} & 0.361 & 0.322 & \textbf{0.331} & \textbf{0.226} & 0.113 \\
VCAM1 & 1 & 0.375 & \textbf{0.387} & 0.325 & \textbf{0.362} & \textbf{0.280} & 0.045 \\
FSP1\_S100A4 & 2 & \textbf{0.535} & 0.425 & \textbf{0.536} & 0.439 & \textbf{0.356} & -0.067 \\
ICOS & 2 & 0.386 & \textbf{0.419} & 0.339 & \textbf{0.352} & \textbf{0.233} & 0.200 \\
CD34 & 1 & \textbf{0.424} & 0.416 & 0.418 & \textbf{0.453} & \textbf{0.271} & 0.146 \\
CD20 & 8 & \textbf{0.506} & 0.461 & \textbf{0.463} & 0.434 & \textbf{0.333} & 0.224 \\
IDO & 3 & \textbf{0.353} & 0.326 & \textbf{0.380} & 0.344 & \textbf{0.251} & 0.164 \\
CD303 & 1 & \textbf{0.481} & 0.461 & 0.509 & \textbf{0.512} & \textbf{0.302} & 0.276 \\
Ki-67 & 8 & \textbf{0.473} & 0.378 & \textbf{0.457} & 0.410 & \textbf{0.401} & 0.173 \\
PAX5 & 1 & \textbf{0.398} & 0.204 & \textbf{0.388} & 0.227 & \textbf{0.216} & 0.138 \\
Podoplanin & 2 & \textbf{0.462} & 0.456 & 0.417 & \textbf{0.430} & \textbf{0.289} & 0.038 \\
CD38 & 3 & 0.461 & \textbf{0.463} & \textbf{0.394} & 0.378 & \textbf{0.281} & 0.208 \\
CXCL13 & 1 & \textbf{0.460} & 0.398 & \textbf{0.412} & 0.391 & \textbf{0.270} & 0.224 \\
CD15 & 4 & 0.326 & \textbf{0.352} & 0.292 & \textbf{0.339} & \textbf{0.301} & 0.135 \\
Cytokeratin 5 & 5 & \textbf{0.417} & 0.372 & 0.316 & \textbf{0.332} & \textbf{0.290} & 0.099 \\
CD134 & 1 & 0.111 & \textbf{0.120} & 0.019 & \textbf{0.045} & \textbf{0.138} & 0.089 \\
LAG3 & 2 & \textbf{0.310} & 0.204 & \textbf{0.308} & 0.217 & \textbf{0.207} & 0.049 \\
GITR & 1 & \textbf{0.401} & 0.380 & \textbf{0.338} & 0.324 & \textbf{0.204} & 0.118 \\
CCL21\_6Ckine & 1 & \textbf{0.325} & 0.307 & \textbf{0.248} & 0.248 & \textbf{0.217} & 0.108 \\
CD56 & 2 & \textbf{0.217} & 0.186 & \textbf{0.228} & 0.204 & \textbf{0.198} & 0.080 \\
p75\_CD271 & 1 & \textbf{0.314} & 0.274 & \textbf{0.251} & 0.240 & \textbf{0.268} & 0.114 \\
cleaved PARP & 3 & \textbf{0.492} & 0.476 & 0.456 & \textbf{0.486} & \textbf{0.365} & 0.125 \\
pErk1/2 & 1 & 0.315 & \textbf{0.395} & 0.395 & \textbf{0.410} & \textbf{0.382} & -0.039 \\
MPO & 2 & 0.208 & \textbf{0.294} & 0.099 & \textbf{0.114} & \textbf{0.199} & 0.185 \\
CK17 & 1 & \textbf{0.237} & 0.188 & \textbf{0.266} & 0.185 & \textbf{0.221} & 0.188 \\
Calprotectin & 1 & 0.090 & \textbf{0.157} & 0.013 & \textbf{0.077} & \textbf{0.389} & 0.024 \\
ChyTr & 1 & 0.050 & \textbf{0.164} & -0.033 & \textbf{0.089} & \textbf{0.557} & 0.011 \\
\midrule
\textbf{Mean} &  & \textbf{0.516} & 0.495 & \textbf{0.475} & 0.462 & \textbf{0.353} & 0.157 \\
\bottomrule
\end{tabular}
\label{suppl.tab:mx_17_cohorts_breast}
\end{table}
\pagebreak
\begin{table}[H]
\centering
\caption{\new{Multiplex virtual staining performance on CRC cohorts (6 datasets, 87 markers).
M2 denotes VirTues-M2 and VT denotes VirTues. Metrics are averaged over tiles within a
dataset, then over datasets with equal weight; $n$ is the number of datasets in which the
marker is measured. Bold marks the better model.}}
\scriptsize
\setlength{\tabcolsep}{2.5pt}
\begin{tabular}[t]{@{}l r@{\hskip 5pt} r r@{\hskip 5pt} r r@{\hskip 5pt} r r@{}}
\toprule
Protein & $n$ & \multicolumn{2}{c}{Pearson $r$} & \multicolumn{2}{c}{Spearman $r$} & \multicolumn{2}{c}{SSIM} \\
\cmidrule(lr){3-4}\cmidrule(lr){5-6}\cmidrule(lr){7-8}
 &  & M2 & VT & M2 & VT & M2 & VT \\
\midrule
Actin & 1 & \textbf{0.828} & 0.477 & \textbf{0.894} & 0.708 & \textbf{0.768} & 0.121 \\
CHROMOGRANIN A & 1 & \textbf{0.761} & 0.517 & \textbf{0.567} & 0.392 & \textbf{0.498} & 0.136 \\
CD5 & 1 & \textbf{0.758} & 0.666 & \textbf{0.780} & 0.696 & \textbf{0.533} & 0.328 \\
CD123 & 1 & \textbf{0.746} & 0.388 & \textbf{0.785} & 0.481 & \textbf{0.485} & -0.125 \\
CD2 & 1 & 0.741 & \textbf{0.748} & \textbf{0.738} & 0.706 & \textbf{0.504} & 0.041 \\
Cytokeratin 8/18 & 1 & \textbf{0.800} & 0.732 & \textbf{0.647} & 0.549 & \textbf{0.789} & -0.436 \\
CD45RO & 4 & \textbf{0.742} & 0.739 & \textbf{0.771} & 0.743 & \textbf{0.571} & 0.458 \\
Histone H3 & 6 & \textbf{0.677} & 0.514 & \textbf{0.717} & 0.504 & \textbf{0.471} & 0.225 \\
IRF4 & 1 & \textbf{0.710} & 0.641 & \textbf{0.728} & 0.652 & \textbf{0.459} & 0.190 \\
CD25 & 1 & \textbf{0.680} & 0.584 & \textbf{0.734} & 0.625 & \textbf{0.486} & 0.416 \\
CD14 & 1 & 0.719 & \textbf{0.729} & \textbf{0.718} & 0.715 & \textbf{0.518} & 0.367 \\
MELAN A & 1 & \textbf{0.697} & 0.552 & \textbf{0.668} & 0.516 & \textbf{0.460} & 0.092 \\
MMP12 & 1 & \textbf{0.684} & 0.556 & \textbf{0.671} & 0.556 & \textbf{0.421} & 0.177 \\
E-cadherin & 3 & 0.775 & \textbf{0.777} & 0.752 & \textbf{0.757} & \textbf{0.624} & 0.090 \\
CD30 & 1 & \textbf{0.679} & 0.661 & 0.636 & \textbf{0.684} & \textbf{0.459} & 0.368 \\
CD4 & 5 & \textbf{0.755} & 0.738 & \textbf{0.774} & 0.654 & \textbf{0.573} & 0.353 \\
CD45 & 4 & 0.703 & \textbf{0.725} & \textbf{0.712} & 0.644 & \textbf{0.537} & 0.195 \\
EpCAM & 1 & \textbf{0.670} & 0.453 & \textbf{0.728} & 0.545 & \textbf{0.481} & 0.192 \\
beta-tubulin & 1 & \textbf{0.673} & 0.638 & \textbf{0.675} & 0.657 & \textbf{0.451} & 0.180 \\
HEP-PAR-1 & 1 & \textbf{0.671} & 0.603 & 0.526 & \textbf{0.577} & \textbf{0.420} & 0.303 \\
B2M & 1 & \textbf{0.662} & 0.643 & 0.686 & \textbf{0.691} & \textbf{0.446} & 0.211 \\
MMP9 & 1 & \textbf{0.719} & 0.688 & \textbf{0.579} & 0.553 & \textbf{0.486} & 0.178 \\
EGFR & 1 & \textbf{0.742} & 0.680 & \textbf{0.751} & 0.697 & \textbf{0.488} & 0.431 \\
CCR4 & 1 & \textbf{0.675} & 0.601 & \textbf{0.751} & 0.690 & \textbf{0.460} & 0.294 \\
CD16 & 1 & 0.684 & \textbf{0.760} & 0.568 & \textbf{0.599} & \textbf{0.442} & 0.207 \\
GATA3 & 1 & \textbf{0.645} & 0.621 & \textbf{0.646} & 0.596 & \textbf{0.436} & 0.429 \\
HLA-ABC & 1 & \textbf{0.625} & 0.616 & \textbf{0.659} & 0.651 & \textbf{0.447} & 0.356 \\
b-Catenin & 1 & \textbf{0.770} & 0.523 & \textbf{0.783} & 0.503 & \textbf{0.445} & 0.213 \\
Cytokeratin 7 & 1 & \textbf{0.637} & 0.608 & \textbf{0.542} & 0.452 & \textbf{0.484} & 0.298 \\
CD3 & 5 & 0.696 & \textbf{0.763} & \textbf{0.689} & 0.593 & \textbf{0.560} & 0.169 \\
CD107a & 1 & \textbf{0.617} & 0.564 & \textbf{0.853} & 0.803 & \textbf{0.473} & -0.022 \\
GFAP & 1 & \textbf{0.608} & 0.428 & \textbf{0.583} & 0.466 & \textbf{0.400} & -0.080 \\
NA-K-ATPASE & 1 & \textbf{0.599} & 0.554 & 0.586 & \textbf{0.606} & \textbf{0.422} & 0.163 \\
Keratin 14 (KRT14) & 1 & 0.810 & \textbf{0.824} & 0.838 & \textbf{0.842} & \textbf{0.622} & 0.419 \\
CD44 & 1 & 0.602 & \textbf{0.611} & 0.627 & \textbf{0.637} & \textbf{0.448} & -0.038 \\
CD45RA & 2 & \textbf{0.531} & 0.463 & \textbf{0.539} & 0.505 & \textbf{0.373} & 0.258 \\
CD40 & 1 & \textbf{0.585} & 0.556 & \textbf{0.552} & 0.527 & \textbf{0.367} & 0.256 \\
HLA-DR & 4 & \textbf{0.594} & 0.539 & \textbf{0.619} & 0.566 & \textbf{0.406} & 0.132 \\
Bcl-2 & 1 & \textbf{0.620} & 0.534 & 0.597 & \textbf{0.618} & \textbf{0.446} & 0.177 \\
CD71 & 1 & \textbf{0.578} & 0.437 & \textbf{0.665} & 0.521 & \textbf{0.388} & -0.115 \\
CD11c & 4 & \textbf{0.550} & 0.481 & \textbf{0.625} & 0.578 & \textbf{0.460} & 0.288 \\
CD27 & 1 & 0.572 & \textbf{0.596} & 0.560 & \textbf{0.567} & \textbf{0.347} & 0.316 \\
MUC-1 & 1 & \textbf{0.567} & 0.536 & 0.542 & \textbf{0.603} & \textbf{0.404} & 0.032 \\
CD163 & 3 & 0.527 & \textbf{0.582} & \textbf{0.521} & 0.483 & \textbf{0.558} & 0.452 \\
\bottomrule
\end{tabular}\hspace{0.9em}%
\begin{tabular}[t]{@{}l r@{\hskip 5pt} r r@{\hskip 5pt} r r@{\hskip 5pt} r r@{}}
\toprule
Protein & $n$ & \multicolumn{2}{c}{Pearson $r$} & \multicolumn{2}{c}{Spearman $r$} & \multicolumn{2}{c}{SSIM} \\
\cmidrule(lr){3-4}\cmidrule(lr){5-6}\cmidrule(lr){7-8}
 &  & M2 & VT & M2 & VT & M2 & VT \\
\midrule
CD68 & 5 & 0.545 & \textbf{0.590} & \textbf{0.665} & 0.606 & \textbf{0.537} & 0.428 \\
panCK & 2 & \textbf{0.475} & 0.469 & \textbf{0.454} & 0.443 & \textbf{0.422} & 0.178 \\
CD117 & 1 & \textbf{0.654} & 0.617 & \textbf{0.729} & 0.675 & \textbf{0.458} & -0.041 \\
CD33 & 1 & \textbf{0.553} & 0.551 & \textbf{0.625} & 0.623 & \textbf{0.384} & 0.330 \\
CD21 & 2 & \textbf{0.709} & 0.676 & \textbf{0.700} & 0.509 & \textbf{0.469} & 0.413 \\
p53 & 1 & \textbf{0.572} & 0.564 & \textbf{0.670} & 0.618 & \textbf{0.384} & 0.197 \\
CD11b & 1 & \textbf{0.625} & 0.575 & \textbf{0.737} & 0.616 & \textbf{0.465} & 0.384 \\
Vimentin & 2 & \textbf{0.416} & 0.363 & \textbf{0.458} & 0.368 & \textbf{0.311} & 0.048 \\
SMA & 3 & 0.455 & \textbf{0.461} & 0.418 & \textbf{0.420} & \textbf{0.443} & 0.010 \\
Granzyme B & 2 & \textbf{0.615} & 0.589 & 0.553 & \textbf{0.556} & \textbf{0.442} & 0.282 \\
CD206 & 1 & 0.545 & \textbf{0.587} & \textbf{0.469} & 0.466 & \textbf{0.425} & 0.376 \\
PD-L1 & 3 & \textbf{0.596} & 0.544 & \textbf{0.609} & 0.579 & \textbf{0.466} & 0.231 \\
CD31 & 4 & 0.537 & \textbf{0.557} & \textbf{0.619} & 0.512 & \textbf{0.579} & 0.500 \\
COLLAGEN IV & 1 & \textbf{0.526} & 0.493 & \textbf{0.538} & 0.527 & \textbf{0.434} & 0.154 \\
CD7 & 2 & 0.544 & \textbf{0.551} & 0.527 & \textbf{0.558} & \textbf{0.414} & 0.238 \\
PDGFRB & 1 & 0.459 & \textbf{0.537} & 0.484 & \textbf{0.517} & \textbf{0.295} & 0.263 \\
VISTA & 2 & \textbf{0.533} & 0.519 & \textbf{0.625} & 0.586 & \textbf{0.428} & 0.298 \\
FOXP3 & 3 & \textbf{0.534} & 0.454 & \textbf{0.531} & 0.354 & \textbf{0.513} & 0.374 \\
TCF1 & 1 & 0.517 & \textbf{0.537} & \textbf{0.564} & 0.556 & \textbf{0.344} & 0.300 \\
CD138 & 1 & \textbf{0.508} & 0.496 & \textbf{0.518} & 0.489 & \textbf{0.345} & 0.162 \\
CD8 & 5 & 0.547 & \textbf{0.616} & \textbf{0.529} & 0.376 & \textbf{0.551} & 0.480 \\
CD57 & 1 & \textbf{0.606} & 0.482 & \textbf{0.641} & 0.495 & \textbf{0.443} & 0.275 \\
SYNAPTOPHYSIN & 1 & \textbf{0.498} & 0.449 & \textbf{0.526} & 0.445 & \textbf{0.367} & 0.097 \\
TBET & 1 & \textbf{0.669} & 0.626 & \textbf{0.678} & 0.642 & \textbf{0.449} & 0.359 \\
Carbonic Anhydrase IX & 1 & \textbf{0.459} & 0.424 & \textbf{0.570} & 0.520 & \textbf{0.314} & 0.211 \\
PD-1 & 3 & \textbf{0.660} & 0.648 & \textbf{0.683} & 0.666 & \textbf{0.523} & 0.356 \\
ICOS & 2 & 0.567 & \textbf{0.577} & \textbf{0.563} & 0.542 & \textbf{0.361} & 0.346 \\
CD34 & 1 & 0.429 & \textbf{0.530} & 0.397 & \textbf{0.443} & \textbf{0.397} & 0.233 \\
CD20 & 5 & \textbf{0.572} & 0.466 & \textbf{0.531} & 0.420 & \textbf{0.512} & 0.358 \\
IDO & 2 & \textbf{0.596} & 0.496 & \textbf{0.643} & 0.519 & \textbf{0.386} & 0.206 \\
CD303 & 1 & \textbf{0.481} & 0.461 & 0.509 & \textbf{0.512} & \textbf{0.302} & 0.276 \\
Ki-67 & 4 & 0.484 & \textbf{0.520} & \textbf{0.561} & 0.512 & \textbf{0.557} & 0.363 \\
PAX5 & 2 & \textbf{0.545} & 0.320 & \textbf{0.533} & 0.333 & \textbf{0.289} & -0.010 \\
Podoplanin & 2 & \textbf{0.467} & 0.404 & \textbf{0.609} & 0.601 & \textbf{0.449} & 0.233 \\
CD38 & 2 & \textbf{0.506} & 0.489 & \textbf{0.551} & 0.488 & \textbf{0.353} & 0.164 \\
CDX2 & 1 & \textbf{0.452} & 0.300 & \textbf{0.498} & 0.343 & \textbf{0.311} & -0.228 \\
CD79A & 1 & \textbf{0.434} & 0.303 & \textbf{0.271} & 0.203 & \textbf{0.333} & 0.229 \\
CD15 & 2 & 0.457 & \textbf{0.542} & 0.455 & \textbf{0.487} & \textbf{0.367} & 0.144 \\
CD134 & 1 & \textbf{0.716} & 0.505 & \textbf{0.732} & 0.580 & \textbf{0.459} & 0.304 \\
LAG3 & 2 & \textbf{0.546} & 0.429 & \textbf{0.522} & 0.413 & \textbf{0.358} & 0.278 \\
CD56 & 2 & \textbf{0.432} & 0.332 & \textbf{0.481} & 0.383 & \textbf{0.346} & 0.250 \\
cleaved PARP & 1 & \textbf{0.281} & 0.274 & 0.236 & \textbf{0.265} & \textbf{0.253} & 0.103 \\
MPO & 1 & 0.301 & \textbf{0.415} & 0.174 & \textbf{0.203} & 0.238 & \textbf{0.240} \\
\midrule
\textbf{Mean} &  & \textbf{0.601} & 0.552 & \textbf{0.609} & 0.548 & \textbf{0.446} & 0.219 \\
\bottomrule
\end{tabular}
\label{suppl.tab:mx_17_cohorts_crc}
\end{table}
\pagebreak
\begin{table}[H]
\centering
\caption{\new{Multiplex virtual staining performance on Lung cohorts (7 datasets, 87 markers).
M2 denotes VirTues-M2 and VT denotes VirTues. Metrics are averaged over tiles within a
dataset, then over datasets with equal weight; $n$ is the number of datasets in which the
marker is measured. Bold marks the better model.}}
\scriptsize
\setlength{\tabcolsep}{2.5pt}
\begin{tabular}[t]{@{}l r@{\hskip 5pt} r r@{\hskip 5pt} r r@{\hskip 5pt} r r@{}}
\toprule
Protein & $n$ & \multicolumn{2}{c}{Pearson $r$} & \multicolumn{2}{c}{Spearman $r$} & \multicolumn{2}{c}{SSIM} \\
\cmidrule(lr){3-4}\cmidrule(lr){5-6}\cmidrule(lr){7-8}
 &  & M2 & VT & M2 & VT & M2 & VT \\
\midrule
CD183 & 1 & \textbf{0.855} & 0.752 & \textbf{0.859} & 0.755 & \textbf{0.521} & 0.379 \\
CD63 & 1 & \textbf{0.727} & 0.664 & \textbf{0.791} & 0.741 & \textbf{0.438} & 0.286 \\
Cytokeratin 8/18 & 2 & \textbf{0.707} & 0.508 & \textbf{0.692} & 0.528 & \textbf{0.498} & 0.018 \\
S100A8 & 1 & \textbf{0.720} & 0.586 & \textbf{0.739} & 0.668 & \textbf{0.560} & 0.239 \\
CD45RO & 4 & \textbf{0.700} & 0.674 & \textbf{0.673} & 0.655 & \textbf{0.489} & 0.289 \\
Histone H3 & 6 & \textbf{0.747} & 0.691 & \textbf{0.801} & 0.767 & \textbf{0.541} & 0.223 \\
CD25 & 1 & \textbf{0.739} & 0.464 & \textbf{0.756} & 0.502 & \textbf{0.466} & 0.335 \\
CD14 & 3 & \textbf{0.719} & 0.700 & \textbf{0.751} & 0.722 & \textbf{0.480} & 0.303 \\
MMP11 & 1 & \textbf{0.693} & 0.689 & \textbf{0.726} & 0.726 & \textbf{0.448} & 0.278 \\
Arginase1 & 1 & \textbf{0.684} & 0.559 & \textbf{0.645} & 0.593 & \textbf{0.370} & 0.158 \\
E-cadherin & 3 & 0.632 & \textbf{0.652} & 0.686 & \textbf{0.699} & \textbf{0.391} & 0.124 \\
CD4 & 6 & 0.600 & \textbf{0.624} & 0.537 & \textbf{0.556} & \textbf{0.368} & 0.182 \\
CD45 & 2 & 0.718 & \textbf{0.735} & \textbf{0.644} & 0.602 & \textbf{0.483} & -0.002 \\
B2M & 1 & \textbf{0.662} & 0.643 & 0.686 & \textbf{0.691} & \textbf{0.446} & 0.211 \\
MMP9 & 1 & \textbf{0.682} & 0.655 & \textbf{0.715} & 0.697 & \textbf{0.457} & 0.241 \\
CD248 & 1 & \textbf{0.663} & 0.610 & \textbf{0.728} & 0.689 & \textbf{0.445} & 0.264 \\
CCR4 & 1 & \textbf{0.647} & 0.585 & \textbf{0.655} & 0.632 & \textbf{0.312} & 0.234 \\
TIM3 & 1 & \textbf{0.785} & 0.763 & \textbf{0.802} & 0.781 & \textbf{0.389} & 0.271 \\
CD16 & 2 & 0.686 & \textbf{0.718} & 0.613 & \textbf{0.630} & \textbf{0.416} & 0.221 \\
CAV1 & 1 & \textbf{0.706} & 0.575 & \textbf{0.744} & 0.625 & \textbf{0.462} & 0.328 \\
RORgammaT & 1 & \textbf{0.818} & 0.492 & \textbf{0.852} & 0.538 & \textbf{0.481} & 0.348 \\
Cadherin-11 & 1 & \textbf{0.607} & 0.573 & \textbf{0.606} & 0.599 & \textbf{0.397} & 0.201 \\
PF4 & 1 & \textbf{0.639} & 0.569 & \textbf{0.649} & 0.616 & \textbf{0.332} & 0.177 \\
CD146 & 1 & \textbf{0.664} & 0.639 & \textbf{0.674} & 0.620 & \textbf{0.441} & 0.399 \\
CD31$+$vWF & 1 & 0.653 & \textbf{0.688} & \textbf{0.668} & 0.666 & \textbf{0.380} & 0.310 \\
C1Qa & 1 & \textbf{0.618} & 0.604 & 0.592 & \textbf{0.625} & \textbf{0.356} & 0.214 \\
CD3 & 6 & 0.585 & \textbf{0.644} & 0.500 & \textbf{0.527} & \textbf{0.372} & 0.125 \\
Collagen Ty1 & 4 & \textbf{0.602} & 0.389 & \textbf{0.662} & 0.222 & \textbf{0.410} & -0.023 \\
CD107a & 1 & 0.600 & \textbf{0.646} & \textbf{0.696} & 0.691 & \textbf{0.451} & -0.083 \\
CXCL12 & 1 & \textbf{0.563} & 0.556 & 0.438 & \textbf{0.498} & \textbf{0.441} & -0.007 \\
FAP & 1 & \textbf{0.655} & 0.571 & \textbf{0.701} & 0.634 & \textbf{0.422} & -0.009 \\
Keratin 14 (KRT14) & 3 & \textbf{0.611} & 0.520 & \textbf{0.626} & 0.582 & \textbf{0.420} & 0.253 \\
CD44 & 3 & \textbf{0.493} & 0.446 & \textbf{0.491} & 0.469 & \textbf{0.337} & 0.120 \\
CD45RA & 4 & \textbf{0.601} & 0.518 & \textbf{0.595} & 0.562 & \textbf{0.413} & 0.219 \\
CD40 & 1 & \textbf{0.585} & 0.556 & \textbf{0.552} & 0.527 & \textbf{0.367} & 0.256 \\
HLA-DR & 5 & \textbf{0.597} & 0.580 & \textbf{0.652} & 0.620 & \textbf{0.446} & 0.183 \\
CD10 & 1 & \textbf{0.671} & 0.614 & \textbf{0.747} & 0.709 & \textbf{0.496} & 0.258 \\
CD11c & 4 & \textbf{0.646} & 0.550 & \textbf{0.633} & 0.571 & \textbf{0.391} & 0.307 \\
CD27 & 1 & 0.572 & \textbf{0.596} & 0.560 & \textbf{0.567} & \textbf{0.347} & 0.316 \\
CD141 & 1 & \textbf{0.567} & 0.361 & \textbf{0.533} & 0.355 & \textbf{0.441} & 0.234 \\
CD163 & 4 & \textbf{0.613} & 0.584 & \textbf{0.447} & 0.446 & \textbf{0.444} & 0.319 \\
CD68 & 6 & 0.602 & \textbf{0.616} & \textbf{0.663} & 0.658 & \textbf{0.505} & 0.373 \\
CD117 & 1 & \textbf{0.454} & 0.350 & \textbf{0.369} & 0.286 & \textbf{0.404} & 0.187 \\
CD33 & 1 & \textbf{0.553} & 0.551 & \textbf{0.625} & 0.623 & \textbf{0.384} & 0.330 \\
\bottomrule
\end{tabular}\hspace{0.9em}%
\begin{tabular}[t]{@{}l r@{\hskip 5pt} r r@{\hskip 5pt} r r@{\hskip 5pt} r r@{}}
\toprule
Protein & $n$ & \multicolumn{2}{c}{Pearson $r$} & \multicolumn{2}{c}{Spearman $r$} & \multicolumn{2}{c}{SSIM} \\
\cmidrule(lr){3-4}\cmidrule(lr){5-6}\cmidrule(lr){7-8}
 &  & M2 & VT & M2 & VT & M2 & VT \\
\midrule
LYVE-1 & 1 & \textbf{0.650} & 0.524 & \textbf{0.670} & 0.568 & \textbf{0.405} & 0.174 \\
CD21 & 1 & \textbf{0.396} & 0.375 & \textbf{0.210} & 0.171 & \textbf{0.407} & 0.239 \\
CD11b & 2 & \textbf{0.419} & 0.406 & 0.248 & \textbf{0.300} & \textbf{0.315} & 0.137 \\
Vimentin & 5 & \textbf{0.636} & 0.570 & \textbf{0.668} & 0.583 & \textbf{0.455} & -0.006 \\
CD66a & 1 & \textbf{0.540} & 0.506 & 0.633 & \textbf{0.633} & \textbf{0.390} & 0.259 \\
CD73 & 1 & \textbf{0.606} & 0.566 & \textbf{0.645} & 0.626 & \textbf{0.394} & 0.178 \\
SMA & 5 & \textbf{0.563} & 0.509 & \textbf{0.606} & 0.474 & \textbf{0.453} & -0.089 \\
Granzyme B & 5 & \textbf{0.464} & 0.391 & \textbf{0.389} & 0.358 & \textbf{0.366} & 0.217 \\
CD206 & 2 & 0.511 & \textbf{0.545} & 0.398 & \textbf{0.414} & \textbf{0.376} & 0.317 \\
PD-L1 & 2 & \textbf{0.580} & 0.477 & \textbf{0.601} & 0.499 & \textbf{0.379} & 0.157 \\
CDH6 & 1 & \textbf{0.532} & 0.492 & 0.448 & \textbf{0.490} & \textbf{0.339} & 0.125 \\
CD31 & 3 & \textbf{0.515} & 0.492 & \textbf{0.428} & 0.424 & \textbf{0.345} & 0.311 \\
CD7 & 1 & 0.516 & \textbf{0.532} & 0.498 & \textbf{0.538} & \textbf{0.371} & 0.236 \\
PDGFRB & 2 & \textbf{0.580} & 0.545 & \textbf{0.600} & 0.570 & \textbf{0.366} & 0.177 \\
Siglec8 & 1 & \textbf{0.518} & 0.510 & \textbf{0.499} & 0.488 & \textbf{0.414} & -0.029 \\
VISTA & 1 & \textbf{0.509} & 0.437 & \textbf{0.558} & 0.476 & \textbf{0.370} & 0.179 \\
FOXP3 & 6 & \textbf{0.529} & 0.444 & \textbf{0.478} & 0.423 & \textbf{0.331} & 0.199 \\
TCF1 & 3 & \textbf{0.558} & 0.506 & \textbf{0.560} & 0.538 & \textbf{0.376} & 0.224 \\
CD8 & 6 & 0.501 & \textbf{0.592} & 0.453 & \textbf{0.477} & \textbf{0.387} & 0.294 \\
Carbonic Anhydrase IX & 2 & \textbf{0.474} & 0.428 & \textbf{0.547} & 0.515 & \textbf{0.336} & 0.196 \\
PD-1 & 3 & \textbf{0.466} & 0.442 & 0.407 & \textbf{0.435} & \textbf{0.298} & 0.153 \\
VCAM1 & 1 & \textbf{0.614} & 0.566 & \textbf{0.653} & 0.640 & \textbf{0.404} & 0.042 \\
FSP1\_S100A4 & 1 & \textbf{0.448} & 0.315 & \textbf{0.441} & 0.350 & \textbf{0.280} & -0.016 \\
ICOS & 1 & 0.511 & \textbf{0.583} & 0.510 & \textbf{0.538} & \textbf{0.321} & 0.282 \\
CD34 & 2 & \textbf{0.601} & 0.521 & \textbf{0.559} & 0.494 & \textbf{0.447} & 0.088 \\
CD20 & 6 & \textbf{0.373} & 0.336 & 0.272 & \textbf{0.277} & \textbf{0.347} & 0.145 \\
IDO & 2 & \textbf{0.496} & 0.442 & \textbf{0.560} & 0.497 & \textbf{0.361} & 0.221 \\
CD303 & 1 & \textbf{0.481} & 0.461 & 0.509 & \textbf{0.512} & \textbf{0.302} & 0.276 \\
Ki-67 & 6 & \textbf{0.479} & 0.403 & \textbf{0.453} & 0.431 & \textbf{0.388} & 0.143 \\
Podoplanin & 2 & \textbf{0.456} & 0.454 & 0.456 & \textbf{0.485} & \textbf{0.401} & 0.173 \\
CD38 & 2 & \textbf{0.415} & 0.344 & \textbf{0.347} & 0.256 & \textbf{0.344} & 0.208 \\
SPP1 & 1 & \textbf{0.447} & 0.409 & \textbf{0.360} & 0.319 & \textbf{0.385} & 0.149 \\
CD197 & 1 & \textbf{0.447} & 0.271 & \textbf{0.323} & 0.279 & \textbf{0.402} & 0.192 \\
PGP9.5 & 1 & \textbf{0.430} & 0.258 & \textbf{0.409} & 0.308 & \textbf{0.332} & 0.203 \\
CD15 & 3 & 0.486 & \textbf{0.495} & \textbf{0.551} & 0.510 & \textbf{0.405} & 0.217 \\
LAG3 & 1 & \textbf{0.363} & 0.235 & \textbf{0.334} & 0.228 & \textbf{0.255} & 0.077 \\
CCL21\_6Ckine & 1 & \textbf{0.409} & 0.398 & 0.279 & \textbf{0.375} & \textbf{0.342} & 0.130 \\
CD56 & 1 & \textbf{0.448} & 0.372 & \textbf{0.241} & 0.184 & \textbf{0.436} & 0.217 \\
p75\_CD271 & 1 & \textbf{0.413} & 0.369 & 0.279 & \textbf{0.351} & \textbf{0.276} & 0.086 \\
cleaved PARP & 1 & \textbf{0.281} & 0.274 & 0.236 & \textbf{0.265} & \textbf{0.253} & 0.103 \\
CD95 & 1 & \textbf{0.348} & 0.238 & \textbf{0.196} & 0.193 & \textbf{0.233} & 0.109 \\
MPO & 2 & 0.335 & \textbf{0.420} & 0.170 & \textbf{0.219} & 0.246 & \textbf{0.257} \\
CD103 & 1 & \textbf{0.252} & 0.216 & 0.126 & \textbf{0.188} & \textbf{0.191} & 0.086 \\
\midrule
\textbf{Mean} &  & \textbf{0.566} & 0.514 & \textbf{0.547} & 0.510 & \textbf{0.392} & 0.191 \\
\bottomrule
\end{tabular}
\label{suppl.tab:mx_17_cohorts_lung}
\end{table}

\begin{figure}[H]
    \centering
    \includegraphics[width=0.9\linewidth,page = 15]{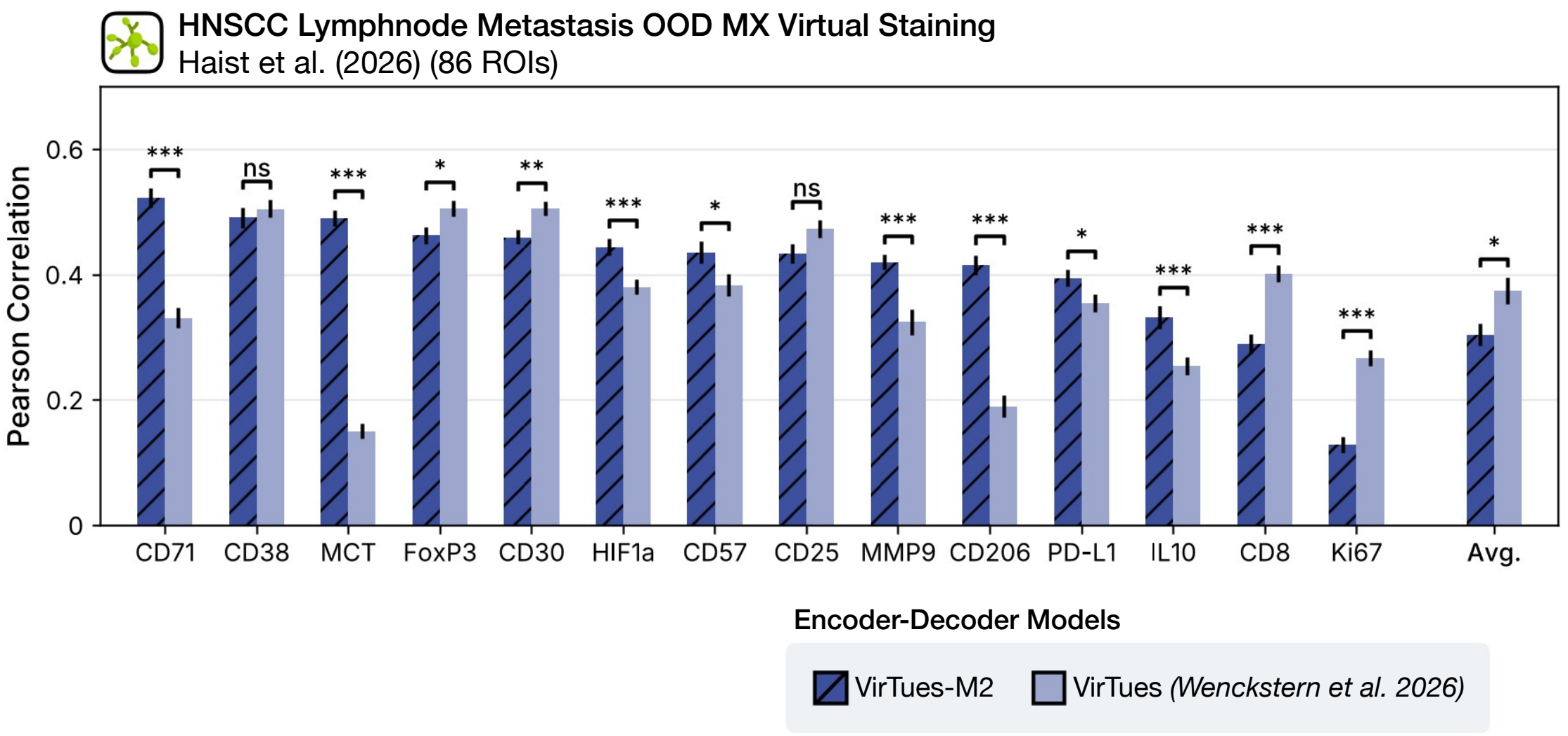}
    \caption{\new{MX-based virtual staining performance (Pearson correlation) on the \texttt{haist2026lymph} OOD cohort.}
}\label{suppl:fig:vs_mx_ood}
\end{figure}

\pagebreak

\begin{figure}[H]
    \centering
    \includegraphics[width=0.7\linewidth,page = 9]{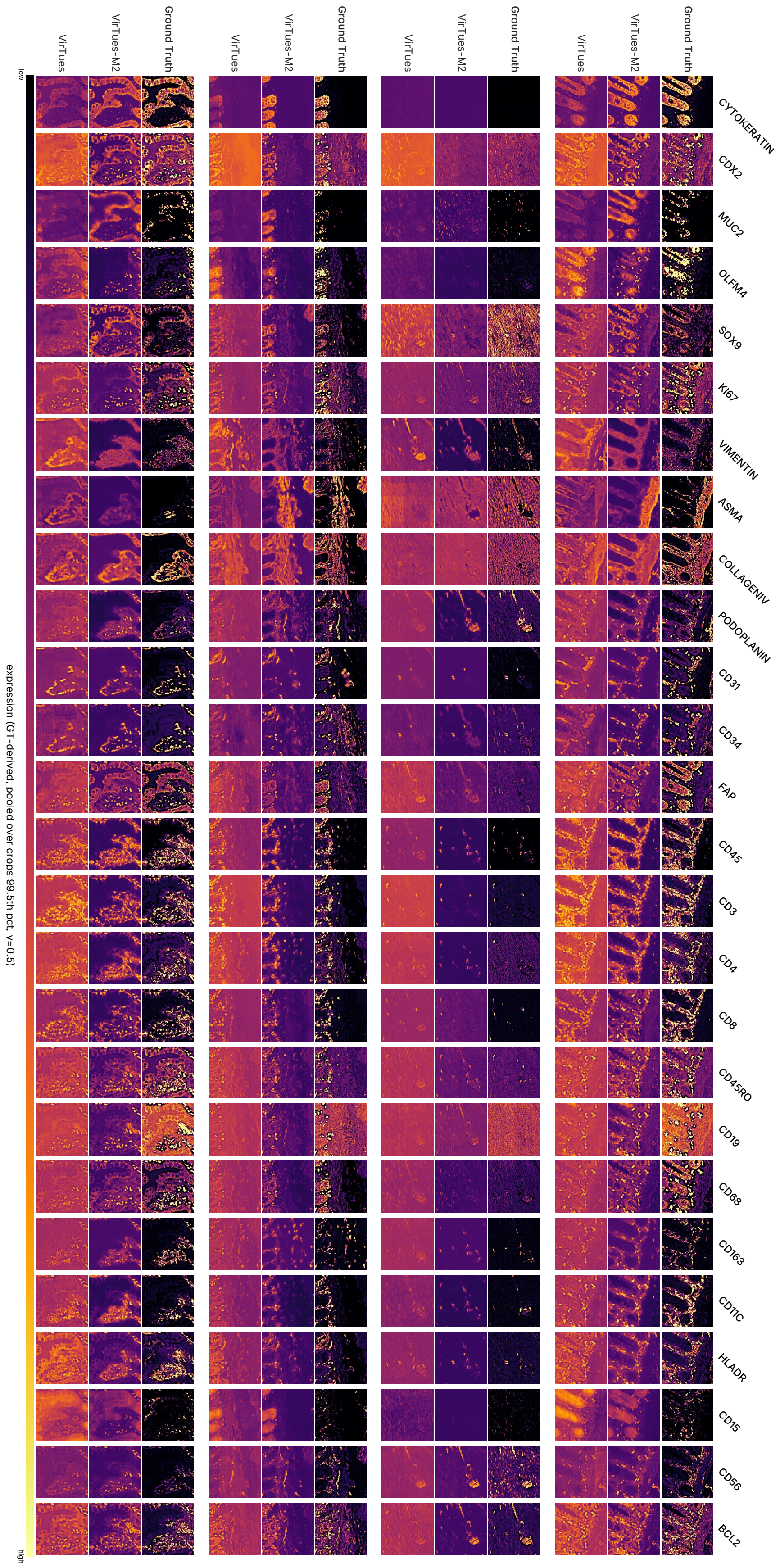}
    \caption{\new{Qualitative virtual staining examples for four crop level samples on 26 clinical relevant markers. We observe that VirTues-M2 achieves more sharp prediction, presumably as it is not only trained on a masked inpainting but full inpainting task.}
}\label{suppl:fig:mx_inpainting}
\end{figure}

\begin{table}[H]
\centering
\caption{\new{Pearson correlation for H\&E-based virtual staining across 11 markers. Scores are averaged over tiles within each of the 4 cohorts, then over cohorts with equal weight; $n$ is the number of contributing cohorts. Bold marks the best model per marker.}}
\small
\setlength{\tabcolsep}{5pt}
\begin{tabular}{lr@{\hskip 8pt}rrrrr}
\toprule
Protein & $n$ & VirTues-M2 & VM2 with UNI & ROSIE & HistoPlexer & HEX \\
\midrule
\multicolumn{7}{l}{\textit{Easy}} \\
Hoechst & 4 & \textbf{0.907} & 0.892 & 0.657 & 0.700 & 0.217 \\
PanCK & 4 & \textbf{0.633} & 0.597 & 0.429 & 0.126 & 0.140 \\
\addlinespace[2pt]
\multicolumn{7}{l}{\textit{Moderate}} \\
CD31 & 4 & \textbf{0.383} & 0.308 & 0.170 & 0.031 & 0.061 \\
CD45 & 4 & \textbf{0.571} & 0.508 & 0.437 & 0.209 & 0.177 \\
CD45RO & 4 & \textbf{0.466} & 0.422 & 0.340 & 0.201 & 0.105 \\
CD68 & 4 & \textbf{0.345} & 0.293 & 0.219 & 0.073 & 0.064 \\
CD3 & 4 & \textbf{0.510} & 0.461 & 0.363 & 0.216 & 0.083 \\
\addlinespace[2pt]
\multicolumn{7}{l}{\textit{Hard}} \\
CD4 & 4 & \textbf{0.446} & 0.381 & 0.313 & 0.183 & 0.083 \\
CD8 & 4 & \textbf{0.237} & 0.208 & 0.132 & 0.076 & 0.025 \\
CD163 & 4 & \textbf{0.294} & 0.224 & 0.140 & 0.052 & 0.035 \\
KI-67 & 4 & \textbf{0.314} & 0.277 & 0.158 & 0.074 & 0.020 \\
\midrule
\textbf{Mean} &  & \textbf{0.464} & 0.416 & 0.305 & 0.176 & 0.092 \\
\bottomrule
\end{tabular}
\label{suppl.tab:he_all4_pearson}
\end{table}

\begin{table}[H]
\centering
\caption{\new{Spearman correlation for H\&E-based virtual staining across 11 markers. Scores are averaged over tiles within each of the 4 cohorts, then over cohorts with equal weight; $n$ is the number of contributing cohorts. Bold marks the best model per marker.}}
\small
\setlength{\tabcolsep}{5pt}
\begin{tabular}{lr@{\hskip 8pt}rrrrr}
\toprule
Protein & $n$ & VirTues-M2 & VM2 with UNI & ROSIE & HistoPlexer & HEX \\
\midrule
\multicolumn{7}{l}{\textit{Easy}} \\
Hoechst & 4 & \textbf{0.863} & 0.855 & 0.664 & 0.576 & 0.000 \\
PanCK & 4 & 0.492 & \textbf{0.518} & 0.404 & 0.116 & 0.000 \\
\addlinespace[2pt]
\multicolumn{7}{l}{\textit{Moderate}} \\
CD31 & 4 & \textbf{0.292} & 0.288 & 0.197 & 0.043 & 0.001 \\
CD45 & 4 & \textbf{0.551} & 0.541 & 0.412 & 0.184 & 0.000 \\
CD45RO & 4 & \textbf{0.498} & 0.470 & 0.329 & 0.190 & 0.000 \\
CD68 & 4 & \textbf{0.485} & 0.477 & 0.343 & 0.110 & 0.000 \\
CD3 & 4 & \textbf{0.448} & 0.443 & 0.311 & 0.188 & 0.000 \\
\addlinespace[2pt]
\multicolumn{7}{l}{\textit{Hard}} \\
CD4 & 4 & \textbf{0.486} & 0.444 & 0.324 & 0.185 & 0.000 \\
CD8 & 4 & 0.252 & \textbf{0.255} & 0.114 & 0.064 & 0.000 \\
CD163 & 4 & \textbf{0.254} & 0.226 & 0.148 & 0.048 & 0.000 \\
KI-67 & 4 & \textbf{0.274} & 0.237 & 0.131 & 0.062 & 0.000 \\
\midrule
\textbf{Mean} &  & \textbf{0.445} & 0.432 & 0.307 & 0.161 & 0.000 \\
\bottomrule
\end{tabular}
\label{suppl.tab:he_all4_spearman}
\end{table}

\begin{table}[H]
\centering
\caption{\new{SSIM for H\&E-based virtual staining across 11 markers. Scores are averaged over tiles within each of the 4 cohorts, then over cohorts with equal weight; $n$ is the number of contributing cohorts. Bold marks the best model per marker.}}
\small
\setlength{\tabcolsep}{5pt}
\begin{tabular}{lr@{\hskip 8pt}rrrrr}
\toprule
Protein & $n$ & VirTues-M2 & VM2 with UNI & ROSIE & HistoPlexer & HEX \\
\midrule
\multicolumn{7}{l}{\textit{Easy}} \\
Hoechst & 4 & \textbf{0.780} & 0.752 & 0.437 & 0.513 & 0.244 \\
PanCK & 4 & 0.618 & \textbf{0.650} & 0.590 & 0.500 & 0.257 \\
\addlinespace[2pt]
\multicolumn{7}{l}{\textit{Moderate}} \\
CD31 & 4 & 0.626 & \textbf{0.665} & 0.607 & 0.563 & 0.334 \\
CD45 & 4 & 0.508 & \textbf{0.527} & 0.443 & 0.469 & 0.287 \\
CD45RO & 4 & \textbf{0.484} & 0.479 & 0.414 & 0.292 & 0.382 \\
CD68 & 4 & 0.522 & \textbf{0.538} & 0.489 & 0.352 & 0.369 \\
CD3 & 4 & 0.522 & \textbf{0.531} & 0.435 & 0.307 & 0.417 \\
\addlinespace[2pt]
\multicolumn{7}{l}{\textit{Hard}} \\
CD4 & 4 & 0.487 & \textbf{0.490} & 0.382 & 0.281 & 0.360 \\
CD8 & 4 & 0.455 & \textbf{0.517} & 0.418 & 0.318 & 0.311 \\
CD163 & 4 & 0.516 & \textbf{0.535} & 0.487 & 0.428 & 0.337 \\
KI-67 & 4 & 0.533 & \textbf{0.553} & 0.485 & 0.286 & 0.386 \\
\midrule
\textbf{Mean} &  & 0.550 & \textbf{0.567} & 0.471 & 0.392 & 0.335 \\
\bottomrule
\end{tabular}
\label{suppl.tab:he_all4_ssim}
\end{table}

\begin{figure}[H]
    \centering
    \includegraphics[width=\linewidth,page = 10]{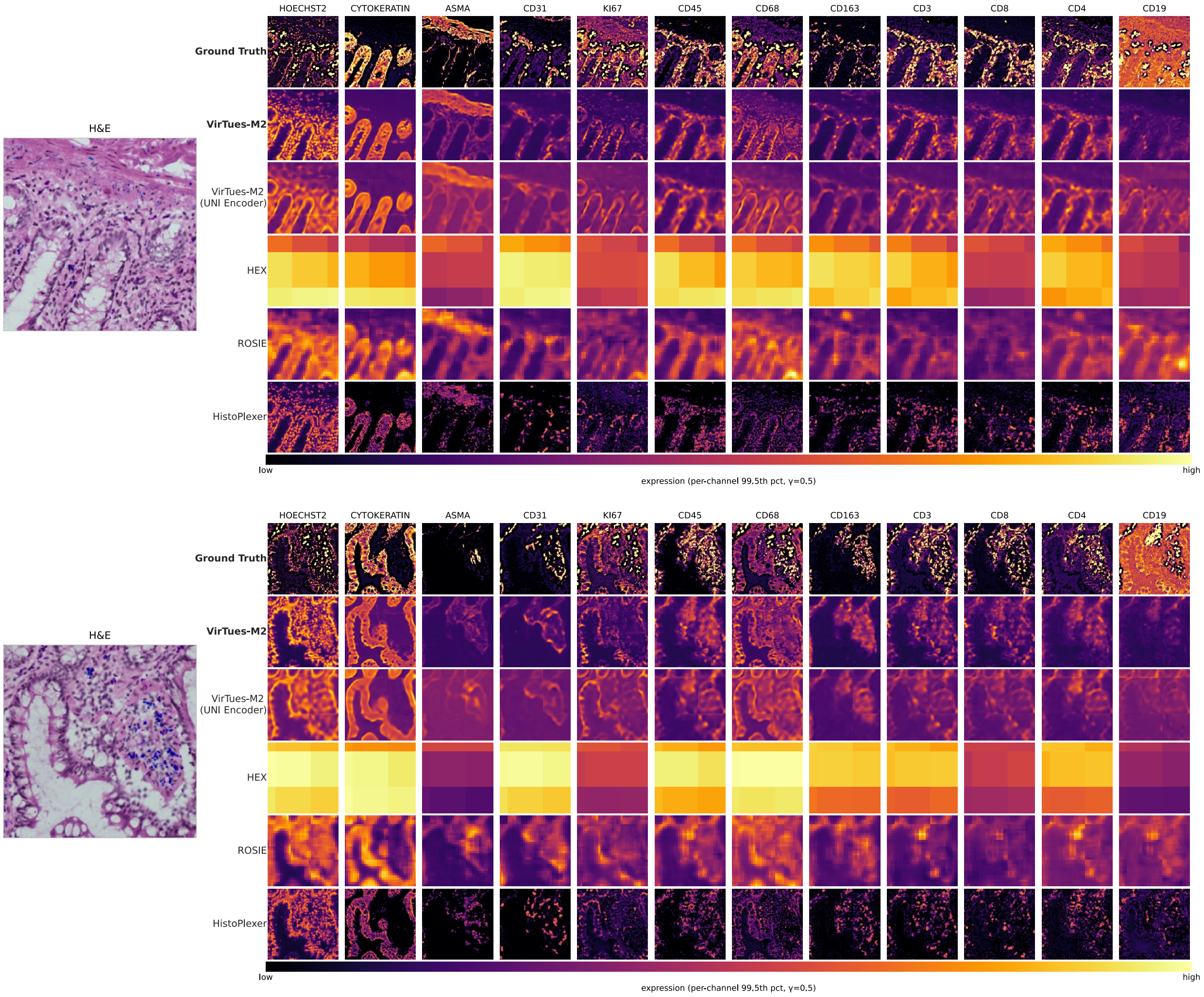}
    \caption{Qualitative virtual staining examples for two crop-size samples on 12 markers. We observe strong differences between the models. The results based on ROSIE and UNI are less sharp and in case of ROSIE often predict intensity at the wrong location. While HistoPlexer can produce sharp, pixel level virtual stainings, often the predicted location of marker intensity is wrong. We observe these errors mainly as the methods can not well generalize across datasets and organs. \new{The HEX predictions (with 50\% overlap during inference) show the crop based mean marker expressions.}
}\label{suppl:fig:he_inpainting}
\end{figure}

\begin{table}[H]
\centering
\caption{\new{Pearson correlation for H\&E-based virtual staining on the \texttt{hickey2023organization} cohort across 12 markers. Scores are means over tiles; $n$ is the tile count. Bold marks the best model per marker.}}
\small
\setlength{\tabcolsep}{5pt}
\begin{tabular}{lr@{\hskip 8pt}rrrrr}
\toprule
Protein & $n$ & VirTues-M2 & UNI & ROSIE & HistoPlexer & HEX \\
\midrule
\multicolumn{7}{l}{\textit{Easy}} \\
Hoechst & 4 & \textbf{0.923} & 0.866 & 0.697 & 0.881 & 0.425 \\
$\alpha$SMA & 4 & \textbf{0.577} & 0.527 & 0.350 & 0.325 & 0.174 \\
PanCK & 4 & \textbf{0.785} & 0.774 & 0.569 & 0.598 & 0.366 \\
\addlinespace[2pt]
\multicolumn{7}{l}{\textit{Moderate}} \\
CD31 & 4 & \textbf{0.411} & 0.391 & 0.252 & 0.159 & 0.127 \\
CD45 & 4 & \textbf{0.556} & 0.496 & 0.442 & 0.337 & 0.278 \\
CD68 & 4 & \textbf{0.495} & 0.441 & 0.332 & 0.262 & 0.293 \\
CD3 & 4 & \textbf{0.534} & 0.446 & 0.342 & 0.312 & 0.219 \\
CD19 & 4 & \textbf{0.212} & 0.140 & 0.130 & 0.160 & 0.027 \\
\addlinespace[2pt]
\multicolumn{7}{l}{\textit{Hard}} \\
CD4 & 4 & \textbf{0.430} & 0.368 & 0.315 & 0.229 & 0.128 \\
CD8 & 4 & \textbf{0.389} & 0.243 & 0.112 & 0.145 & 0.035 \\
CD163 & 4 & \textbf{0.432} & 0.311 & 0.224 & 0.171 & 0.108 \\
KI-67 & 4 & \textbf{0.381} & 0.301 & 0.206 & 0.204 & 0.025 \\
\midrule
\textbf{Mean} &  & \textbf{0.511} & 0.442 & 0.331 & 0.315 & 0.184 \\
\bottomrule
\end{tabular}
\label{suppl.tab:he_hickey2023organization_pearson}
\end{table}

\begin{table}[H]
\centering
\caption{\new{Spearman correlation for H\&E-based virtual staining on the \texttt{hickey2023organization} cohort across 12 markers. Scores are means over tiles; $n$ is the tile count. Bold marks the best model per marker.}}
\small
\setlength{\tabcolsep}{5pt}
\begin{tabular}{lr@{\hskip 8pt}rrrrr}
\toprule
Protein & $n$ & VirTues-M2 & UNI & ROSIE & HistoPlexer & HEX \\
\midrule
\multicolumn{7}{l}{\textit{Easy}} \\
Hoechst & 4 & \textbf{0.869} & 0.716 & 0.813 & 0.740 & 0.558 \\
$\alpha$SMA & 4 & \textbf{0.450} & 0.261 & 0.312 & 0.341 & 0.309 \\
PanCK & 4 & \textbf{0.637} & 0.221 & 0.535 & 0.482 & 0.480 \\
\addlinespace[2pt]
\multicolumn{7}{l}{\textit{Moderate}} \\
CD31 & 4 & 0.158 & 0.100 & 0.239 & 0.006 & \textbf{0.309} \\
CD45 & 4 & 0.494 & 0.285 & \textbf{0.495} & 0.292 & 0.401 \\
CD68 & 4 & \textbf{0.611} & 0.364 & 0.507 & 0.228 & 0.408 \\
CD3 & 4 & \textbf{0.315} & 0.123 & 0.270 & 0.122 & 0.243 \\
CD19 & 4 & 0.085 & -0.012 & -0.017 & \textbf{0.117} & 0.018 \\
\addlinespace[2pt]
\multicolumn{7}{l}{\textit{Hard}} \\
CD4 & 4 & 0.314 & 0.141 & \textbf{0.354} & 0.130 & 0.202 \\
CD8 & 4 & \textbf{0.223} & 0.024 & 0.047 & 0.001 & 0.002 \\
CD163 & 4 & 0.180 & 0.137 & 0.176 & 0.032 & \textbf{0.184} \\
KI-67 & 4 & \textbf{0.442} & 0.205 & 0.343 & 0.230 & 0.109 \\
\midrule
\textbf{Mean} &  & \textbf{0.398} & 0.214 & 0.340 & 0.227 & 0.269 \\
\bottomrule
\end{tabular}
\label{suppl.tab:he_hickey2023organization_spearman}
\end{table}

\begin{table}[H]
\centering
\caption{\new{SSIM for H\&E-based virtual staining on the \texttt{hickey2023organization} cohort across 12 markers. Scores are means over tiles; $n$ is the tile count. Bold marks the best model per marker.}}
\small
\setlength{\tabcolsep}{5pt}
\begin{tabular}{lr@{\hskip 8pt}rrrrr}
\toprule
Protein & $n$ & VirTues-M2 & UNI & ROSIE & HistoPlexer & HEX \\
\midrule
\multicolumn{7}{l}{\textit{Easy}} \\
Hoechst & 4 & \textbf{0.787} & 0.738 & 0.477 & 0.748 & 0.265 \\
$\alpha$SMA & 4 & 0.569 & 0.509 & 0.559 & \textbf{0.696} & 0.169 \\
PanCK & 4 & 0.682 & 0.608 & 0.462 & \textbf{0.733} & 0.403 \\
\addlinespace[2pt]
\multicolumn{7}{l}{\textit{Moderate}} \\
CD31 & 4 & 0.659 & 0.704 & 0.648 & \textbf{0.737} & 0.476 \\
CD45 & 4 & 0.614 & 0.648 & 0.605 & \textbf{0.725} & 0.235 \\
CD68 & 4 & \textbf{0.522} & 0.466 & 0.451 & 0.519 & 0.370 \\
CD3 & 4 & 0.539 & \textbf{0.641} & 0.418 & 0.550 & 0.582 \\
CD19 & 4 & 0.398 & 0.509 & 0.340 & 0.344 & \textbf{0.546} \\
\addlinespace[2pt]
\multicolumn{7}{l}{\textit{Hard}} \\
CD4 & 4 & 0.584 & \textbf{0.694} & 0.533 & 0.571 & 0.532 \\
CD8 & 4 & 0.444 & 0.338 & 0.458 & \textbf{0.611} & 0.195 \\
CD163 & 4 & 0.595 & 0.589 & 0.602 & \textbf{0.766} & 0.315 \\
KI-67 & 4 & 0.495 & \textbf{0.590} & 0.452 & 0.515 & 0.487 \\
\midrule
\textbf{Mean} &  & 0.574 & 0.586 & 0.501 & \textbf{0.626} & 0.381 \\
\bottomrule
\end{tabular}
\label{suppl.tab:he_hickey2023organization_ssim}
\end{table}

\begin{figure}[H]
    \centering
    \includegraphics[width=0.9\linewidth,page = 11]{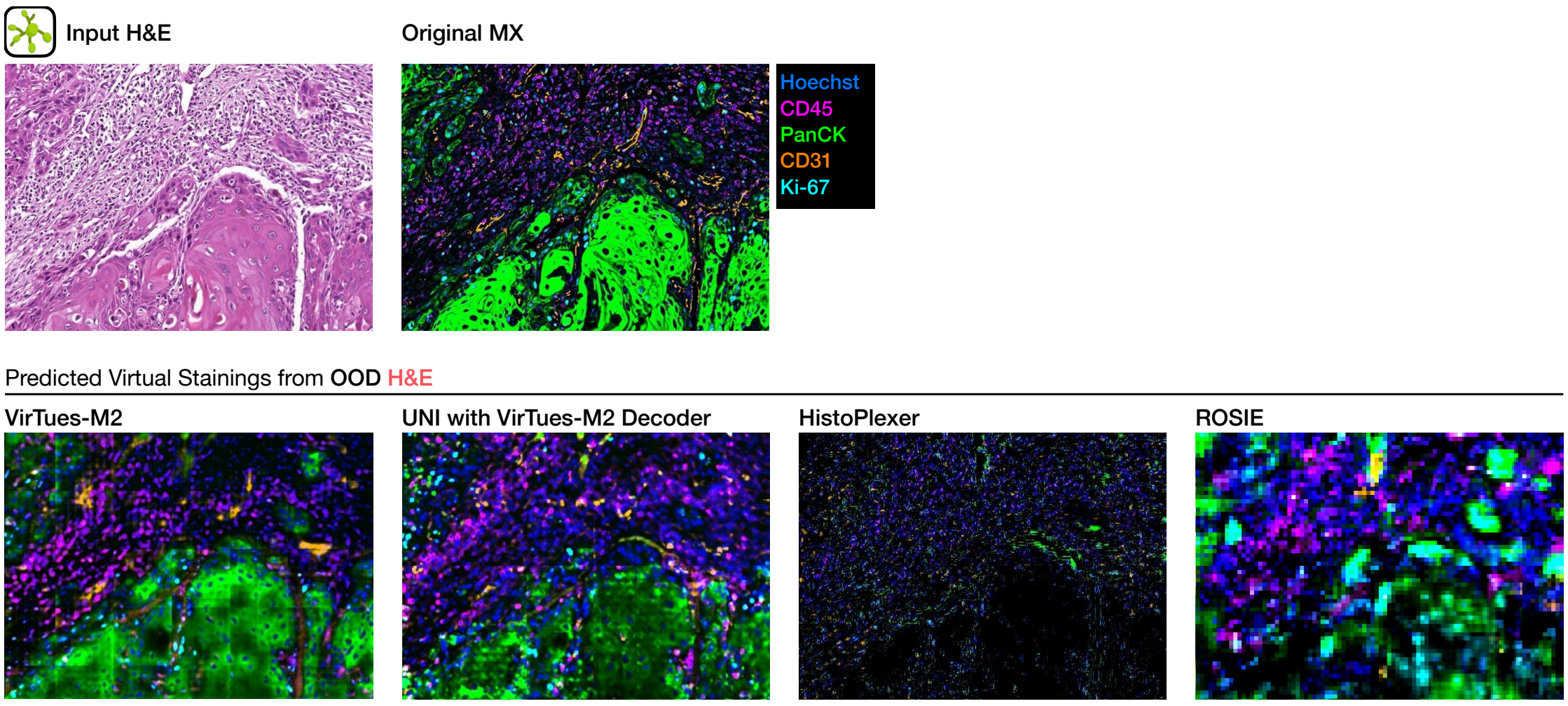}
    \caption{\new{Qualitative virtual staining on an out-of-distribution HNSCC lymphnode cohort. None of the models encountered HNSCC or lymphnode tissue during training, making this a strong test of generalization. based on qualitative evaluation, VirTues-M2 and VirTues-M2 Decoder with UNI as the backbone yield the most robust predictions on this cohort.}
}\label{suppl:fig:vs_he_ood_examples}
\end{figure}

\begin{table}[H]
\centering
\caption{\new{Pearson correlation for H\&E-based virtual staining on the OOD \texttt{haist2026lymph} cohort across 10 markers. Scores are means over tiles; $n$ is the tile count. Bold marks the best model per marker.}}
\small
\setlength{\tabcolsep}{5pt}
\begin{tabular}{lr@{\hskip 8pt}rrrrr}
\toprule
Protein & $n$ & VirTues-M2 & UNI & ROSIE & HistoPlexer & HEX \\
\midrule
\multicolumn{7}{l}{\textit{Easy}} \\
Hoechst & 86 & \textbf{0.793} & 0.708 & 0.446 & 0.658 & 0.047 \\
PanCK & 86 & \textbf{0.291} & 0.181 & 0.061 & 0.012 & 0.030 \\
\addlinespace[2pt]
\multicolumn{7}{l}{\textit{Moderate}} \\
CD31 & 86 & \textbf{0.214} & 0.112 & 0.065 & 0.017 & 0.018 \\
CD45 & 86 & \textbf{0.302} & 0.177 & 0.175 & 0.131 & 0.040 \\
CD68 & 86 & \textbf{0.155} & 0.076 & 0.076 & 0.029 & 0.012 \\
CD3 & 86 & \textbf{0.340} & 0.203 & 0.134 & 0.124 & 0.018 \\
\addlinespace[2pt]
\multicolumn{7}{l}{\textit{Hard}} \\
CD4 & 86 & \textbf{0.285} & 0.199 & 0.180 & 0.132 & 0.014 \\
CD8 & 86 & \textbf{0.125} & 0.055 & 0.061 & 0.051 & 0.001 \\
CD163 & 86 & \textbf{0.101} & 0.049 & 0.020 & 0.010 & 0.017 \\
KI-67 & 86 & \textbf{0.193} & 0.073 & 0.065 & 0.028 & 0.011 \\
\midrule
\textbf{Mean} &  & \textbf{0.280} & 0.183 & 0.128 & 0.119 & 0.021 \\
\bottomrule
\end{tabular}
\label{suppl.tab:he_haist2026lymph_pearson}
\end{table}

\begin{table}[H]
\centering
\caption{\new{Spearman correlation for H\&E-based virtual staining on the OOD \texttt{haist2026lymph} cohort across 10 markers. Scores are means over tiles; $n$ is the tile count. Bold marks the best model per marker.}}
\small
\setlength{\tabcolsep}{5pt}
\begin{tabular}{lr@{\hskip 8pt}rrrrr}
\toprule
Protein & $n$ & VirTues-M2 & UNI & ROSIE & HistoPlexer & HEX \\
\midrule
\multicolumn{7}{l}{\textit{Easy}} \\
Hoechst & 86 & \textbf{0.707} & 0.633 & 0.405 & 0.521 & 0.043 \\
PanCK & 86 & \textbf{0.268} & 0.216 & 0.054 & 0.024 & 0.028 \\
\addlinespace[2pt]
\multicolumn{7}{l}{\textit{Moderate}} \\
CD31 & 86 & \textbf{0.127} & 0.113 & 0.048 & 0.044 & 0.002 \\
CD45 & 86 & \textbf{0.283} & 0.210 & 0.167 & 0.084 & 0.033 \\
CD68 & 86 & \textbf{0.244} & 0.150 & 0.154 & 0.048 & 0.002 \\
CD3 & 86 & \textbf{0.262} & 0.204 & 0.101 & 0.094 & 0.005 \\
\addlinespace[2pt]
\multicolumn{7}{l}{\textit{Hard}} \\
CD4 & 86 & \textbf{0.252} & 0.191 & 0.208 & 0.115 & 0.007 \\
CD8 & 86 & \textbf{0.131} & 0.036 & 0.103 & 0.018 & 0.025 \\
CD163 & 86 & \textbf{0.069} & 0.064 & -0.020 & 0.031 & 0.005 \\
KI-67 & 86 & \textbf{0.135} & 0.018 & 0.048 & 0.014 & 0.002 \\
\midrule
\textbf{Mean} &  & \textbf{0.248} & 0.184 & 0.127 & 0.099 & 0.015 \\
\bottomrule
\end{tabular}
\label{suppl.tab:he_haist2026lymph_spearman}
\end{table}

\begin{table}[H]
\centering
\caption{\new{SSIM for H\&E-based virtual staining on the OOD \texttt{haist2026lymph} cohort across 10 markers. Scores are means over tiles; $n$ is the tile count. Bold marks the best model per marker.}}
\small
\setlength{\tabcolsep}{5pt}
\begin{tabular}{lr@{\hskip 8pt}rrrrr}
\toprule
Protein & $n$ & VirTues-M2 & UNI & ROSIE & HistoPlexer & HEX \\
\midrule
\multicolumn{7}{l}{\textit{Easy}} \\
Hoechst & 86 & \textbf{0.433} & 0.346 & 0.175 & 0.430 & 0.079 \\
PanCK & 86 & 0.326 & 0.346 & 0.322 & \textbf{0.367} & 0.162 \\
\addlinespace[2pt]
\multicolumn{7}{l}{\textit{Moderate}} \\
CD31 & 86 & 0.401 & 0.528 & 0.465 & \textbf{0.553} & 0.168 \\
CD45 & 86 & \textbf{0.305} & 0.299 & 0.191 & 0.283 & 0.237 \\
CD68 & 86 & 0.311 & \textbf{0.368} & 0.359 & 0.330 & 0.212 \\
CD3 & 86 & 0.248 & \textbf{0.323} & 0.304 & 0.248 & 0.203 \\
\addlinespace[2pt]
\multicolumn{7}{l}{\textit{Hard}} \\
CD4 & 86 & 0.252 & \textbf{0.288} & 0.219 & 0.128 & 0.182 \\
CD8 & 86 & 0.240 & 0.290 & \textbf{0.330} & 0.247 & 0.193 \\
CD163 & 86 & 0.251 & 0.367 & 0.267 & \textbf{0.373} & 0.172 \\
KI-67 & 86 & 0.350 & \textbf{0.353} & 0.310 & 0.176 & 0.274 \\
\midrule
\textbf{Mean} &  & 0.312 & \textbf{0.351} & 0.294 & 0.313 & 0.188 \\
\bottomrule
\end{tabular}
\label{suppl.tab:he_haist2026lymph_ssim}
\end{table}

\section{Biomarker discovery}
\begin{figure}[H]
    \centering
    \includegraphics[width=0.8\linewidth,page = 8]{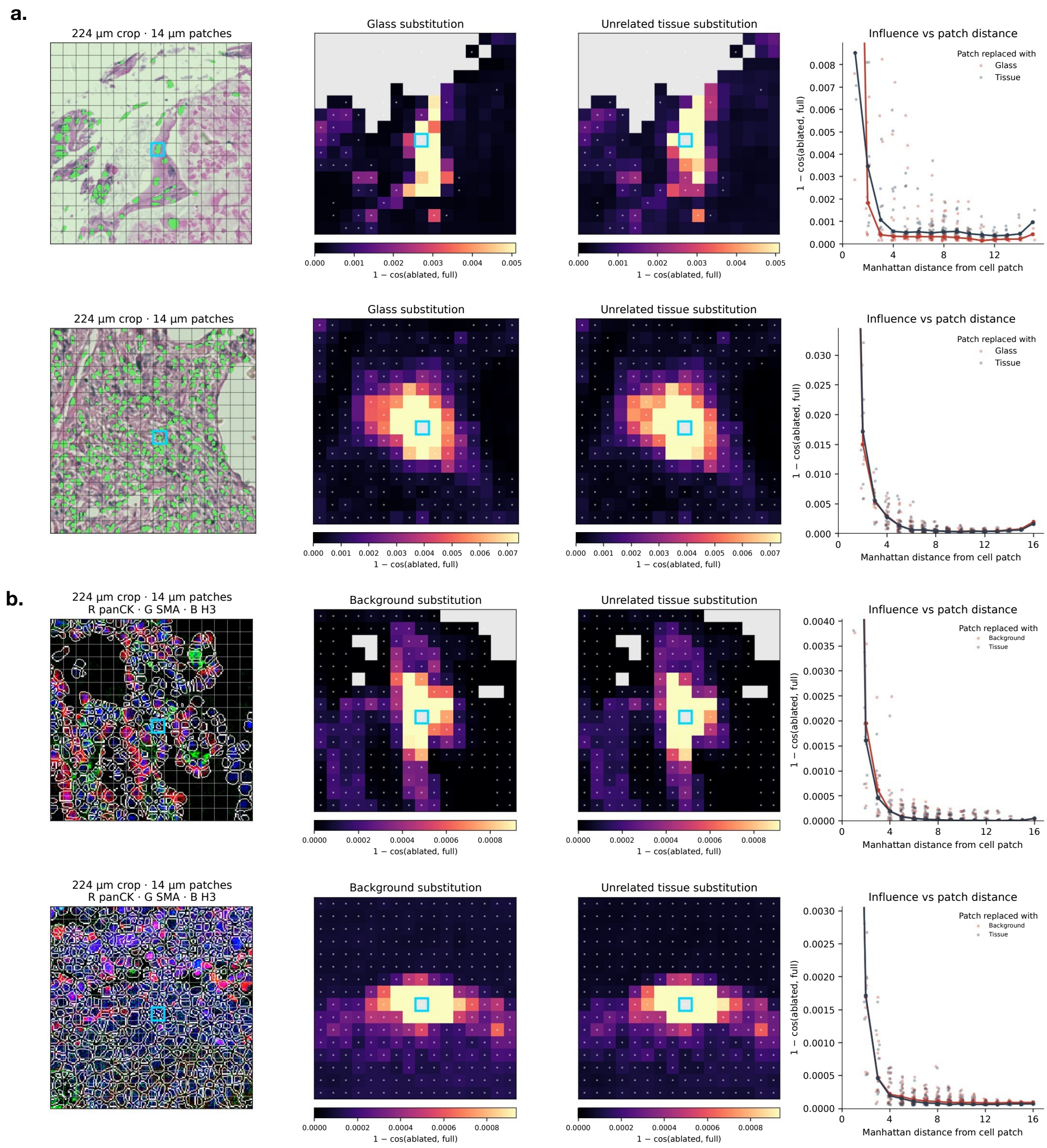}
    \caption{\new{\textbf{Cell tokens are shaped by their local tissue context.} Leave-one-patch-out ablation for two representative cells per modality, in \textbf{a} H\&E and \textbf{b} multiplex imaging. Left: 224 µm crop with the $16\times16$ patch grid and segmentation masks, target patch outlined in blue. Center: change in the cell token, $1 - \cos(\text{ablated}, \text{full})$, when each patch is replaced with background or with tissue from an unrelated core. Gray patches lie outside the tissue mask. Right: influence versus Manhattan distance from the target patch, points showing individual patches and lines the median.}
}\label{suppl:fig:cell_tokens}
\end{figure}
\vspace{-1.5cm}
\new{
\paragraph{Cell tokens encode local microenvironment.} 
VirTues-M2 is pretrained with a patch-level iBOT objective, which predicts masked patches in latent space, and a crop-level DINO objective over all patch tokens. Because the model is a vision transformer, every patch representation is computed with attention over all patches in the crop and is therefore contextual by construction. The masked-prediction objective further rewards patch representations that are predictive of their surroundings.
}

\new{
We exploit this to define cell tokens. Images are resampled such that a patch spans approximately the scale of a single cell ($14$µm at $1.0$µm/px), and for each segmented cell we extract a 224px crop positioned so that the cell centroid falls at the center of patch (8, 8) of the $16\times16$ grid. The representation of that patch, computed with attention over the full crop, is taken as the cell token. Since a patch is larger than a typical nucleus and is contextualized by the surrounding tissue, the resulting token encodes both the properties of the cell itself and those of its local microenvironment.
}

\new{
We verify this by leave-one-patch-out ablation in \autoref{suppl:fig:cell_tokens}. 
For a given cell token (blue patch), each of the remaining 255 patches is replaced in turn, either with background (glass for H\&E, off-tissue signal for multiplex) or with tissue drawn from an unrelated core, and the resulting change in the token is measured as $1 - \cos(\text{ablated}, \text{full})$. 
In both modalities the influence is concentrated on the patches immediately adjacent to the cell and decays to baseline within approximately five patches, corresponding to a radius of roughly $55$µm. 
The two substitution types produce near-identical decay profiles, indicating that the effect reflects the removal of tissue context rather than the particular fill value.
}

\begin{figure}[H]
    \centering
    \includegraphics[width=\linewidth,page = 12]{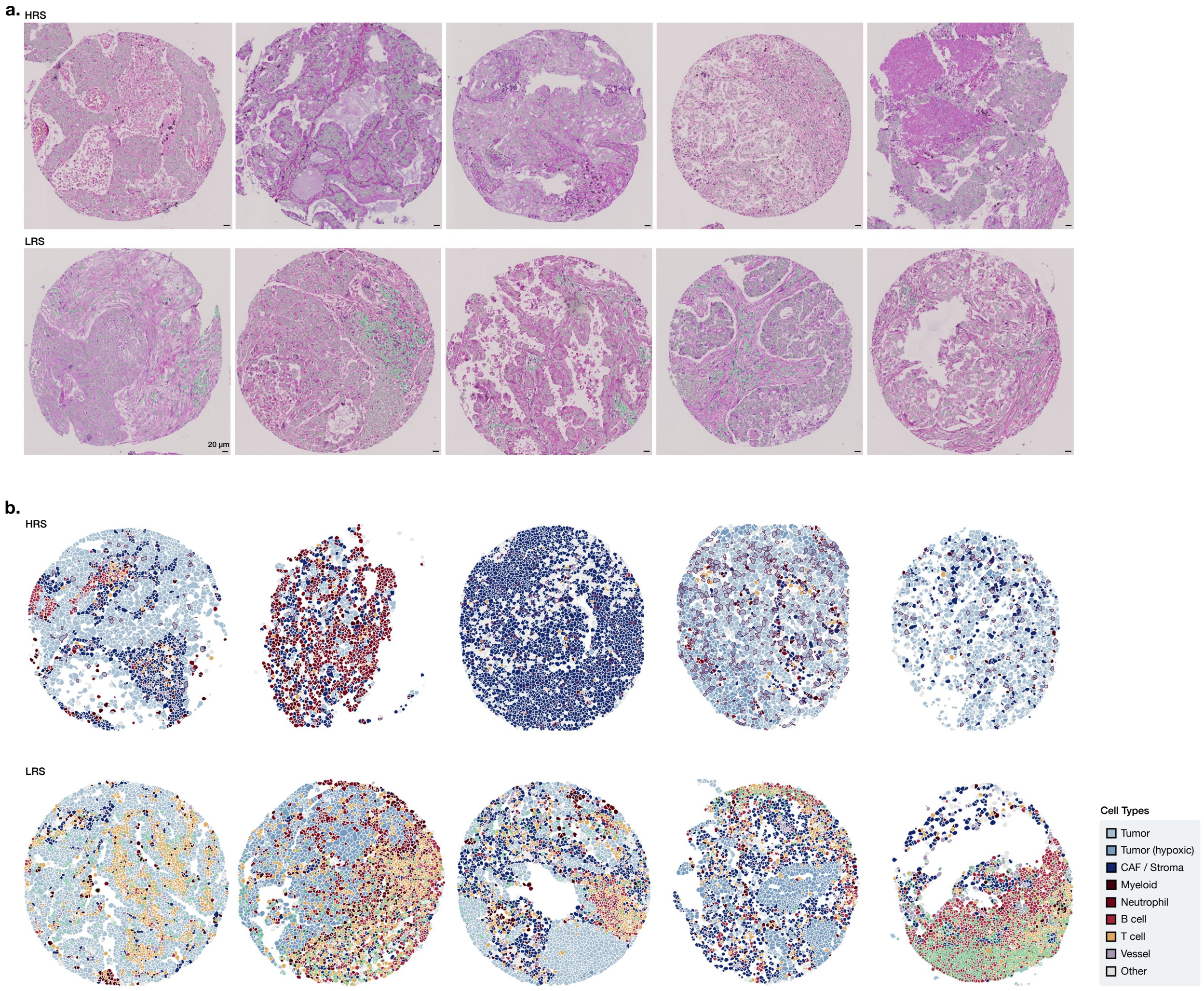}
    \caption{\textbf{Exemplary tumor cores of high- and low-risk signatures.} Five representative cores per signature. \textbf{a}, H\&E cores with signature cells outlined in color and all other cells in gray. \textbf{b}, multiplex cores with cell masks colored by annotated cell type and signature cells outlined. Low-risk cores contain dense lymphoid regions, whereas high-risk cores are dominated by tumor and stroma.
}\label{suppl:fig:biomarker_discovery}
\end{figure}

\begin{figure}[H]
    \centering
    \includegraphics[width=\linewidth,page = 13]{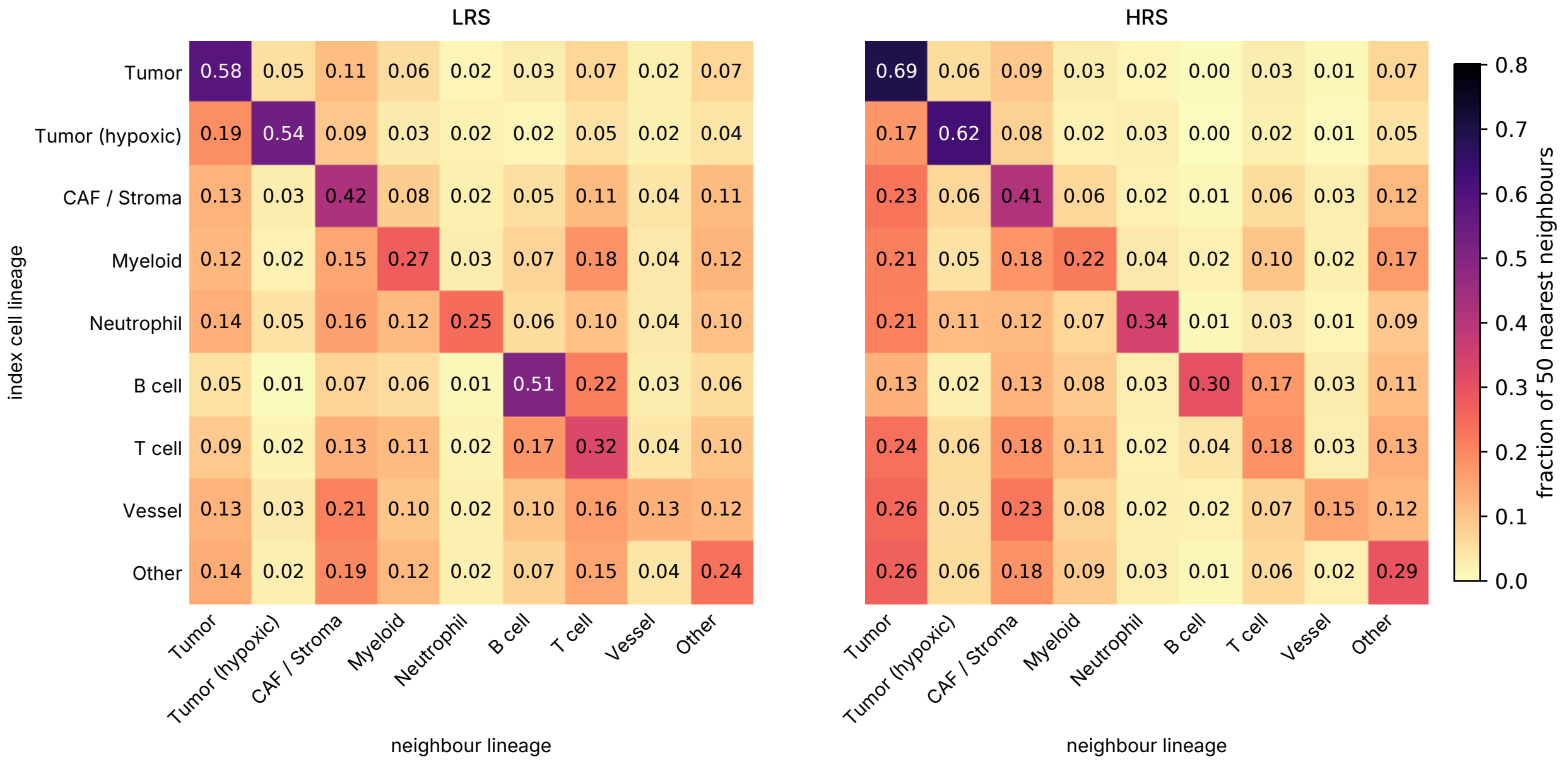}
    \caption{\new{\textbf{Absolute neighborhood composition of the low- and high-risk signatures.} Mean fraction of the 50 nearest neighbors belonging to each cell type (columns) for cells of each lineage (rows), in LRS (left) and HRS (right). Rows sum to one. Ratios in \autoref{fig:biomarker}f involving rare contacts (i.e., low absolute values in both plots) should be interpreted with care.}
}\label{suppl:fig:biomarker_cell_neighbors}
\end{figure}

\section{VirTues-M2 datasets overview}

\begin{sidewaystable}[!htbp]
\centering
\scriptsize
\setlength{\tabcolsep}{2.2pt}
\renewcommand{\arraystretch}{1.2}
\caption{Overview of cohorts included in the VirTues-M2 pretraining. mx = multiplex spatial proteomics images. Alignment indicates whether MX and H\&E modalities are registered at the pixel level or unregistered but from matched samples. Specimen type: WSI = whole-slide image, TC = tumor core.}
\label{suppl.tab:dataset_overview}

\newcolumntype{L}[1]{>{\raggedright\arraybackslash}p{#1}}

\begin{adjustbox}{max width=\textheight, max totalheight=\textwidth, keepaspectratio}
\begin{tabular}{@{}%
L{5.5cm}  
L{2.2cm}  
L{1.8cm}  
r         
L{0.8cm}  
L{2.35cm} 
L{2.45cm} 
r         
r         
r         
r         
L{2.5cm}  
@{}}

\toprule
\textbf{Dataset} &
\textbf{Modalities} &
\textbf{Alignment} &
\textbf{\# Patients} &
\textbf{\makecell[l]{Image\\Type}} &
\textbf{Organ} &
\textbf{\makecell[l]{Primary\\Cancer}} &
\textbf{\makecell{\#MX\\imgs}} &
\textbf{\makecell{\#MX\\crops}} &
\textbf{\makecell{\#H\&E\\imgs}} &
\textbf{\makecell{\#H\&E\\crops}} &
\textbf{License} \\
\midrule
\texttt{allam2022spatially} \citep{allam2022spatially} & IMC & - & 26 & TC & Lung, Lymphnode & Lung & 26 & 619 & 0 & 0 & CC BY 4.0 \\
\texttt{cho2025modeling} \citep{cho2025modeling} & IMC & - & 12 & TC & Pancreas, Liver & Pancreatic & 58 & 1,085 & 0 & 0 & CC BY 4.0 \\
\texttt{cords2023cancer} \citep{cords2023cancer} & IMC & - & 12 & TC & Breast & Breast & 109 & 1,882 & 0 & 0 & CC BY 4.0 \\
\texttt{cords2024cancer} \citep{cords2024cancer} & IMC, H\&E & Matched & 1,361 & TC & Lung & Lung & 2,040 & 21,251 & 2,328 & 26,507 & CC BY 4.0 \\
\texttt{damond2019map} \citep{damond2019map} & IMC & - & 12 & TC & Pancreas & Other disease & 845 & 4,308 & 0 & 0 & CC BY 4.0 \\
\texttt{danenberg2022breast} \citep{danenberg2022breast} & IMC & - & 718 & TC & Breast & Breast & 794 & 5,799 & 0 & 0 & CC BY 4.0 \\
\texttt{ehret2025shaking} \citep{ehret2025shaking} & IMC & - & 55 & TC & Colon & Colorectal & 108 & 1,656 & 0 & 0 & CC BY 4.0 \\
\texttt{fischer2023multiplex} \citep{fischer2023multiplex} & IMC & - & 892 & TC & Breast & Breast & 1,252 & 11,247 & 0 & 0 & CC BY 4.0 \\
\texttt{greenwald2026temporal} \citep{greenwald2026temporal} & MIBI & - & 113 & TC & Breast & Breast & 1,258 & 33,273 & 0 & 0 & CC SA 4.0 \\
\texttt{gupta2026single} \citep{gupta2026single} & IMC & - & 527 & TC & Breast & No disease & 1,492 & 47,407 & 0 & 0 & CC BY 4.0 \\
\texttt{hickey2023organization} \citep{hickey2023organization} & PhenoCycler, H\&E & Registered & 8 & WSI & Intestine & Other disease & 64 & 9,011 & 19 & 2,076 & CC 0 1.0\\
\texttt{hoch2022multiplexed} \citep{hoch2022multiplexed} & IMC & - & 35 & TC & Skin, Lymphnode & Skin (melanoma) & 50 & 858 & 0 & 0 & CC BY 4.0 \\
\texttt{hoefflin2026spatial} \citep{hoefflin2026spatial} & PhenoCycler, H\&E & Registered & 25 & WSI & Brain & Brain & 30 & 22,364 & 30 & 22,364 & CC BY 4.0 \\
\texttt{hu2023single} \citep{hu2023single} & IMC & - & 12 & TC & Lung, Head \& Neck & Lung & 51 & 1,226 & 0 & 0 & CC BY 4.0 \\
\texttt{hubmap2019human} \citep{hubmap2019human} & PhenoCycler & - & 15 & WSI & Spleen, Lymphnode & Other disease & 80 & 17,553 & 0 & 0 & CC BY 4.0 \\
\texttt{jackson2020single} \citep{jackson2020single} & IMC & - & 285 & TC & Breast & Breast & 723 & 5,722 & 0 & 0 & CC BY 4.0 \\
\texttt{lin2023high} \citep{lin2023high} & H\&E, CYCIF (Orion) & Registered & 40 & WSI & Colon & Colorectal & 41 & 153,807 & 41 & 153,807 & CC BY 4.0 \\
\texttt{lin2023multiplexed} \citep{lin2023multiplexed} & PhenoCycler & - & 6 & WSI & Colon, Head and Neck & Colorectal & 13 & 8,074 & 0 & 0 & CC BY 4.0 \\
\texttt{liu2022reproducible} \citep{liu2022reproducible} & MIBI & - & 28 & TC & Lymphnode, Tonsil, Soft Tissue, Bladder, Breast, Thymus, Spleen, Colon & Breast, Colorectal & 164 & 738 & 0 & 0 & CC BY 4.0 \\
\texttt{malchiodi2024natural} \citep{malchiodi2024natural} & IMC & - & 63 & TC & Pancreas & Pancreatic, Other disease & 164 & 2,817 & 0 & 0 & CC BY 4.0 \\
\texttt{meyer2025stratification} \citep{meyer2025stratification} & IMC & - & 215 & TC & Breast & Breast & 450 & 4,571 & 0 & 0 & CC BY 4.0 \\
\texttt{moldoveanu2022spatially} \citep{moldoveanu2022spatially} & IMC & - & 90 & TC & Lymphnode, Skin & Skin (melanoma) & 90 & 1,466 & 0 & 0 & CC BY 4.0 \\
\texttt{monkman2024spatial} \citep{monkman2024spatial} & PhenoCycler & - & 42 & TC & Lung & Lung & 42 & 795 & 0 & 0 & CC BY 4.0 \\
\texttt{phillips2021immune} \citep{phillips2021immune} & PhenoCycler, H\&E & Registered & 14 & TC & Skin & Lymphoma & 70 & 523 & 70 & 523 & CC BY 4.0 \\
\texttt{rigamonti2024integrating} \citep{rigamonti2024integrating} & H\&E, IMC & Matched & 108 & TC & Lung & Lung & 158 & 4,228 & 210 & 4,614 & Access by request \\
\texttt{ruf2023tumor} \citep{ruf2023tumor} & PhenoCycler & - & 17 & WSI & Liver & Liver & 17 & 16,104 & 0 & 0 & CC BY 4.0 \\
\texttt{salie2025spatial} \citep{salie2025spatial} & IMC & - & 96 & WSI & Liver & Liver & 194 & 5,355 & 0 & 0 & CC BY 4.0 \\
\texttt{schulz2018simultaneous} \citep{schulz2018simultaneous} & IMC & - & 79 & TC & Breast & Breast & 79 & 612 & 0 & 0 & CC BY 4.0 \\
\texttt{schulz2024immucan} \citep{eling2025multi} & IMC & - & 179 & TC & Lung, Head \& Neck & Colorectal, Lung, Head \& Neck, Breast & 179 & 1,583 & 0 & 0 & CC BY 4.0 \\
\texttt{schurch2020coordinated} \citep{schurch2020coordinated} & H\&E, PhenoCycler & Registered & 35 & TC & Colon, Other & Colorectal & 251 & 1,699 & 250 & 1,692 & CC BY 4.0 \\
\texttt{shapira2022understanding} \citep{shapira2022understanding} & IMC, PhenoCycler & - & 186 & WSI & Pancreas & Other disease & 945 & 480,931 & 0 & 0 & Access by request \\
\texttt{tupro2025aml} \citep{wegmann2024single} & IMC & - & 14 & WSI & Bone Marrow & Myeloid & 178 & 1,901 & 0 & 0 & TuPro Consortium \\
\texttt{tupro2025melanoma} \citep{miglino2025feasibility} & IMC & - & 116 & WSI & Brain, Lymphnode, Other, Liver, Soft Tissue & Skin (melanoma) & 1,314 & 16,869 & 0 & 0 & TuPro Consortium \\
\texttt{tupro2025ovarian} \citep{jacob2026tumor} & H\&E, IMC & - & 46 & WSI & Ovaries & Ovarian & 284 & 4,802 & 0 & 0 & TuPro Consortium \\
\texttt{wang2023spatial} \citep{wang2023spatial} & IMC & - & 279 & TC & Breast & Breast & 1,855 & 14,794 & 0 & 0 & Access by request \\
\texttt{xiao2022multiplexed} \citep{xiao2022multiplexed} & IMC & - & 26 & WSI & Lymphnode, Soft Tissue, Other & Skin (melanoma) & 158 & 1,585 & 0 & 0 & CC BY 4.0 \\
\texttt{zhuhuman2025spatial} \citep{zhu2025spatial} & IMC & - & 62 & TC & Lung & Lung & 1,649 & 24,387 & 0 & 0 & CC BY 4.0 \\
\hline \\
\textbf{Total} & & & 5,849 & & & & 17,275 & 932,902 & 2,948 & 211,583 & \\

\bottomrule
\end{tabular}
\end{adjustbox}
\end{sidewaystable}
\begin{table}[H]
\centering
\caption{Overview of the curated H\&E-only dataset constructed using a hierarchical clustering.}
\label{suppl.tab:dataset_overview_he}
\begin{tabular}{l| r r r r}
\hline
Dataset & Patients & \#WSIs & \#Crops pre-curation & \#Crops post-curation \\
\hline
GTEx        & 949   & 9,645 & 15,000,436 & 1,724,380    \\     
TCGA-BRCA   & 1,055 & 1,126 & 2,814,584  & 188,296     \\     
TCGA-KIRC   & 513   & 519   & 1,815,869  & 150,038     \\     
TCGA-LUSC   & 478   & 512   & 1,576,557  & 67,335      \\      
TCGA-LUAD   & 468   & 531   & 1,622,014  & 104,264     \\     
TCGA-COAD   & 433   & 441   & 901,257   & 105,240     \\     
TCGA-SKCM   & 433   & 475   & 1,446,136  & 199,004     \\     
TCGA-STAD   & 375   & 400   & 1,191,710  & 79,568      \\      
TCGA-KIRP   & 273   & 297   & 939,600   & 137,220     \\     
TCGA-PAAD   & 182   & 208   & 662,273   & 77,680      \\      
TCGA-READ   & 156   & 157   & 253,185   & 39,258      \\      
TCGA-KICH   & 109   & 121   & 436,962   & 63,490      \\      
TCGA-MESO   & 75    & 87    & 225,082   & 9,670       \\       
TCGA-CHOL   & 39    & 39    & 195,945   & 4,557       \\       

\hline
\hline
Total: & 5,538 & 14,558 & 29,081,610 & 2,950,000
\end{tabular}
\end{table}

\section{VirTues-M2 hyperparameter}

\begin{table}[H]
\centering
\caption{Hyperparameters used for VirTues-M2 and the multi-modal DINOv2 pretraining.}
\label{suppl.tab:dinov2_hparams}
\begin{tabular}{l|l}
\hline
Hyperparameter & Value \\
\hline
Mx tok. layers              & 2         \\
Mx tok. heads               & 8         \\
Mx tok. embedding dim.      & 1024      \\
Mx tok. feedforward dim.    & 2048       \\
\hline
Layers                       & 24        \\
Heads                        & 16        \\
Patch size                   & 14        \\
FFN layer                    & SwiGLU       \\
Head activation              & SwiGLU      \\
Embedding dimension          & 1024      \\
Feedforward dimension          & 5472      \\
Stochastic dropout rate      & 0.1       \\
Number of register tokens      & 4       \\
\hline
Global crop scale            & 0.48, 1.0 \\
Global crop number \& size   & 2, 224    \\
Local crop scale             & 0.16, 0.48 \\
Local crop number \& size    & 8, 98     \\
Min masking ratio            & 0.1       \\
Max masking ratio            & 0.5       \\
Gradient clipping max norm   & 3.0       \\
Normalize last layer         & \ding{51} \\
Shared head                  & \ding{51}     \\
\hline
Multiplex subset local crops & $\left[0.75, 0.9\right]$ \\
Channel subset global crops & $\left[0.75, 0.9\right]$ \\
MM local view dropout       & 0.7       \\
MM min masking ratio        & 0.6       \\
MM max masking ratio         & 0.9       \\
Dataset balancing $\alpha$ & 0.25   \\
\hline
AdamW $\beta$                & (0.9, 0.999) \\
Effective batch size                   & 1664      \\
Freeze last layer iterations & 1250      \\
Warmup iterations            & 12500     \\
Warmup teacher temperature iterations & 37500 \\
High-resolution finetuning iterations & 12500 \\
Max Iterations               & 125000    \\
Learning rate schedule       & Cosine    \\
Learning rate (start)        & 0         \\
Learning rate (post warmup)  & 0.0053      \\
Learning rate (final)        & 1e-6      \\
Teacher temperature (start)  & 0.04      \\
Teacher temperature (final)  & 0.4       \\
Teacher momentum (start)     & 0.992     \\
Teacher momentum (final)     & 1.000     \\
Weight decay (start)         & 0.04      \\
Weight decay (end)           & 0.4       \\
Automatic mixed precision    & FP16      \\
\hline
\end{tabular}
\end{table}

\section{Related literature} 

\paragraph{Foundation models for histopathology.}
Self-supervised learning on large-scale histopathology datasets has produced foundation models that capture generalizable morphological features from H\&E-stained tissue. 
UNI \citep{chen2024uni} introduced the first large-scale general-purpose foundation model for computational pathology, training a Vision Transformer on 100 million patches from over 100,000 diagnostic WSIs using DINOv2 \citep{oquab2023dinov2}, with UNI2 subsequently expanding to 350,000 slides including immunohistochemistry. 
Virchow \citep{vorontsov2024foundation} achieved clinical-grade performance by training on 1.5 million H\&E slides, demonstrating pan-cancer detection capabilities. 
Prov-GigaPath \citep{xu2024whole} was the first WSI foundation model, processing 1.3 billion tiles using a two-stage curriculum combining DINOv2 for tile-level pretraining with a LongNet-based \citep{ding2023longnet} masked autoencoder for slide-level context. 
Additional models including CTransPath \citep{wang2022transformer}, Phikon \citep{filiot2023scaling}, and Hibou \citep{nechaev2024hibou} have contributed open-source alternatives with competitive performance. 
These models have become standard feature extractors for computational pathology, yet they remain fundamentally unimodal-learning from morphology alone without direct integration of molecular measurements.

\paragraph{Foundation models for spatial proteomics.}
Foundation models for multiplex imaging must address heterogeneous marker panels across technologies and smaller training datasets compared to histopathology. 
VirTues \citep{wenckstern2026virtual} introduced marker-aware tokenization using protein language model embeddings, enabling pretraining across diverse multiplex panels without assuming fixed channel orderings. 
The model generates hierarchical representations spanning patch, cell, niche, and tissue scales. 
CA-MAE \citep{kraus2024masked} introduced channel-agnostic masked autoencoders for microscopy, trained on 93 million images, demonstrating scaling laws for microscopy foundation models. 
KRONOS \citep{shaban2025foundation} adopted a DINOv2-based approach on 47 million patches from 175 protein markers.

\paragraph{Multimodal learning in computational pathology.}
Recent work has begun exploring multimodal integration, though predominantly through late fusion or vision-language alignment. 
MUSK \citep{xiang2025vision} introduced unified masked modeling for multimodal pathology, training on 50 million images and 1 billion text tokens using a mixture-of-modality-experts architecture that leverages unpaired data. 
CONCH \citep{lu2024visual} trained on 1.17 million pathology image-caption pairs using contrastive captioning, achieving strong zero-shot performance including rare disease subtyping. 
TANGLE \cite{jaume2024transcriptomics} was the first method for slide representation learning using transcriptomics, aligning H\&E slides with bulk RNA-seq profiles. 
BiomedCLIP \citep{zhang2025multimodal} provides broader biomedical coverage trained on 15 million PubMed image-text pairs. 
These approaches demonstrate the value of multimodal information but do not enable the cross-modal structure learning that emerges from joint pretraining of histology and spatial proteomics within a unified architecture.
In addition, a separate line of work combines H\&E with \textit{predicted} molecular readouts (e.g., virtual stains) using light cross-attention or late fusion, typically as task-tuned pipelines rather than as general-purpose multimodal foundation models \citep{li2026ai,valanarasu2025multimodal}.

\paragraph{Virtual staining and cross-modal prediction.}
Virtual staining, i.e., predicting molecular marker expression from H\&E, offers a route to expand information content of routine slides. 
ROSIE \citep{wu2025rosie} achieved prediction of 50 protein biomarkers from H\&E using a ConvNeXt architecture trained on over 16 million cells, demonstrating cross-platform generalization. 
HEX \citep{li2026ai} targets virtual staining for a fixed 40-plex panel by extracting features from H\&E tiles with a pretrained pathology foundation model backbone (e.g., MUSK) and using a lightweight MLP regression head to jointly predict the expression of all 40 markers. As a second step, \citet{li2026ai} introduce a multimodal integration module with separate unimodal encoders for H\&E and (virtual) spatial proteomics, coupled via co-attention layers and fine-tuned for downstream tasks.
Similarly, GigaTIME \citep{valanarasu2025multimodal} learns a cross-modal translator from paired H\&E and mIF to generate virtual mIF images for a fixed 21-marker panel, and uses the resulting virtual protein maps to analyze associations with clinical biomarkers, cancer stage, and survival.
HistoPlexer \citep{andani2025histopathology} predicts 11 IMC markers simultaneously using conditional GANs, improving survival prediction and immune subtype classification. 
DeepLIIF \citep{ghahremani2022deep} provides simultaneous stain deconvolution, segmentation, and scoring. 
VirtualMultiplexer \citep{pati2024accelerating} introduced unpaired translation eliminating pixel-wise registration requirements. 
These approaches require fixed marker sets and separate models, whereas foundation model embeddings offer more flexible prediction as a downstream capability.
Because these methods impute a \textit{fixed, predefined} marker panel and are typically optimized for specific downstream endpoints, they are best viewed as cross-modal prediction or task-specialized pipelines rather than multimodal foundation models that learn a shared representation through joint pretraining on both modalities.

\paragraph{Cell segmentation and phenotyping.}
Instance segmentation and phenotyping are typically addressed through staged pipelines. 
Cellpose \citep{stringer2021cellpose} introduced gradient flow representations and became the standard for general cell segmentation with human-in-the-loop fine-tuning. 
HoVer-Net \citep{graham2019hover} introduced horizontal-vertical distance maps enabling simultaneous segmentation and classification. 
StarDist \citep{schmidt2018} models cells as star-convex polygons for computational efficiency. 
Mesmer \citep{greenwald2022whole} achieved human-level whole-cell segmentation using TissueNet with over 1 million labeled cells. 
CellViT \citep{horst2024cellvit} applied Vision Transformers to nuclear segmentation, demonstrating foundation model transferability. 
These staged approaches can propagate errors and do not exploit complementarity between morphological and molecular signals for joint inference.

\paragraph{Spatial proteomics technologies.}
Multiplexed spatial proteomics technologies enable in situ quantification of dozens of proteins while preserving tissue architecture. 
Iterative multiplexed immunofluorescence (mIF) methods achieve high-plex imaging through repeated staining and imaging cycles: CODEX (now PhenoCycler) \citep{black2021codex} uses DNA-barcoded antibodies with fluorescent oligonucleotide hybridization; Orion \citep{lin2023high} employs staining with fluorophore inactivation; and cyclic immunofluorescence provides an accessible public-domain alternative using standard fluorescence microscopy; all achieving 50–60+ markers at subcellular resolution on FFPE tissues. 
Mass spectrometry-based approaches offer complementary capabilities: imaging mass cytometry (IMC) \citep{giesen2014highly} couples immunohistochemistry with laser ablation and CyTOF detection of 40+ metal-conjugated antibodies at ${\sim}1 \mu m$ resolution, while MIBI \citep{angelo2014multiplexed} and MIBI-TOF \citep{keren2019mibi} use secondary ion mass spectrometry for up to 100 markers with exceptional sensitivity and dynamic range.

\paragraph{Protein language models.}
Protein language models provide semantic embeddings enabling marker-aware tokenization across heterogeneous antibody panels. 
ESM-2 \citep{lin2023evolutionary} demonstrated emergence of structural information at scale, with the 650M parameter model providing embeddings that capture biochemical properties and enable downstream applications. 
ProtTrans \citep{elnaggar2022prottrans} systematically compared transformer architectures for proteins. 
These embeddings enable VirTues-M2 to generalize across different marker configurations by encoding marker identity through biological meaning rather than channel position.

\end{document}